\documentclass[10pt]{IEEEtran}

\usepackage{amssymb}
\usepackage[cmex10]{amsmath}
\usepackage{graphicx}
\usepackage{cite}
\usepackage{cases}
\usepackage{color}
\usepackage{enumerate}

\makeatletter
\let\MYcaption\@makecaption
\makeatother

\usepackage[font=footnotesize]{subcaption}

\makeatletter
\let\@makecaption\MYcaption
\makeatother

\usepackage{glossaries}
\usepackage{todonotes}
\usepackage{enumitem}

\usepackage{tikz}
\usetikzlibrary{dsp,chains}
\usepackage[utf8]{inputenc}
\usepackage{pgfplots} 
\usepackage{pgfgantt}
\usepackage{pdflscape}
\pgfplotsset{compat=newest} 
\pgfplotsset{plot coordinates/math parser=false}

\graphicspath{{figures/}}

\newacronym{ild}{ILD}{interaural level difference}%
\newacronym{itd}{ITD}{interaural time difference}%
\newacronym{hrtf}{HRTF}{head related transfer function}%
\newacronym{wfs}{WFS}{wave-field synthesis}%
\newacronym{hoa}{HOA}{higher-order ambisonics}%
\newacronym{oct}{OCT}{optimized cardioid triangle}%
\newacronym{psr}{PSR}{perceptual sound-field reconstruction}%
\newacronym{vbap}{VBAP}{vector-base amplitude panning}%
\newacronym{idir}{ID}{intensity directivity}%
\newacronym{ictd}{ICTD}{inter-channel time difference}%
\newacronym{icld}{ICLD}{inter-channel level difference}%
\newacronym{rictd}{RICTD}{relative inter-channel time difference}%
\newacronym{ricld}{RICLD}{relative inter-channel level difference}%
\newacronym{tild}{TILD}{time-intensity linear directivity}%
\newacronym{tid}{TID}{time-intensity directivity}%
\newacronym{erb}{ERB}{equivalent rectangular bandwidth}%
\newacronym{wng}{WNG}{white noise gain}%
\newacronym{snr}{SNR}{signal to noise ratio}%
\newacronym{rir}{RIR}{room impulse response}%
\newacronym{sdn}{SDN}{scattering delay network}%
\newacronym{dwm}{DWM}{digital waveguide mesh}%
\newacronym{dwn}{DWN}{digital waveguide network}%
\newacronym{fdn}{FDN}{feedback delay network}%
\newacronym{ism}{ISM}{image-source method}%
\newacronym{ml}{ML}{maximum likelihood}

\newacronym{ortf}{ORTF}{office de radiodiffusion t\'el\'evision fran\c{c}aise}%z

\newcommand{\ild}{\text{ILD}}
\newcommand{\itd}{\text{ITD}}
\newcommand{\fild}{\text{FILD}}
\newcommand{\fitd}{\text{FITD}}
\newcommand{\nild}{\overline{\text{ILD}}}
\newcommand{\nitd}{\overline{\text{ITD}}}
\newcommand{\nfild}{\overline{\text{FILD}}}
\newcommand{\nfitd}{\overline{\text{FITD}}}

\newcommand{\icld}{\text{ICLD}}
\newcommand{\ictd}{\text{ICTD}}
\newcommand{\ricld}{\text{RICLD}}
\newcommand{\rictd}{\text{RICTD}}

\begin{document}

\definecolor{skin}{RGB}{248,198,135}
\definecolor{mygreen}{RGB}{66,213,0}

\newcounter{speaker}

\tikzset{
  PlaneWave/.pic={
  	\draw (-1.5,0) -- (1.5,0);
  	\draw (-1.5,0.4) -- (1.5,0.4);
  	\draw (-1.5,0.8) -- (1.5,0.8);
  	\draw[thick,->] (0, 1.4) -- (0,-0.5);
  },
  Mic/.pic={
    \filldraw[fill=gray!40,pic actions] 
    	(-0.15,0) rectangle (0.15,-0.8);
    \filldraw[fill=gray] 
      (0,0) ellipse [x radius=0.25cm, y radius=0.25cm];
  },
  Speaker/.pic={
    \filldraw[fill=gray!40,pic actions] 
    (-15pt,0) -- 
      coordinate[midway] (-front) 
    (15pt,0) -- 
    ++([shift={(-6pt,8pt)}]0pt,0pt) coordinate (aux1) -- 
    ++(-18pt,0) coordinate (aux2) 
    -- cycle 
    (aux1) -- ++(0,6pt) -- coordinate[midway] (-back) ++(-18pt,0) -- (aux2);
  },
  Human/.pic={
    \filldraw[fill=skin] 
      (0,0.7cm) ellipse [x radius=7pt,y radius=3pt]
      (0,-0.7cm) ellipse [x radius=7pt,y radius=3pt];    
    \filldraw[fill=skin] 
      (0.7,4pt) -- (0.7,-4pt) -- (0.95,0pt) -- cycle;    
    \filldraw[fill=skin] 
      (0,0) ellipse [x radius=0.8cm, y radius=0.7cm];
    \filldraw[fill=brown!85!black] 
      (0.5,0.55) arc[start angle=50,end angle=310,x radius=0.8cm, y radius=0.7cm]
      to[out=60,in=230] cycle;
  }
}

\title{Localization Uncertainty in Time-Amplitude Stereophonic Reproduction} % in Stereophonic Reproduction
% \title{Localization Uncertainty Modelling and Stability of Stereophonic Reproduction}
\author{Enzo De Sena,~\textit{Senior Member,~IEEE}, 
		Zoran Cvetkovi\'c,~\textit{Senior Member,~IEEE}, 
		H\"{u}seyin~Hac\i{}habibo\u{g}lu,~\textit{Senior Member,~IEEE}, 
        Marc Moonen,~\textit{Fellow,~IEEE}, 
		Toon van Waterschoot,~\textit{Member,~IEEE}% <-this % stops a space
\thanks{Copyright \copyright~2020 IEEE. Personal use of this material is permitted. However, permission to use this material for any other purposes must be obtained from the IEEE by sending a request to pubs-permissions@ieee.org. https://ieeexplore.ieee.org/document/9004547

Manuscript received July 18, 2019; revised January 13, 2020; accepted February 12, 2020. Date of publication February 20, 2020.  
The work reported in this paper was partially funded by (i) EPSRC Grant EP/F001142/1, (ii) Turkish Scientific and Technological Research Council (TUBITAK) Research Grant 119U254 “Audio Signal Processing for Six Degrees of Freedom (6DOF) Immersive Media”,  (iii) KU Leuven internal funds C2-16-00449 ``Distributed Digital Signal Processing for Ad-hoc Wireless Local Area Audio Networking'', (iv) European Commission under Grant Agreement no. 316969 within the FP7-PEOPLE Marie Curie Initial Training Network ``Dereverberation and Reverberation of Audio, Music, and Speech (DREAMS)" (v) European Research Council under the European Union's Horizon 2020 research and innovation program / ERC Consolidator Grant: SONORA (no. 773268). 
This paper reflects only the authors' views, and the European Union is not liable for any use that may be made of the contained information.
Parts of this work were previously presented in \cite{De-Sena:2013}, \cite{desena:thesis} and~\cite{desena:2010aes}. 
The associate editor coordinating the review of this manuscript and approving it for publication was Prof. Stefan Bilbao. (Corresponding author: Enzo De Sena.)

Enzo De Sena is with the Institute of Sound Recording at the University of Surrey (UK) (e.desena@surrey.ac.uk). Zoran Cvetkovi\'c is with the Department of Engineering at King's College London (UK). H\"{u}seyin Hac\i{}habibo\u{g}lu is with the Graduate School of Informatics, Middle East Technical University (METU) (Turkey). Marc Moonen and Toon van Waterschoot are with the Department of Electrical Engineering at KU Leuven (Belgium).
}}
%\thanks{Publisher Item Identifier XX.XXXX/XXXXX.XXXX.XXXXXX}}

\IEEEpubid{Copyright \copyright~2020 IEEE}%
\markboth{IEEE/ACM TRANSACTIONS ON AUDIO, SPEECH, AND LANGUAGE PROCESSING, VOL. 28, 2020}{De Sena \MakeLowercase{\textit{et al.}}: Localization Uncertainty in Time-Amplitude Stereophonic Reproduction}%
\maketitle %\normalsize

\setlength{\arraycolsep}{0.0em}

\begin{abstract}

This paper studies the effects of inter-channel time and level differences in stereophonic reproduction on perceived localization uncertainty, which is defined as how difficult it is for a listener to tell where a sound source is located.
Towards this end, a computational model of localization uncertainty is proposed first.
The model calculates inter-aural time and level difference cues, and compares them to those associated to free-field point-like sources. 
The comparison is carried out using a particular distance functional that replicates the increased uncertainty observed experimentally with inconsistent inter-aural time and level difference cues.
The model is validated by formal listening tests, achieving a Pearson correlation of $0.99$.
The model is then used to predict localization uncertainty for stereophonic setups and a listener in central and off-central positions.
Results show that amplitude methods achieve a slightly lower localization uncertainty for a listener positioned exactly in the center of the sweet spot. 
As soon as the listener moves away from that position, the situation reverses, with time-amplitude methods achieving a lower localization uncertainty. 
\end{abstract}

\section{Introduction}
\label{sec:compmodel}
\IEEEPARstart{D}{espite} significant advancements in the field of multichannel audio~\cite{hacihabiboglu2017perceptual}, the most common reproduction system in use today remains the two-channel stereophonic system.
% The ITU standard recommends the loudspeakers be positioned in directions $\pm 30^\circ$
% as referred to the listener's look direction~\cite{itu775}.
In typical stereophonic panning, the two loudspeakers are positioned at $\pm 30^\circ$ with respect to the listener's look direction and reproduce delayed and attenuated versions of the same signals. %~\cite{itu775}
The differences in time and level are typically frequency-independent, and are referred to as \gls{ictd} and \gls{icld}, respectively. 
For \glspl{ictd} smaller than $1$~ms, 
the listener does not perceive the two loudspeaker signals as separate,
but rather a single fused auditory event, 
often referred to as ``phantom source''.
The perceived location of the phantom source depends on both the \gls{ictd} and \gls{icld}.
This psychoacoustic effect is called ``summing localization'', and is at the basis of stereophonic panning~\cite{BlauertBook}. 
% Summing localization and the law of the first front are collectively referred to as ``precedence effect"~\cite{BlauertBook}.
The phantom source can be moved using \glspl{icld} alone, i.e. amplitude panning, \glspl{ictd} alone, i.e. time panning, or both \glspl{icld} and \glspl{ictd}, i.e. time-amplitude panning. 

Panning is inherently linked to recording. 
Consider a plane wave impinging on two microphones, each connected to a loudspeaker without mixing.
The distance between the microphones dictates the \glspl{ictd}, while the ratio between the two directivity patterns, e.g. cardioid or figure-8, dictates the \glspl{icld}.
% Therefore, as long as the microphones are approximately frequency-independent, panning is equivalent to recording of a plane wave.
% When recording, one is seldom in the presence of a single source but multiple ones, either uncorrelated sources, or correlated ones (e.g. room reflections). 
% These can be approximated using a summation of plane waves. 
Recording with coincident microphones is equivalent to amplitude panning, while recording with non-coincident omnidirectional microphones is equivalent to time panning.
Recording with microphones that are neither omnidirectional nor coincident is equivalent to time-amplitude panning. For point-like sources, distance attenuation means that small, distance-dependent level differences will also be observed at non-coincident microphones.

% Sound mixing uses almost exclusively intensity panning method, usually the sine or tangent law. 
Amplitude panning is widely used in sound mixing, with most mixing desks and digital audio workstation (DAW) software implementing a version of the sine/tangent law~\cite{rumsey2001spatial,Pulkki:1997jt}. 
Amplitude recording methods are used in a wide variety of methods, e.g. the original Blumlein pair~\cite{eargle2004microphone},  Ambisonics~\cite{daniel1998ambisonics}, and the spatial decomposition method~\cite{tervo2013spatial}. 
While it is possible to pan a stereophonic image using time panning for certain signals, it provides higher localization uncertainty for sustained higher frequency stimuli~\cite{lee2013level}.

\IEEEpubidadjcol

Time-amplitude recording methods~\cite{eargle2004microphone,rumsey2001spatial}, on the other hand, are popular within the audio engineering community, 
mainly for their strong sense of spaciousness,
which may be attributed to the higher decorrelation 
between the microphone signals~\cite{eargle2004microphone,rumsey2001spatial,BlauertBook}.
Widely-spaced microphones with inter-microphone distances in the order of meters, while extensively used, are often criticized for 
their unstable imaging~\cite{lipshitz1986stereo} and irregular distribution of reproduced auditory events~\cite{eargle2004microphone}. 
% Similarly, time panning ~\cite{lee2013level}.
Near-coincident stereophonic microphones such as the ORTF (\textit{Office de Radiodiffusion T\'el\'evision Fran\c{c}aise}), NOS (\textit{Nederlandse Omroep Stichting}) and DIN (\textit{Deutsches Institut für Normung})  pairs have also been used widely in practice~\cite{rumsey2014sound}. % 
These arrays are preferred by practitioners for providing a stable and natural stereophonic image. 
ORTF was shown to have a localization curve most similar to a binaural recording~\cite{plewa2011choosing}. %, for which the spacing between two cardioid microphones is $17$ cm, 
A 3D extension to ORTF was recently proposed~\cite{wittek2017development} and was shown to provide a good overall localization and auditory spaciousness in comparison with a coincident recording setup~\cite{riitano2018comparison}.

More recent work on  time-amplitude recording techniques generally consider the problem from a constrained perspective, by using standard microphone directivities, by evaluating spatial acuity only at the sweet spot, or both. 
For example, a psychoacoustical evaluation of equal segment microphone array (ESMA)~\cite{williamsbook}, revealed that the selection of the recording array dimensions has a distinct effect on the localization accuracy for an ESMA using cardioid microphones~\cite{lee2019capturing}. 
In another subjective study, first-order Ambisonics with \textit{max-rE} encoding  provided lower stereophonic image shifts in comparison with ESMA for a central listening position \cite{millns2018investigation}.
% A notable example of non-coincident methods 
% is the \emph{decca tree}~\cite{eargle2004microphone,rumsey2001spatial}.

% Among time-amplitude recording methods are the \gls{oct}
% Theile proposes  a class of non-coincident microphone arrays 
% for recording frontal scenes in 3/2-stereo ITU-R standard format, which is known as \emph{optimized cardioid triangle} (OCT)~\cite[p.255]{eargle2004microphone},\cite{theile2001multichannel}.
% Williams describes more general guidelines 
% for appropriately arranging standard first-order studio microphones~\cite{eargle2004microphone}
% given a loudspeaker layout and a desired coverage angle~\cite{williams1999microphone}. 
% Johnston {\sl et al.} propose the {\sl perceptual sound field reconstruction} (PSR)~\cite{Johnston:2000oe,Johnston:2005sf,gail} scheme, 
% which aims to render convincing auditory perspective by capturing interaural time and level differences.
% %and provides the starting point for the work presented in this paper.
% All these methods are still mainly a result of empirical observation and hands-on tuning.
% This work  develops further insight into the underlying  
% physical and perceptual phenomena, and based on that refines Johnston's PSR approach, and extends it to a more general and systematic framework.

While the design of microphone arrays for recording spatial audio has traditionally been an \textit{ad hoc} process driven by practical evidence that is not necessarily objectively validated, systematic approaches have also been proposed. A recently proposed design tool called microphone array recording and reproduction simulator (MARRS) allows the designer to design a stereophonic microphone pair using standard microphone directivity patterns and to observe the performance of the design both by visualization of the resulting localization curves and auralization of the simulation~\cite{lee2017interactive}.  
% A similar tool called \textit{Image Assistant} allows the calculation of different stereophonic image parameters including localization curves, ICLD, ICTD, coherence in the diffuse field and loudness, both at the sweet-spot and at a shifted position~\cite{helmutwittek2002}. 

A similar systematic framework for the design 
of amplitude and time-amplitude circular multichannel recording and reproduction systems was proposed in~\cite{desena2013}, based on earlier work~\cite{Johnston:2000oe, Hacihabiboglu:2009vo}.
An objective analysis based on active intensity fields
showed that for stable rendition of plane waves it is beneficial to render each such wave by no more than two loudspeakers, thus re-framing the multichannel problem as a stereophonic one.
% Based on that finding, we propose a methodology for the design of  circular microphone arrays, 
% in the same configuration as the corresponding loudspeaker system, 
% which aims to capture inter-channel time and intensity  differences that ensure accurate rendition of the auditory perspective. 
% The methodology is applicable to regular and irregular microphone/speaker layouts, and a wide range of microphone array radii,
% including the special case of coincident arrays which corresponds to intensity-based systems.
Using available psychoacoustic curves, a family of optimal microphone directivity patterns was obtained, parametrized by the array radius. 
The obtained directivity patterns are too spatially selective to be implemented using first-order microphone patterns but can be implemented using higher-order microphones, e.g. differential microphones~\cite{Elko:2004th,desena:diff2011}. % or phase-mode beamformers. 
Formal listening experiments were carried out for a microphone array with $15.5$~cm radius~\cite{desena2013}.
Results showed a significantly improved localization accuracy with respect to Johnston's array, and comparable to that of \gls{vbap} and Ambisonics when the listener is in the center of the loudspeaker array.
The experiments also assessed the localization uncertainty, defined as how difficult it is for the listener to tell where the sound source is located. 
Results showed an improvement in localization uncertainty with respect to \gls{vbap} and Ambisonics when the listener is in a position $30$~cm off-center.

This paper explains why that is the case and shows that this is a more general characteristic of time-amplitude methods. 
% \cite{Johnston:2000oe} % PSR JJ
% \cite{desena2013} % PSR taslp
% \cite{desena:2010aes} % Radius
% \cite{hacihabiboglu2017perceptual}
Towards this end, this paper proposes a computational model of localization uncertainty. 
The model also allows prediction of the perceived direction, but this is left for future work.
% Although subjective listening experiments are the most reliable method for studying perceptual phenomena, 
% they require very careful design
% and carrying them out is expensive and time-consuming~\cite{Bech:2006bx}.

Formal listening experiments require very careful design, 
and carrying them out is expensive and time-consuming~\cite{Bech:2006bx}.
Computational models provide a fast and repeatable alternative. 
% Devising reliable models, however, is challenging due
% to the complexity of the human auditory system.
% Modelling of spatial hearing presents many challenges due
% to the considerable complexity of the underlying psychoacoustic phenomena. 
Spatial hearing involves several stages of processing 
of sound waves impinging on the listener's head, which have been the subject of intensive study over the years.
Well-established models now exist for the early stages of the mechanisms of spatial hearing, such as the effects of head diffraction, cochlear filtering and neural transduction.
On the other hand, the higher levels of processing, where the spatial cues are combined, are not well understood yet.
Various models have been proposed in the literature, but none has proved to be capable of predicting all characteristics of human hearing. 

This paper proposes a model that first calculates \gls{itd} and \gls{ild} cues that arise from stereophonic reproduction and then compares them to the ones associated to point-like sources. 
The comparison is carried out using a distance function which replicates the auditory event splitting observed with inconsistent \gls{itd}-\gls{ild} cues.
It is shown that predictions of localization uncertainty based on this model are highly correlated with subjective scores. 
This model is then applied to generic \gls{ictd}-\gls{icld} values in central and off-central positions to assess how time-amplitude stereophony affects localization uncertainty. 
% \textbf{}
%\todozc{in this paper we focus on the issue of naturalness... especially relevant now with the advent of XR technologies. the issues is no longer to create some pleasing effect, but a natural and convincing experience of being transported to an alternate reality.}

%The psychoacoustic curves in \figurename~\ref{fig:franssenalone} reveal that it 
%is also possible to obtain full displacement with simple omnidirectional microphones.
%This approach is still popular within the audio engineering community, 
%mainly for their strong sense of spaciousness,
%which may be attributed to the higher decorrelation 
%between the microphone signals~\cite{eargle2004microphone,rumsey2001spatial,BlauertBook}.
%However, widely spaced arrays of this type are often criticized for 
%their unstable imaging~\cite{lipshitz1986stereo} 
%and irregular distribution of reproduced auditory events~\cite{eargle2004microphone}.
%A notable example of non-coincident methods 
%is the \emph{decca tree}~\cite{eargle2004microphone,rumsey2001spatial}.

The paper is organized as follows. 
Section~\ref{sec:background} provides the background on time-amplitude stereophony and on auditory system modelling.
Section~\ref{sec:proposedmodel} presents the proposed model to predict localization uncertainty. % (a reader not interested in how the localization estimates are obtained can skip this section without compromising the understanding of subsequent sections). 
Section~\ref{sec:ti} discusses how time-amplitude panning affects localization uncertainty in stereophonic reproduction.
Section~\ref{sec:titrading} narrows the focus on a specific family of time-amplitude panning curves proposed in~\cite{desena2013}. 
Section~\ref{sec:conclusions} concludes the paper.

\section{Background}
\label{sec:background}

\IEEEpubidadjcol

\begin{figure}[tb]
    \centering
    \scalebox{0.95}{
    \begin{tikzpicture}
\node at (0,0) (origin) {};
\node at (1.2,1) (human) {};

\renewcommand{\dspfilterwidth}{20mm}
\node at (0,8.1) (input) {Input};

\node[coordinate, below=of input, below=3ex] (inputconn) {};
\node[coordinate, right=of inputconn, right=17ex] (inputconn2) {};
\node[dspfilter,below=of inputconn2, below=3ex]  (delayright) {Delay $\tau_R$};
\node[dspmultiplier, dsp/label=right, below=of delayright, below=3ex] (multright) {$g_R$};
\node[coordinate, below=of multright, below=2ex] (loudspright) {};

\begin{scope}[start chain]
	\chainin (input);
	\chainin (inputconn) [join=by dspline];
	\chainin (inputconn2) [join=by dspline];
	\chainin (delayright) [join=by dspconn];
	\chainin (multright) [join=by dspconn];
	\chainin (loudspright) [join=by dspconn];
\end{scope}

\node[coordinate, left=of inputconn, left=17ex] (inputconn2) {};
\node[dspfilter,below=of inputconn2, below=3ex]  (delayright) {Delay $\tau_L$};
\node[dspmultiplier, dsp/label=right, below=of delayright, below=3ex] (multright) {$g_L$};
\node[coordinate, below=of multright, below=2ex] (loudspright) {};

\begin{scope}[start chain]
	\chainin (input);
	\chainin (inputconn) [join=by dspline];
	\chainin (inputconn2) [join=by dspline];
	\chainin (delayright) [join=by dspconn];
	\chainin (multright) [join=by dspconn];
	\chainin (loudspright) [join=by dspconn];
\end{scope}

\draw[thick,->] (origin.center) -- (3.5,0); % node[anchor=north west] {x};
\draw[thick,->] (origin.center) -- (0,4.5); % node[anchor=south east] {y};

\draw[dotted] (0,1) node[anchor=east] {$y$} -- (human.center);
\draw[dotted] (1.2,0) node[anchor=north] {$x$} -- (human.center);

\draw[dashed] (human.center) -- node[anchor=north west] {$d_R$} ++(1.3,3.33);
\draw[dashed] (human.center) -- node[anchor=north east] {$d_L$} ++(-3.7,3.33);

\foreach \Angle [count=\xi] in {60,120}
  \pic[rotate=\Angle-90] (sp\xi) at (\Angle:5cm) {Speaker};
  
\draw[dotted] (0,0) -- (60:5cm);
\draw[dotted] (0,0) -- node[anchor=north east] {$r_l$} ++ (120:5cm);
\draw[->] (0,3) node[anchor=south west] {$\phi_0$} arc (90:60:3cm);
\draw[->] (0,3) node[anchor=south west] {$\phi_0$} arc (90:120:3cm);
\pic[rotate=90,scale=0.5] at (human.center) {Human};

\end{tikzpicture}
    }
    \caption{Reference system for the stereophonic reproduction system.}
    \label{fig:refsystemrelative}
\end{figure}
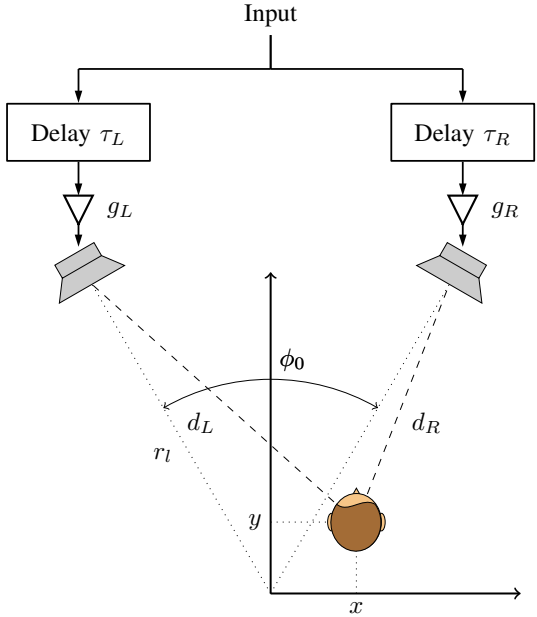

% In order to calculate the deviation between the natural \gls{itd}-\gls{ild} pairs and the 
% measured ones, the $0.5$-norm distance is used,
% and this choice is justified on psychoacoustic grounds.
% On the basis of this distance, the model yields predictions of localization uncertainty.

\subsection{Stereophonic Reproduction}

Consider a stereophonic reproduction setup as shown in \figurename~\ref{fig:refsystemrelative} with base angle $\phi_0$ and loudspeaker distance $r_l$. 
The two loudspeakers are reproducing delayed and attenuated versions of the same signal. 
The gains applied to the left and right loudspeaker are denoted as $g_L$ and $g_R$, respectively, while the delays are $\tau_L$ and $\tau_R$.
The \gls{icld} is defined as $\icld=20\log_{10}\frac{g_L}{g_R}=G_L-G_R$, where $G_L=20\log_{10}g_L$ and $G_R=20\log_{10}g_R$. 
The \gls{ictd}, on the other hand, is defined as $\tau_R-\tau_L$.
Notice how these definitions are given such that whenever \gls{ictd} and \gls{icld} have the same sign, their effect is consistent to one another. 
For instance, when both are positive, the left loudspeaker is louder and its signal also arrives earlier. 

If the \gls{ictd} is below the \emph{echo threshold},
the listener will perceive a single fused auditory event~\cite{BlauertBook}.
The echo threshold is strongly stimulus-dependent, 
and varies between $2$~ms (for clicks) and $40$~ms (for speech)~\cite{BlauertBook}.
For \glspl{ictd} between $1$~ms and the echo threshold, 
the auditory event is localised at the 
loudspeaker whose signal arrives first~\cite{BlauertBook,Litovsky:1999hc}.
This effect is called ``law of the first wavefront"~\cite{gardner1968historical}.
For \glspl{ictd} smaller than $1$~ms, 
a single fused ``phantom'' sound source is localised in a position that depends on both \gls{ictd} and \gls{icld}.
This psychoacoustic effect is called ``summing localization"~\cite{BlauertBook}. 
% and is the phenomenon at the basis of most stereophonic systems
In the literature, summing localization and the law of the first wavefront 
are collectively referred to as ``precedence effect"~\cite{BlauertBook}.

\begin{figure}[tb]
\centering
\includegraphics[width=\columnwidth]{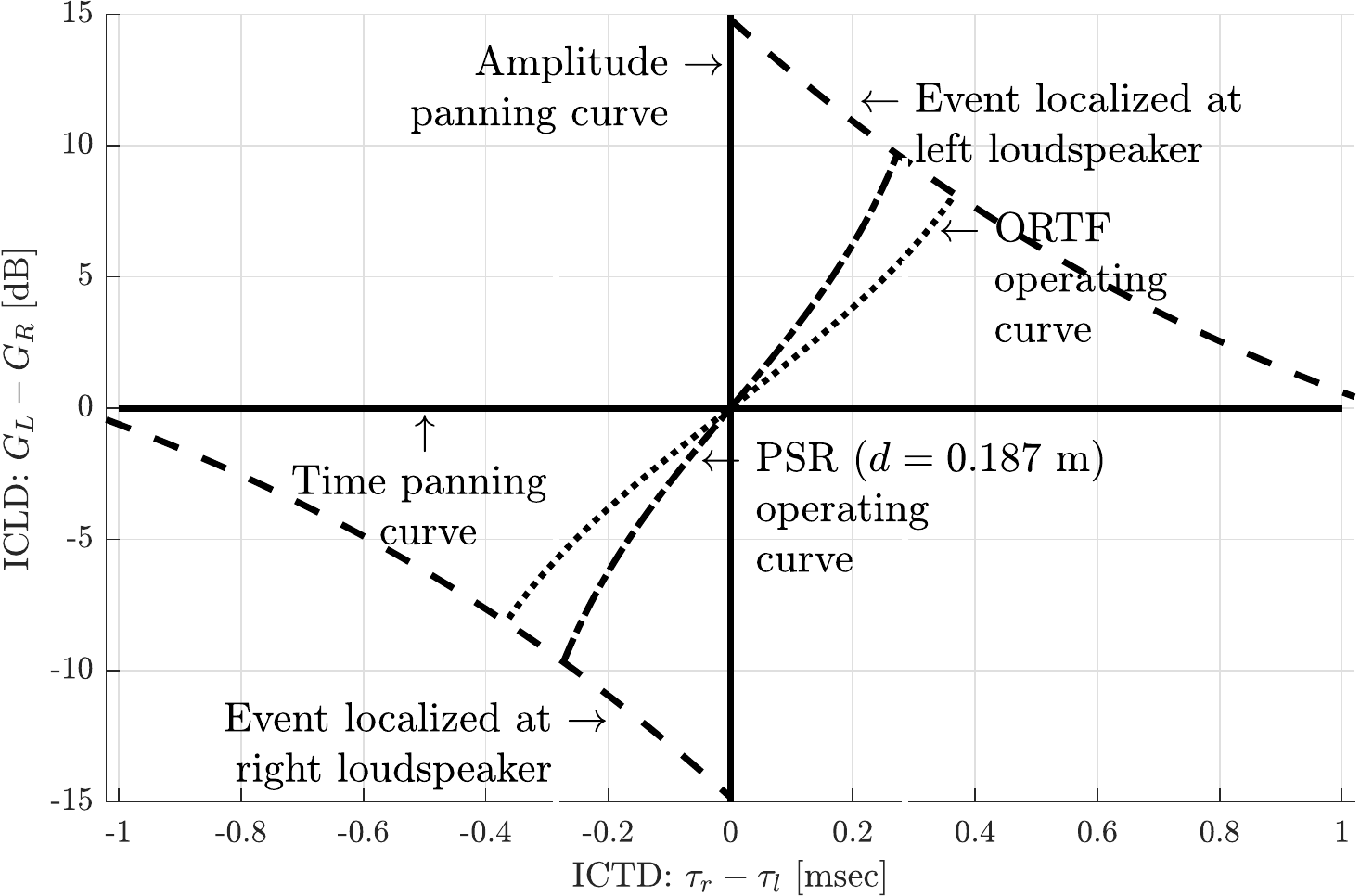}
\caption{Williams psychoacoustic curves, together with 
the panning curve associated to the \gls{psr} method with $18.7$~cm inter-microphone distance and to the ORTF microphone pair.
Amplitude methods are associated to points on the y-axis ($\ictd=0$~ms).}
\label{fig:williamsonly}
\end{figure}

\figurename~\ref{fig:williamsonly} shows Williams time-amplitude psychoacoustic curves, 
which represent all the \gls{icld}-\gls{ictd} pairs that render the phantom source in the direction of the left and right loudspeaker~\cite{williams1999microphone}.
Other time-amplitude psychoacoustic curves are also available in the literature~\cite{Franssen:1964lp,rumsey2001spatial,simon2010time}. 

The psychoacoustic curves show that if one wishes to render a phantom source in the direction of the left (respectively, right) loudspeaker it is possible to use an \gls{icld} of about $15$ dB (respectively, $-15$~dB) without any \gls{ictd} (amplitude panning).
It is then possible to continuously pan between the two loudspeakers using \glspl{icld} that vary between these two extremes (notice that the psychoacoustic curves do not give information on how to do this exactly, and only provide information about the extreme directions).
% The curve connecting these two points is referred to as  
% Panning that uses no \glspl{ictd} in this way is typically referred to as \emph{intensity panning}. 
\figurename~\ref{fig:williamsonly} also shows that it is possible to render a sound source in the direction of the left loudspeaker by reducing the \gls{icld} but increasing the \gls{ictd} (time-amplitude panning). 
% Panning that uses both \glspl{ictd} and \glspl{icld} is typically referred to as \emph{time-intensity panning}. 
If the \gls{icld} is reduced to zero (time panning), it is still possible to displace the phantom source all the way to the left (respectively, right) loudspeaker with an \gls{ictd} of about $1$~ms (respectively, $-1$~ms), which corresponds to the onset of the law of the first wavefront. 
% Panning that uses no \glspl{icld} is typically referred to as \emph{time panning}. 

% <HHO>
Note that most of the available stereophonic panning curves assume the standard stereophonic setup involving a phantom source panned between two loudspeakers positioned symmetrically to the left and right of the listener at a specified distance and can only provide suboptimal performance for panning lateral sources.

%However, in a practical multichannel reproduction situation with symmetrically positioned loudspeakers such as perceptual soundfield reconstruction (cf.~Sec.~\ref{sec:psr}) a listener is allowed to freely rotate their orientation with respect to a reference reproduction axis. Such freedom also means that the front direction of the listener would likely be between a pair of consecutive speakers except when the listener is directly facing a loudspeaker. In order to provide an orientation-independent homogeneous localization for such symmetric setups, stereophonic pairwise panning laws can be used as a starting point despite the fact that these laws are also likely to provide suboptimal localization performance for lateral sources. This way, regardless of the direction that the listener is facing, sources that are positioned in the central and near-peripheral azimuths are rendered accurately.
% </HHO>

\subsection{Stereophonic recording}

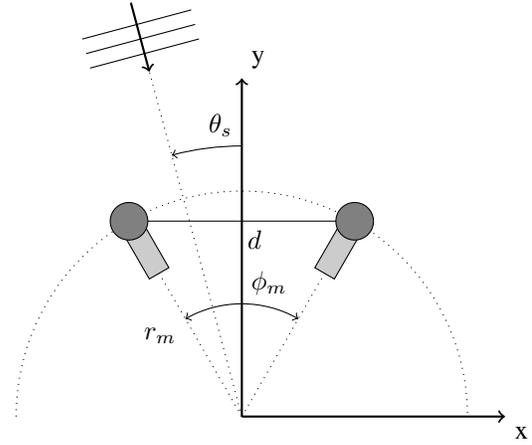
\begin{figure}[t]
    \centering
    % \scalebox{0.95}{
    \begin{tikzpicture}
\node at (0,0) (origin) {};
\node at (1.2,1) (human) {};

\draw[thick,->] (origin.center) -- (3.5,0) node[anchor=north west] {x};
\draw[thick,->] (origin.center) -- (0,4.5) node[anchor=south west] {y};

\draw[dotted] (-3,0) {} arc (180:0:3cm);

\foreach \Angle [count=\xi] in {60}
	\draw[dotted] (0,0) -- (\Angle:3cm);
	
\draw[dotted] (0,0) -- node[anchor=north east] {$r_m$} ++ (120:3cm);
% \draw[dotted] (0,0) -- node[anchor=north west] {$r_m$} ++ (60:3cm);

\draw (-1.7,2.6) -- (1.7,2.6) node[pos=0.55,below] {$d$};

\foreach \Angle [count=\xi] in {60,120}
  \pic[rotate=\Angle-90] (sp\xi) at (\Angle:3cm) {Mic};
  
\foreach \Angle [count=\xi] in {105} {
 \pic[rotate=\Angle-90,scale=0.5] (sp\xi) at (\Angle:5cm) {PlaneWave};
 \draw[dotted] (0,0) --(\Angle:5cm);
\draw[->] (0,3.6) node[anchor=south east] {$\theta_s$} arc (90:\Angle:3.6cm);
 }
 
\draw[->] (0,1.5) node[anchor=south west] {$\phi_m$} arc (90:60:1.5cm);
\draw[->] (0,1.5) node[anchor=south west] {} arc (90:120:1.5cm);

\end{tikzpicture}
    % }
    \caption{Reference system for the stereophonic recording system with incoming plane wave.}
    \label{fig:refsystemmics}
\end{figure}

Consider the reference system in \figurename~\ref{fig:refsystemmics}. 
Here, the microphones are positioned on a circle with radius $r_m$ facing outwards.
The inter-microphone distance is denoted by  $d$ and is related to the array radius by 
\begin{equation}
d=2r_m\sin\left(\frac{\phi_m}{2}\right)~,    
\label{eq:drm}
\end{equation}
where $\phi_m$ is the angle separating the two microphones, typically referred to as microphone base-angle. 
The value $\theta_s$ denotes the angle of an incoming plane wave.
The arrangement of the two microphones on a circle (as opposed to positioning them on the x-axis) is preferred in this paper because it facilitates the extension to the multichannel case, and aids the comparison with \gls{psr}, as discussed later. 

The first systematic approach to the problem of two-channel stereophonic recording
is attributed to Blumlein~\cite{eargle2004microphone,rumsey2001spatial}. 
The Blumlein pair consists of two figure-8 microphones positioned 
orthogonally to each other ($\phi_m=90^\circ$) in the same position ($d=0$~cm).
Here, each microphone is connected to the corresponding loudspeaker without mixing, which means that the inter-microphone time and level differences are identical to \glspl{ictd} and \glspl{icld}, respectively. 
Systems without mixing are also the focus of the rest of this paper.
In the specific case of the Blumlein pair, the \glspl{ictd} are zero, since the microphone pair is coincident.

% One of the advantages of the Blumlein pair is that 
% the total signal power picked up by the two microphones is independent of the source direction.
% This ensures that all sound sources are reproduced with similar loudness.
% One of the limitations of the Blumlein pair is that rear sources blend with the frontal ones
% and are rendered left-right reversed~\cite{rumsey2001spatial}.
% Furthermore, sources on the sides are picked up out-of-phase by the two microphones,
% which may cause additional confusion or, in extreme cases, inside-the-head locatedness~\cite[p. 136]{BlauertBook}.
% In order to overcome these limitations, 
% the figure-8 microphones can be replaced by first-order cardioids,
% forming the so-called ``crossed cardioid array"~\cite{eargle2004microphone}.
% Since cardioid patterns do not have phase reversals, this new configuration 
% does not suffer from the out-of-phase issue.
% However, the appealing property of constant signal power is lost, 
% and poorer localization accuracy is obtained.
% More specifically, the cardioid pattern produces \glspl{icld} that are not sufficient to fully displace 
% the phantom source in the direction
% of the loudspeakers, thus resulting in auditory events concentrated 
% in a narrow region around the midline between the loudspeakers~\cite{eargle2004microphone}.

A variety of other stereophonic arrangements have been proposed in the past few decades (see e.g.~\cite{eargle2004microphone, rumsey2001spatial}).
These include the ORTF, DIN, NOS and cardioid XY pair, the characteristics of which are summarised in  
Table~\ref{table:mics}. 
The operating curve associated to the ORTF is shown in \figurename~\ref{fig:williamsonly}.

% A notable example of non-coincident methods 
% is the \emph{decca tree}~\cite{eargle2004microphone,rumsey2001spatial}.

\renewcommand{\arraystretch}{1.2}
\begin{table}[tb]
\begin{center}
\begin{tabular}{l|c|c|c|c}
\multicolumn{1}{c|}{\begin{tabular}[c]{@{}c@{}}Microphone\\ arrangement\end{tabular}} 
& \multicolumn{1}{c|}{\begin{tabular}[c]{@{}c@{}}Inter-mic.\\ distance $d$ \end{tabular}} & \multicolumn{1}{c|}{\begin{tabular}[c]{@{}c@{}}Microphone \\ base-angle $\phi_m$\end{tabular}} & \multicolumn{1}{c|}{\begin{tabular}[c]{@{}c@{}}Directivity\\ pattern\end{tabular}}  & \multicolumn{1}{c}{\begin{tabular}[c]{@{}c@{}}Coverage \\ angle\end{tabular}} \\ \hline \hline
% {\bf Technique} & 
% $d$ & 
% $\phi_m$ &
% {\bf Pattern } &
% {\bf Coverage } \\ \hline\hline

PSR~\cite{desena2013}                                         & Adjustable                                                                                                    & $\phi_0$                   & Custom                                   & $\phi_0$  \\ \hline
Blumlein~\cite{eargle2004microphone}                               & $0$~cm                                                                                                     & $90^\circ$                                                                                               & Figure-8                                 & $68^\circ$                                                                              \\\hline
90-deg XY~\cite{eargle2004microphone}                                     & $0$~cm                                                                                                     & $90^\circ$                                                                                               & Cardioid                                 & $176^\circ$                                                                             \\\hline
ORTF~\cite{eargle2004microphone}                                        & $17$~cm                                                                                                  & $110^\circ$                                                                                              & Cardiod                                  & $94^\circ$                                                                              \\\hline
DIN~\cite{eargle2004microphone}                                         & $20$~cm                                                                                                  & $90^\circ$                                                                                               & Cardioid                                 & $100^\circ$                                                                             \\\hline
NOS~\cite{eargle2004microphone}                                         & $30$~cm                                                                                                  & $90^\circ$                                                                                               & Cardioid                                 & $80^\circ$                                                                              \\
\end{tabular}
\end{center}
\caption{Characteristics of popular microphone arrangements. Here, the coverage angle denotes the range of angles that results in $\ictd$-$\icld$ pairs within the Williams curves.}
\label{table:mics}
\end{table}

\glsreset{psr}
\subsection{Perceptual Soundfield Reconstruction}\label{sec:psr}
The methods discussed in the previous subsection were mostly designed on a trial-and-error basis. 
% These include 
A systematic framework for the design of circular microphone arrays was proposed in~\cite{desena2013}, based on earlier work by Johnston and Lam ~\cite{Johnston:2000oe} and termed \gls{psr}.
This section summarizes the design procedure, but for the specific case of a stereophonic setup.
% The objective here is to define a panning curve, which is characterized by the two functions $\ictd(\theta_s)$ and $\icld(\theta_s)$, where $\theta_s$ is the intended direction of the auditory event. 

% Consider \figurename~\ref{fig:refsystemmics}. 
In \gls{psr}, each microphone is connected to a corresponding loudspeaker. 
The microphone base-angle, $\phi_m$, is identical to the loudspeaker base-angle, $\phi_0$, which allows a straightforward extension to full $360^\circ$ multichannel rendering. 
The inter-microphone delay (and thus \gls{ictd}) is: 
\begin{equation}
    \ictd(\theta_s)=\frac{d}{c}\sin\theta_s=\frac{1}{c}2r_m\sin\left(\frac{\phi_0}{2}\right)\sin\theta_s~,
    \label{eq:ictdmic}
\end{equation} 
where $c$ is the speed of sound ($c=343$ m/s in dry air at $20^\circ$). 
Notice that in the \gls{psr} literature, the $\ictd$ is expressed as a function of the array radius, $r_m$.
The remainder of this paper, on the other hand, uses the inter-microphone distance, $d$, which facilitates the comparison with other popular stereophonic microphone arrangements.

% Notice that instead of the array radius as a parameter, one could have used the inter-microphone distance. 
% This is closer to how stereophonic recording is usually described in the audio engineering literature.
% The advantage of the parametrization in terms of array radius  is that it makes the extension to multichannel systems trivial, as will be discussed at the end of this section. 
% One can easily switch between parametrizations by observing that the inter-microphone distance is $2r_m\sin(\phi_0/2)$.

% which is written explicitly as a function of $\theta_s$.
Consider the \gls{ictd} obtained for a plane wave with $\theta_s=\frac{\phi_0}{2}$, i.e. in the direction of the left microphone:
\begin{equation}
    \ictd\left(\phi_0/2\right) = \frac{d}{c}\sin\left(\frac{\phi_0}{2}\right)~.
    \label{ref:ictdmax}
\end{equation}

% In order to find the minimum \gls{icld} necessary to render a phantom source in the direction of the left loudspeaker, one can use the psychoacoustic curves reported by Franssen~\cite{Franssen:1964lp} or Williams~\cite{williams1999microphone}.
% Rather than using Franssen curves as in~\cite{desena2013}, which, however are considered to be not sufficiently precise for quantitative design~\cite{BlauertBook}.
Given the value of $\ictd\left(\phi_0/2\right)$, psychoacoustic time-amplitude curves can be used to find the minimum \gls{icld} necessary to render a phantom source in the direction of the left loudspeaker. 
Rather than using Franssen curves as in~\cite{desena2013}, which are considered to be not sufficiently precise for quantitative design~\cite{BlauertBook}, this paper uses Williams curves.
% The curve associated to the event being localized at the left loudspeaker 
% can be well approximated by the following sigmoid function (obtained using Matlab's curve fitting toolbox):
% \begin{equation}
% 	\icld_{W,L} (\ictd)= 221.59-\frac{230.18}{1+e^{-(1000\ictd+2.1786)}}~.
% 	\label{eg:williams}
% \end{equation}
Setting the $\icld$ to be equal to the Williams value yields the constraint  $\icld\left(\phi_0/2\right)=\icld_{W,L}(\ictd\left(\phi_0/2\right))$, where $\icld_{W,L} (\ictd)$ is the Williams curve associated to the left loudspeaker (the top curve in \figurename~\ref{fig:williamsonly}).
The value $\icld_{W,L}(\ictd\left(\phi_0/2\right))$ will denoted by $\icld_{W}$ in the following.
For $\theta_s=-\frac{\phi_0}{2}$ (i.e. the direction of the right loudspeaker) one has $\icld\left(-\phi_0/2\right)=\icld_{W,R}(\ictd\left(-\phi_0/2\right))$,
which, due to the symmetry of the problem, is equivalent to $\icld\left(-\phi_0/2\right)=-\icld_{W}$. 
%$\ictd\left(-\phi_0/2\right)=-\ictd_\text{max}$ and
% \begin{align}
% 	\ictd\left(-\phi_0/2\right)&=-\ictd_\text{max}~,\\
% 	\icld\left(-\phi_0/2\right)&=-\icld_W~.
% \end{align}
A third trivial point can be added, i.e. $\icld(0)=0$~dB. % (notice that also $\ictd(0)=0$~ms, according to (\ref{eq:ictdmic})).

% So far, the panning functions have been defined for the three directions, $\theta_s=\{\phi_0/2,0,-\phi_0/2}$.
It remains to choose how to connect these three points.
Two choices were explored in \cite{desena2013}: a simple straight line~\cite{Hacihabiboglu:2010fq}, or a modified version of the tangent panning law~ \cite{desena2013}.
While both approaches were shown to lead to a good localization accuracy (i.e. the phantom source was shown to be perceived close to the intended direction $\theta_s$), the latter approach allows to link the design with other methods (e.g. \gls{vbap}~\cite{Pulkki:1997jt}) and consists of the following parametric function: 
\begin{equation}
	\icld(\theta_s)=20\log_{10}\frac{\sin\left(\frac{\phi_0}{2}+\beta+\theta_s\right)}{\sin\left(\frac{\phi_0}{2}+\beta-\theta_s\right)}~,
	\label{eq:tanpangenphi}
\end{equation}
where $\beta$ is a free parameter that is used to satisfy the constraint $\icld(\phi_0/2)=\icld_{W}$ (the other two points are then also satisfied due to symmetry), which results in 
\begin{equation}
	\beta = \arctan\left(\frac{10^{\frac{\icld_{W}}{20}} \sin(\phi_0)}{1-10^{\frac{\icld_{W}}{20}}\cos(\phi_0)}\right)~.
	\label{eq:beta}
\end{equation}

To summarize, in order to obtain the \gls{psr} panning curve one should (a) set the free parameter $d$ (or, equivalently,  $r_m$), (b) obtain $\ictd\left(\phi_0/2\right)$ from (\ref{ref:ictdmax}), (c) obtain $\icld_{W}$ from Williams psychoacoustic curve and $\beta$ from (\ref{eq:beta}) and (d) obtain $\ictd(\theta_s)$ and $\icld(\theta_s)$ from (\ref{eq:ictdmic}) and (\ref{eq:tanpangenphi}), respectively.
% Towards this end it is convenient to choose a certain parametrization for the \gls{ictd} and \gls{icld}.
% Let this parametrization be defined by equations (\ref{eq:ictdmic}) and (\ref{eq:tanpangenphi}), reported here with the explicit dependence on $d$:
% \begin{align}
%     \ictd(\theta_s,d)&=\frac{d}{c}\sin\theta_s~,\\
% 	\icld(\theta_s,d)&=20\log_{10}\frac{\sin\left(\frac{\phi_0}{2}+\beta(d)+\theta_s\right)}{\sin\left(\frac{\phi_0}{2}+\beta(d)-\theta_s\right)}~\label{eq:ictdmic2}.
% \end{align}
% While this is not the only possible parametrization, it is a convenient choice since it allows to extend the conclusions made in the remainder of this section beyond panning, and into recording with actual microphones. 
% Notice that it is possible to ignore the physical nature of equation~(\ref{eq:ictdmic2}) and simply see it as a continuous and monotonic function of the parameter $\theta_s$. 
% The reason why equation~(\ref{eq:ictdmic2}) is a convenient choice is that it allows to extend the conclusions made in the remainder of this section beyond panning, and into recording with actual microphones. 
The result of this procedure is a family of panning curves parametrized by the value of the inter-microphone distance $d$.
In the extreme case $d=0$, one obtains an amplitude-only method ($ \ictd(\theta_s)=0~\forall~\theta_s$).
As the value of $d$ increases, the \glspl{ictd} increase while the \glspl{icld} dictated by~(\ref{eq:tanpangenphi}) decrease, thus achieving stereophonic rendering with a different time/amplitude trading ratio.
\figurename~\ref{fig:williamsonly} shows the panning curve obtained for $d=18.7$~cm.
% In other words, this family of panning functions achieves a different time/amplitude trading for each value of $d$. 

The so-obtained $\icld(\theta_s)$ and $\ictd(\theta_s)$ can be viewed either in the context of stereophonic panning, whereby they can be used to directly control appropriate loudspeaker gains and delays, or in the context of stereophonic recording.
In the latter case, the microphone directivity patterns can be designed so as to emulate $\icld(\theta_s)$.
Thus, one wishes to set $\icld(\theta_s)=20\log_{10}\frac{\Gamma_L(\theta_s)}{\Gamma_R(\theta_s)}$, where $\Gamma_L(\theta_s)$ and $\Gamma_R(\theta_s)$ denote the directivity patterns of the left and right microphones.
In general, first-order microphones (i.e. microphones with $\Gamma(\theta_s)$ of the type $\Gamma(\theta_s)=a_0+a_1\cos(\theta_s)$) are not sufficiently directive to achieve the necessary \glspl{icld}.
Second-order microphones are already sufficient for this purpose~\cite{desena2013}.
The remainder of this paper will refer to PSR \emph{panning/recording} to emphasize the fact that application is possible in both contexts.

Notice that PSR aims at panoramic reproduction where a listener is allowed not only to move their head (e.g.~to reduce localization ambiguities) but also to freely rotate their orientation with respect to an arbitrary, reference reproduction axis. 
In order to provide an orientation-independent homogeneous localization for a symmetric loudspeaker setup, stereophonic pairwise panning laws can be used as a starting point. 
Despite the fact that these laws provide suboptimal localization performance for lateral sources, sources that are positioned within the range of near-peripheral azimuths (i.e.~$\pm{}\phi_0/2$ with respect to the listener's frontal direction) can be rendered accurately. 
From a purely physical point of view, PSR was also shown, via an energetic analysis of the reproduced sound field, to provide a good reproduction of directional properties of the sound field in a wide listening area~\cite{desena2013}.

\subsection{Auditory system modelling}
\label{sec:audmod}

The auditory system estimates the directions of sound sources based on a combination 
of monaural and binaural cues~\cite{BlauertBook}.
Localization in the horizontal plane is mostly reliant on binaural cues, particularly on differences in the time of arrival and on difference in level of a sound wave at the two ears.
\Glspl{itd} are caused by the different time of arrival of sound waves radiated by sources outside the median plane.
At low frequencies the auditory system
analyses the interaural time difference between the signals' fine structure~\cite{BlauertBook}.
At higher frequencies this mechanism becomes ambiguous, 
and the time differences between the signals' envelopes are used instead~\cite{BlauertBook}.
The maximum naturally occurring \gls{itd} is approximately $0.65$~ms~\cite{rumsey2001spatial,BlauertBook}.
\Glspl{ild} are caused by the acoustical shadowing of the head
and are strongly frequency-dependent.
At low frequencies the head is approximately transparent 
to the sound wave and the level differences are small.
As the wavelength approaches the size of the human head, the level differences become sensible.
The highest natural \gls{ild} is in the region of $20$~dB~\cite{BlauertBook}.

% A common confusion with respect to considering 
A common confusion in this context is to assume that \glspl{ictd} are identical to \glspl{itd}, and \glspl{icld} are identical to \glspl{ild}, which is incorrect. 
The wavefronts of each loudspeaker reach both ears and form interference patterns at the position of the ears,
which, in turn, lead to a complex relationship between \gls{ictd}-\gls{icld} and \gls{itd}-\gls{ild}.
Also notice that, as opposed to \glspl{ictd} and \glspl{icld}, \glspl{itd} and \glspl{ild} are frequency-dependent. 

The mechanisms by which the auditory system interprets the \gls{itd} and \gls{ild} cues
are complex and not yet fully understood~\cite{BlauertBook}.
Experimental evidence
suggests that humans use two main mechanisms for source localization, 
and that these mechanisms are to a certain degree independent from one another~\cite[p.173]{BlauertBook}.
The first interprets the interaural time shifts between the signals' fine structure and uses signal components below $1.6$~kHz.
The second interprets the interaural level differences and time shifts of the envelopes \emph{jointly}.
The latter mechanism seems to be dominant for signals with significant
frequency content above $1.6$~kHz~\cite[p.173]{BlauertBook}.
% Furthermore, \glspl{ild} have greater importance for signals with significant frequency content above $1.6$~kHz~\cite[p.166]{BlauertBook}.

A first, most notable attempt to model binaural processing
was made by Jeffress in 1948~\cite{jeffress1948place}, who hypothesised that sound localization is governed by a mechanism of
running cross-correlation between the two channels.
While today this is still considered to be an adequate mean of measuring 
\glspl{itd}, it does not account for the presence of \glspl{ild}~\cite{BlauertBook}.
Lindemann~\cite{lindemann1986extension} 
proposed a model that incorporates this information in the cross-correlation 
mechanism by way of inhibitory elements that are physiologically plausible.
Gaik~\cite{gaik1993combined} extended this model further, 
based on the observation that \glspl{itd} and \glspl{ild} due to point-like sources in free field come in specific pairs.
For instance the \gls{itd} and \gls{ild} values for a source in the median plane are both small.
On the other hand, for a source to the right/left, both \gls{itd} and \gls{ild} are high.
In fact, in these cases the acoustic wave arriving 
at the far ear is both attenuated (because of head shadowing) 
and delayed (because of propagation time).
% These point-like sources in free field will be referred to as ``free- sources".

Gaik observed that when inconsistent \gls{itd}-\gls{ild} pairs (e.g. a left-leading \gls{ild} and a right-leading \gls{itd}) are presented over headphones, the auditory event width increases, and sometimes two separate events appear~\cite{gaik1993combined,BlauertBook}.
In other words, inconsistent \gls{itd}-\gls{ild} pairs cause increased localization uncertainty.
% \subsection{Effect of unnatural \protect\gls{ild}-\protect\gls{itd} pairs on stereophonic systems}
These unnatural conditions can arise also with multiple sources radiating coherent signals, as in stereophonic reproduction.
Indeed, although each loudspeaker acts as a free-field source, the signals due to the different loudspeakers add up at the ears, creating interference phenomena that may result in inconsistent \gls{itd}-\gls{ild} cues.
Quantifying the deviation between the reproduced \gls{itd}-\gls{ild} pairs and
the ones associated to natural sources is therefore useful
to study the localization uncertainty due to different multichannel methods.
A study presented by Pulkki and Hirvonen in~\cite{pulkki2005localization} 
goes in this direction.
For a given multichannel method they 
find the angle of the closest free-field source in terms of \gls{ild} 
and \gls{itd}, separately.
This model gives useful predictions when the angles corresponding to the \gls{ild} and to the \gls{itd} coincide.
However, in most cases the \gls{itd} and \gls{ild} cues provide contradicting information, 
and therefore the model output is hard to interpret~\cite{pulkki2005localization}.
% Also the model proposed in this paper assumes  
% that the auditory system uses the 
% natural \gls{itd}-\gls{ild} pairs as reference points. 
% This paper proposes a novel way of using the \gls{itd} and \gls{ild} cues \emph{jointly}. 
% This is the topic of the next section

\section{Localization Uncertainty Model}
\label{sec:proposedmodel}

% The underlying hypothesis of the proposed model is that the auditory system uses psychoaoustic features of point-like sources in free-field as a dictionary to interpret all other acoustical conditions. 
% More specifically, the model calculates localization uncertainty predictions based on a distance between the \gls{itd}-\gls{ild} pairs associated to point-like free-field sound sources and those associated to the acoustical scene to be estimated. 
% This process is repeated within each critical band of hearing and the information is then fused across critical bands.
The first step of the model is to calculate \gls{itd}-\gls{ild} pairs of single point-like free-field sound sources in a number of directions on the horizontal plane. 
The so-obtained pairs are referred to as \emph{free-field \gls{itd}-\gls{ild} pairs}.
Similarly to~\cite{pulkki2005localization} and~\cite{faller2004source}, it is hypothesized here that the auditory system uses the free-field \gls{itd}-\gls{ild} pairs as a dictionary to interpret all other acoustical conditions. 
The \gls{itd}-\gls{ild} pairs for the acoustical scene to be estimated are calculated and compared to the free-field \gls{itd}-\gls{ild} pairs using a given distance functional.
Finally, the information is combined across critical bands to obtain an overall estimate of the localization uncertainty.

\subsection{Calculation of \glspl{ild} and \glspl{itd}}
\label{ref:calculate}

The \gls{itd} and \gls{ild} values are calculated as follows.
Acoustic sources are modelled as point sources in the free field. 
% The Kemar mannequin measurement from the CIPIC database (subject 25)~\cite{algazi2001cipic} are used in this paper as 
The \glspl{hrtf} are taken as the Kemar mannequin measurements from the CIPIC database (subject 25)~\cite{algazi2001cipic}. 
The sampling frequency is $44.1$~kHz.
The response of the cochlea is modelled using a gammatone filter-bank~\cite{slaney1993efficient} with 24 center frequencies equally spaced on the \gls{erb} scale between $60$~kHz and $15$~kHz~\cite{glasberg1990derivation}.
% Each bandpass signal is processed using Bernstein's model 
% of neural transduction~\cite{bernstein1999normalized}.
As a rough model of the neuron firing probability, the bandpass signals are half-wave rectified below $1.5$~kHz, while above $1.5$~kHz the envelope of each bandpass signal is taken using the discrete Hilbert transform~\cite{gaik1993combined}. 
% The cross-over frequency is taken as $1.5$~kHz~\cite{gaik1993combined}.
The resulting signals are fed to 24 binaural processors that calculate the \gls{itd} and \gls{ild} values independently.
The \gls{itd} is calculated as the location of the maximum of the cross-correlation function evaluated over time lags between $[-0.7, 0.7]$~ms~\cite{gaik1993combined, faller2004source}.
The \gls{ild} is calculated as the energy ratio of the left and right channel~\cite{faller2004source}.
Altogether, the model produces a set of 24 \gls{itd}-\gls{ild} pairs (48 values in total).

% In the next subsection we examine the values of these pairs for input signals
% due to natural sources.

\begin{figure}[tb]
\centering
    \includegraphics[width=0.95\columnwidth]{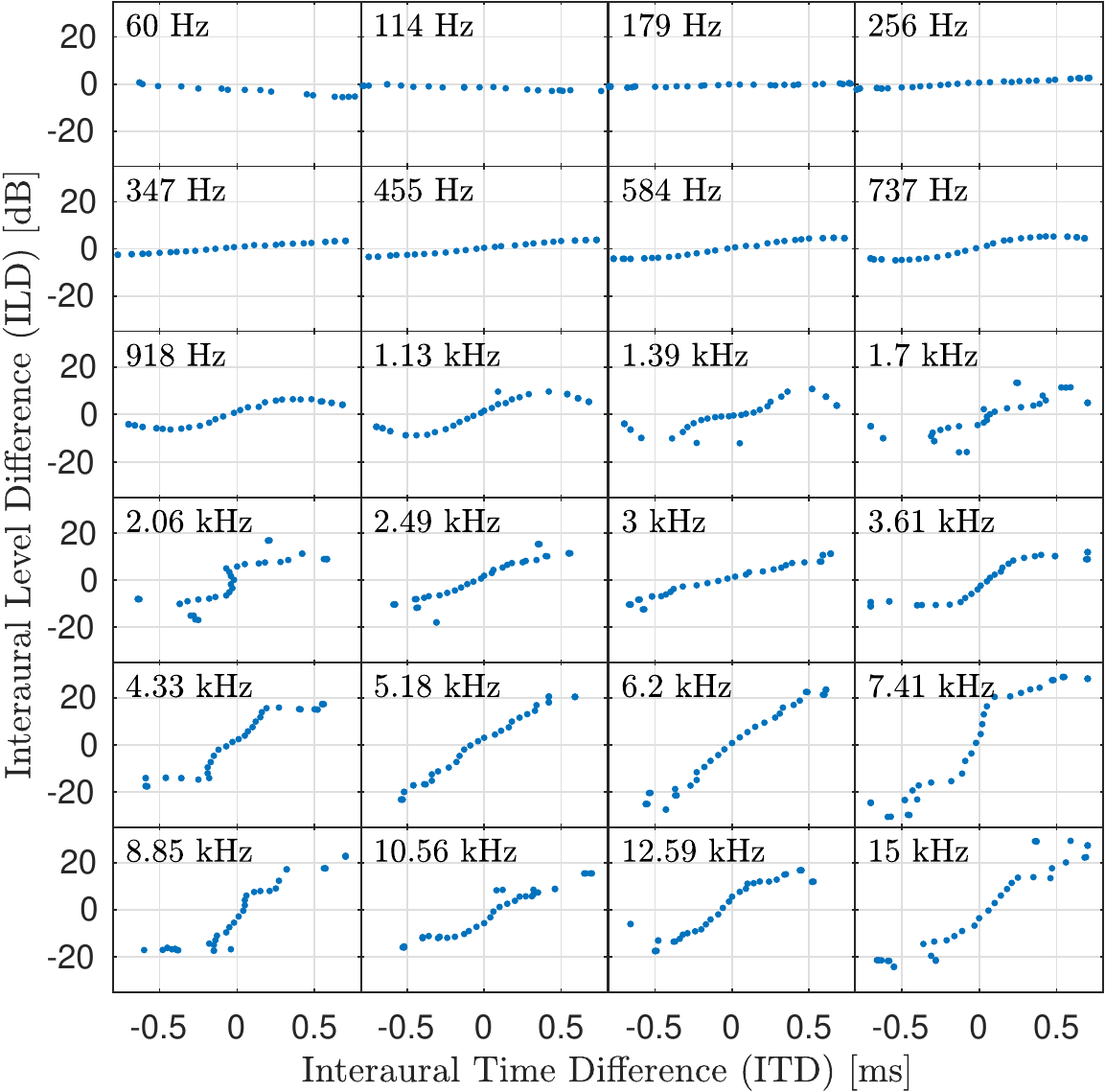}
    \caption{The figure shows the \gls{itd}-\gls{ild} pairs associated to point-like free-field sources in each critical band.}
\label{fig:natural_pairs}
\end{figure}

% \subsection{Free-field \gls{itd}-\gls{ild} pairs}
% \label{sec:natural}
\figurename~\ref{fig:natural_pairs} show the \gls{itd}-\gls{ild} pairs associated to free-field sources in the frontal horizontal plane.
The free-field sources are reproducing $50$~ms long white noise sample, multiplied by a Tukey window with taper parameter $5\%$. % (which is also used in the other simulations in this paper, unless stated otherwise).
Each simulation is repeated ten times to average out the effect of different noise realizations. 
Each point on the plot corresponds to a free field source positioned at angles between $\theta=-90^\circ$ and $\theta=90^\circ$ with an angular resolution of $5^\circ$.
In the remainder of this paper, these values will be denoted as $\fitd_i(\theta)$ and $\fild_i(\theta)$, respectively, where $i$ is the critical band index and $\theta$ is the free-field source angle with respect to the listener's look direction.

% corresponding to $181$ natural sources uniformly spaced 
% in the frontal horizontal plane are shown.
% Two main observations can be made. 
It can be observed in \figurename~\ref{fig:natural_pairs} that the interaural cues are highly correlated, i.e. larger \gls{itd} values are typically associated to larger \gls{ild} values, which is due to the concurrent effect of sound propagation and diffraction around the head~\cite{gaik1993combined}.
Also, the maximum \gls{ild} values increase with frequency, which is due to the increasing head shadowing  associated to decreasing wave lengths~\cite{gaik1993combined}.
Some small asymmetries are observed, which are possibly due to noise or asymmetries in the measurement setup of the \gls{hrtf} dataset. 

% It may be noted that, as compared to the curves generated by Gaik in~\cite{gaik1993combined},
% the curves in \figurename~\ref{fig:natural_pairs} appear 
% compressed along the ordinate axis.
% This is because Gaik calculated the \gls{ild} on the bandpass signals after
% gammatone filtering and before neural transduction.
% The \gls{ild} are calculated here based on the signals 
% after the neural transduction step, 
% as this approach is more physiologically plausible~\cite{faller2004source}.

\subsection{Distance between \gls{itd}-\gls{ild} pairs}
\label{sec:distance}
Let $\ild_i$ and $\itd_i$ denote the \gls{itd} and \gls{ild} values in the $i$-th critical band as observed by a listener under stereophonic reproduction (or other acoustical conditions). 

In order to combine the information of \gls{itd} and \gls{ild} cues across critical bands, it is useful to normalize all quantities to the maximum values of the free-field cues:
\begin{align}
\nitd_i &= \frac{\itd_i}{\max_{\substack{\theta}}|\fitd_i(\theta)|}~, \\
\nild_i&= \frac{\ild_i}{\max_{\substack{\theta}}|\fild_i(\theta)|}~.
\end{align}
The free-field pairs, $\fitd_i(\theta)$ and $\fild_i(\theta)$, 
are normalized in the same way and are
denoted as $\nfitd_i(\theta)$ and $\nfild_i(\theta)$, respectively.

Notice that while $\nfitd_i(\theta)\in[-1,1]$ and $\nfild_i(\theta)\in[-1,1]$,
$\nitd_i$ and $\nild_i$ can in theory be outside that range.
Indeed, there is no guarantee that the acoustical condition being analyzed has \gls{itd}-\gls{ild} values outside the range of values occurring for free-field sources.

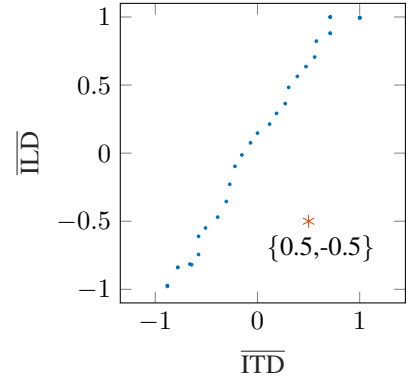
\begin{figure}
\centering 
    \newlength\fheight
    \newlength\fwidth
    \setlength{\fheight}{0.45\columnwidth}
    \setlength{\fwidth}{0.45\columnwidth}
\definecolor{mycolor1}{rgb}{0.00000,0.44700,0.74100}%
\definecolor{mycolor2}{rgb}{0.85000,0.32500,0.09800}%
\begin{tikzpicture}

\begin{axis}[%
width=0.951\fwidth,
height=\fheight,
at={(0\fwidth,0\fheight)},
scale only axis,
xmin=-1.3421498430936,
xmax=1.44721621007613,
xlabel style={font=\color{white!15!black}},
xlabel={$\overline{\text{ITD}}$},
ymin=-1.1,
ymax=1.1,
ylabel style={font=\color{white!15!black}},
ylabel={$\overline{\text{ILD}}$},
axis background/.style={fill=white}
]
\addplot [color=mycolor1, only marks, mark size=0.5pt, mark=*, mark options={solid, mycolor1}, forget plot]
  table[row sep=crcr]{%
1	0.99435113409032\\
1	0.994859434459741\\
1	0.996881904067818\\
1	0.993269654321602\\
0.711864406779661	0.999923901288263\\
0.711864406779661	0.999423893727173\\
0.711864406779661	1\\
0.711864406779661	0.881072835735565\\
0.711864406779661	0.879867451654147\\
0.576271186440678	0.822814881590892\\
0.559322033898305	0.705816039169373\\
0.474576271186441	0.63584083466628\\
0.389830508474576	0.563753890912497\\
0.305084745762712	0.483195599543213\\
0.271186440677966	0.363937210757821\\
0.186440677966102	0.291812705460424\\
0.11864406779661	0.213189077493367\\
0	0.147211006605466\\
-0.0677966101694915	0.0756328510333571\\
-0.152542372881356	-0.0133260551744929\\
-0.220338983050847	-0.0964760930531145\\
-0.271186440677966	-0.228695653446467\\
-0.305084745762712	-0.354956587148391\\
-0.389830508474576	-0.469841739070201\\
-0.508474576271186	-0.549260501012454\\
-0.576271186440678	-0.611051191505823\\
-0.576271186440678	-0.744241010473817\\
-0.661016949152542	-0.815106472341777\\
-0.644067796610169	-0.819497270446529\\
-0.779661016949153	-0.838975613777574\\
-0.779661016949153	-0.839258544885313\\
-0.88135593220339	-0.978396258675131\\
-0.88135593220339	-0.972172749929389\\
-0.898305084745763	-1.13174052875922\\
-0.915254237288136	-1.13277225167795\\
-0.915254237288136	-1.13361470344186\\
-0.898305084745763	-1.13533277820299\\
};
\addplot [color=mycolor2, only marks, mark size=2.5pt, mark=asterisk, mark options={solid, mycolor2}, forget plot]
  table[row sep=crcr]{%
0.5	-0.5\\
};
\node[right, align=left]
at (axis cs:0,-0.7) {\{0.5,-0.5\}};
\end{axis}
\end{tikzpicture}%
\quad\quad\quad\quad
\caption{The figure shows the normalized \gls{itd} and \gls{ild} pairs associated to free-field sources for the $18$th critical band (black dots), and an example of a  $\{\nitd,\nild\}$ point in $\{0.5,-0.5\}$.}
\label{fig:points}
\end{figure}

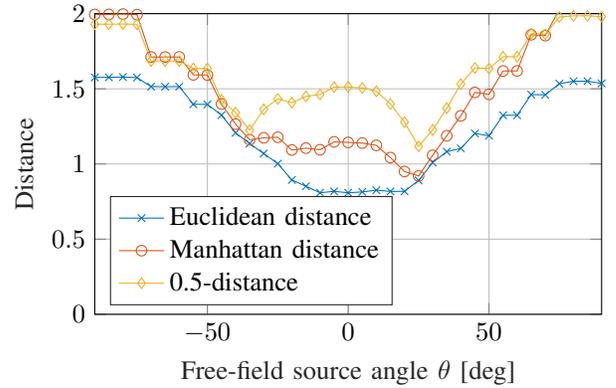
\begin{figure}
\centering 
    \setlength{\fheight}{4cm}
    \setlength{\fwidth}{0.8\columnwidth}
\definecolor{mycolor1}{rgb}{0.00000,0.44700,0.74100}%
\definecolor{mycolor2}{rgb}{0.85000,0.32500,0.09800}%
\definecolor{mycolor3}{rgb}{0.92900,0.69400,0.12500}%
\begin{tikzpicture}

\begin{axis}[%
width=0.951\fwidth,
height=\fheight,
at={(0\fwidth,0\fheight)},
scale only axis,
xmin=-90,
xmax=90,
xlabel style={font=\color{white!15!black}},
xlabel={Free-field source angle $\theta$ [deg]},
ymin=0,
ymax=2,
ylabel style={font=\color{white!15!black}},
ylabel={Distance},
axis background/.style={fill=white},
xmajorgrids,
ymajorgrids,
legend style={at={(0.03,0.03)}, anchor=south west, legend cell align=left, align=left, draw=white!15!black}
]
\addplot [color=mycolor1,mark=x]
  table[row sep=crcr]{%
-90	1.575780857847\\
-85	1.57626289964374\\
-80	1.57818105258101\\
-75	1.57475530179065\\
-70	1.51481293779658\\
-65	1.51431784640479\\
-60	1.51488828857447\\
-55	1.39722893774312\\
-50	1.39603750343413\\
-45	1.3250118885653\\
-40	1.20727437810298\\
-35	1.13612533097461\\
-30	1.06944362043749\\
-25	1.0023300570648\\
-20	0.893724313791239\\
-15	0.851637721665095\\
-10	0.808746565546065\\
-5.00000000000001	0.817852118094256\\
0	0.808545712813293\\
5.00000000000001	0.814041201032019\\
10	0.825663244900975\\
15	0.817517323819128\\
20	0.818045866361693\\
25	0.890341425810589\\
30	1.00967696217421\\
35	1.08198522813266\\
40	1.10363641565495\\
45	1.20301805685998\\
50	1.18784242602427\\
55	1.32379635330938\\
60	1.32386882982305\\
65	1.46185060513989\\
70	1.4598257831705\\
75	1.53439017388109\\
80	1.55027264655467\\
85	1.55061669943921\\
90	1.53587266695028\\
};
\addlegendentry{Euclidean distance}

\addplot [color=mycolor2,mark=o]
  table[row sep=crcr]{%
-90	1.99435113409032\\
-85	1.99485943445974\\
-80	1.99688190406782\\
-75	1.9932696543216\\
-70	1.71178830806792\\
-65	1.71128830050683\\
-60	1.71186440677966\\
-55	1.59293724251523\\
-50	1.59173185843381\\
-45	1.39908606803157\\
-40	1.26513807306768\\
-35	1.16126456347984\\
-30	1.17392338243792\\
-25	1.1781108537805\\
-20	1.09275077007986\\
-15	1.10537202749432\\
-10	1.09454500969676\\
-5.00000000000001	1.14721100660547\\
0	1.14342946120285\\
5.00000000000001	1.13921631770686\\
10	1.12386288999773\\
15	1.0424907872315\\
20	0.950128158614321\\
25	0.919988769404376\\
30	1.05773507728364\\
35	1.1873223779465\\
40	1.32051219691449\\
45	1.47612342149432\\
50	1.4635650670567\\
55	1.61863663072673\\
60	1.61891956183447\\
65	1.85975219087852\\
70	1.85352868213278\\
75	2.03004561350498\\
80	2.04802648896609\\
85	2.04886894072999\\
90	2.03363786294875\\
};
\addlegendentry{Manhattan distance}

\addplot [color=mycolor3,mark=diamond]
  table[row sep=crcr]{%
-90	1.92954333745044\\
-85	1.92975122437506\\
-80	1.930578033022\\
-75	1.9291009114186\\
-70	1.68500111277774\\
-65	1.68479696335258\\
-60	1.68503218034075\\
-55	1.63547786189603\\
-50	1.63496490371896\\
-45	1.42630928226086\\
-40	1.34165770297734\\
-35	1.2252065388715\\
-30	1.36330235226795\\
-25	1.4330542787581\\
-20	1.40782683034896\\
-15	1.4498022631451\\
-10	1.46204545363973\\
-5.00000000000001	1.51160103635899\\
0	1.51222751929688\\
5.00000000000001	1.50542114978675\\
10	1.48396317779625\\
15	1.39904105399053\\
20	1.27811080286456\\
25	1.11696961836699\\
30	1.22617541671139\\
35	1.37067834889074\\
40	1.53164236400453\\
45	1.63884841789568\\
50	1.63485194516151\\
55	1.71343715213087\\
60	1.71368007918412\\
65	1.86697294787022\\
70	1.86245926743697\\
75	1.97732057442546\\
80	1.98511441521828\\
85	1.98564377001894\\
90	1.97957715608348\\
};
\addlegendentry{0.5-distance}

\end{axis}
\end{tikzpicture}%
\caption{Distance between $\{\nitd_{18}, \nild_{18}\}=\{0.5,-0.5\}$ and the free-field point $\{\nfitd_{18}(\theta), \nfild_{18}(\theta)\}$ as a function of direction $\theta$ and for different distance functionals.}
\label{fig:distance_05}
\end{figure}

% \begin{figure*}[t]
% \centering
% \subfloat[]{\includegraphics[height=0.33\textwidth]{norm.pdf}\label{fig:norm}}
% \subfloat[]{\includegraphics[height=0.33\textwidth]{points.pdf}\label{fig:points}}
% \subfloat[]{\includegraphics[height=0.33\textwidth]{distance_05.pdf}\label{fig:distance_05}}
% %\subfloat[]{\includegraphics[height=0.25\textwidth]{distance_00.pdf}}
% \caption{\figurename~\ref{fig:norm} shows the set of points $(x,y)$ with unit distance from the centre for different $p$-norms, with $p=2,1,0.5$.
% The normalised \gls{itd} and \gls{ild} are shown in \figurename~\ref{fig:points} 
% for the $15$th critical band.
% \figurename~\ref{fig:distance_05} shows the distance between the point $\{\delta L_{15}, \delta \tau_{15}\}=\{0.5,-0.5\}$ and the natural points in directions $\theta$, $\{\delta L_{15}(\theta), \delta \tau_{15}(\theta)\}$ for different distance functionals.}
% \label{ref:distancechoices}
% \end{figure*}

The aim is now to select a distance functional between 
$\{\nitd_i, \nild_i\}$ and $\{\nfitd_i(\theta), \nfild_i(\theta)\}$
according to some meaningful psychoacoustic criterion.
% More specifically, trained subjects report that one event (commonly referred to as 
% ``time image") is closely coupled with the \gls{itd} for its direction, 
% while the other (``intensity image") depends on both \gls{itd} and \gls{ild}
% ~\cite[p.170]{BlauertBook}.
Consider the distance defined by the classical $p$-norm:
\begin{equation} \left(\left|\nitd_i-\nfitd_i(\theta)\right|^{p}+\left|\nild_i-\nfild_i(\theta)\right|^{p}\right)^{\frac{1}{p}}~.
\label{eq:distance}
\end{equation}

\figurename~\ref{fig:points} shows an example of an observed $\{\nitd_{15}, \nild_{15}\}$ point positioned at $\{0.5,-0.5\}$ in the $15$th critical band (the one centered at $6.07$~kHz).
The distance between each free-field source $\{\nfitd_{15}(\theta), \nfild_{15}(\theta)\}$ and the $\{\nitd_{15}, \nild_{15}\}=\{0.5,-0.5\}$ point is plotted in \figurename~\ref{fig:distance_05} as a function of the free-field source angle, $\theta$, for different distance functionals.

The experimental evidence shows that subjects presented with contradicting \gls{itd}-\gls{ild} pairs are likely to report split auditory events~\cite{gaik1993combined}.
The Euclidean distance ($p=2$) does not emulate this behaviour, as it leads to a single minimum in $\theta=0$, as shown in \figurename~\ref{fig:distance_05}. 
The Manhattan distance ($p=1$) is nearly constant in the $[-30^\circ,30^\circ]$ angular sector.
The $0.5$-distance, on the other hand,
causes two sharp minima, 
one of which is centred in the direction corresponding to the \gls{itd} cue,
which is compatible with the psychoacoustic evidence~\cite[p.170]{BlauertBook}. 
Other values of $p$ close to $0.5$ would also retain this behaviour.
In the next section it is shown that
the model has very good predictive power even without careful tuning of $p$.

The distance defined in~(\ref{eq:distance}) does not satisfy the triangle inequality for $p<1$.
However, the same distance raised to power $p$, does satisfy all the properties of a distance~\cite{kolmogorov1999elements}.
Hence, the $p$-norm distance used here is:
\begin{equation}
\xi_i\left(\theta|\nitd_i,\nild_i\right) = \left|\nitd_i-\nfitd_i(\theta)\right|^{p}+\left|\nild_i-\nfild_i(\theta)\right|^{p}~.
\label{eq:dpi}
\end{equation}

The objective is to obtain a function of $\theta$ that quantifies the likelihood that a sound is perceived in that direction.
Intuitively, this function should be inversely proportional to $\xi_i(\theta|\nitd_i,\nild_i)$, such that whenever the distance is small, the likelihood is high (and viceversa).
Let this function be
\begin{equation}
f_i\left(\theta|\nitd_i,\nild_i\right) =
% Ke^{-\left(\left|\nild_i-\nfild_i(\theta)\right|^{p}+\left|\nitd_i-\nfitd_i(\theta)\right|^p\right)}~,
Ke^{-\xi_i\left(\theta|\nitd_i,\nild_i\right)}~,
\label{eq:inversefunc}
\end{equation}
where $K$ is a positive constant. 
% Various choices of this functions are available. 
Although other choices are available\cite{De-Sena:2013}, the advantage of~(\ref{eq:inversefunc}) is that it gives the model an explicit statistical interpretation in the \gls{ml} framework, as will be discussed in the next subsection.
% \footnote{For instance, \cite{De-Sena:2013} uses $\max_\theta (\xi_i(\theta))-\xi_i(\theta)$).}

% $$K=\left(\frac{p}{2\Gamma\left(1/p\right)(2\sigma^2)^\frac{1}{p}}\right)^{2}~.$$
The next step is to integrate the information from the different critical bands.
The mechanisms governing this stage of perception are generally regarded 
as complex and not well understood~\cite{pulkki2005localization, BlauertBook}.
Here, the information across critical bands is combined as a loudness-weighted average:
\begin{equation}
f\left(\theta|\nitd,\nild\right) = \frac{K}{N}\sum_{i=1}^N w_i e^{-\xi_i\left(\theta|\nitd_i,\nild_i\right)}~,
\label{eq:like}
\end{equation}
where $w_i$ are the loudness weights, $\nild$ denotes the vector $\nild=\left[\nild_1,...,\nild_N\right]$ and $\nitd=\left[\nitd_1,...,\nitd_N\right]$~, and $N$ is the number of critical bands ($N=24$).
% The loudness weighting is based on the hypothesis that more audible frequency bands have a larger effect~\cite{supper2005onset}. 
The loudness weights are set according to the procedure proposed in~\cite{supper2005onset}, which results in critical bands with a higher signal level weighting more than critical bands with little active content~\cite{supper2005onset}.
The procedure involves (a) calculating the SPL levels in each critical band, (b) converting them to phon levels through the BS ISO 226:2003 equal-loudness contours, and (c) converting them to $w_i$ weights using a function where a $10$ phon reduction leads to a halving of the weight, in line with Stevens's model~\cite{stevens1957psychophysical}.  

% In order to quantify the localization uncertainty we proceed as follows. 
% First, the distance function is flipped:
% \begin{equation}
% \mathcal{L}_i(\theta) = \max_{\substack{\theta}} \left(d_i(\theta)\right) - d_i(\theta)~.
% \label{eq:Gamma}
% \end{equation}
% The function $\mathcal{L}_i(\theta)$ will be referred to as ``score function" of the $i$-th critical band.
% Next, the information due to the different critical bands is combined.
% The mechanisms governing this stage of perception are generally regarded 
% as very complex, and are not well understood~\cite{pulkki2005localization, BlauertBook}.
% For this reason the overall score function is conservatively chosen as 
% the simple average of the individual $\mathcal{L}_i(\theta)$:
% \begin{equation}
% \mathcal{L}(\theta) = \frac{1}{B}\sum_{i=0}^{B-1}\mathcal{L}_i(\theta)~,
% \label{eq:scorefunction}
% \end{equation}
% where $B$ is the number of critical bands.
% In practical cases, the \gls{itd}-\gls{itd} pairs of natural sources will be known for 
% a discrete number of directions.
% Hence, we discretise the problem and denote as $\theta_k$ the $k$-th discrete direction.
% Let the score function be normalised as
% \begin{equation}
% \ell(\theta_k) = \frac{\mathcal{L}(\theta_k)}{\sum_m\mathcal{L}(\theta_m)}~,
% \end{equation}
% such that $\sum_k\ell(\theta_k)=1$.

% The next section shows that the model described so far has an interpretation within the framework of \gls{ml} estimation. 

\subsection{Statistical interpretation}
Suppose that $\{\nitd_i, \nild_i\}$ are noisy observations of a point-like free-field source at angle $\theta_0$ associated to a set of true $\{\nfild_i(\theta_0), \nfitd_i(\theta_0)\}$ values:
\begin{align}
    \begin{split}
        \nitd_i&=\nfitd_i(\theta_0)+u_i~,\\
        \nild_i&=\nfild_i(\theta_0)+v_i~.
    \end{split}
\end{align}
where $u_i$ and $v_i$ are the noise components, which may arise for instance as a consequence of room reverberation or reproduction with multiple loudspeakers.
Let the joint distribution of the noise components be the following mixture of zero-mean bivariate theta-generalized normal distributions~\cite{goodman1973multivariate}:
\begin{equation}
f\left(u_1,...,u_N;v_1,...,v_N\right) = \frac{1}{N}\sum_{i=1}^N Ce^{-\frac{\left|u_i\right|^{p}+
\left|v_i\right|^p}{2\sigma^2}}~,
\label{eq:likestat}
\end{equation}
% where $$K=\left(\frac{p}{2\Gamma\left(1/p\right)(2\sigma^2)^\frac{1}{p}}\right)^{2}~.$$
where $\sigma$ represents a standard deviation and $C$ is a normalization constant. %  with zero mean and covariance matrix $\sigma \Vec{I}_{2\times 2}$ %, where $\Vec{I}$ denotes the identical matrix.  

Notice that if $\frac{1}{2\sigma^2}=1$, equation (\ref{eq:likestat}) becomes identical to (\ref{eq:like}).
Suppose now that the auditory system estimates the true value $\theta_0$ using an \gls{ml} approach. 
Within this context, $f\left(\theta|\nitd,\nild\right)$, seen here as a function of $\theta$,  takes the meaning of a likelihood function. 

Although the objective of the proposed model is not to estimate $\theta$ itself (which is left for future work), the shape of the likelihood function gives information about how difficult it is to estimate it, as will be explained in the next subsection.
% The focus of the next subsection is to quantify 

\subsection{Calculation of the localization uncertainty}
\label{sec:quantunc}

The likelihood function quantifies the probability that a subject would perceive a source in a given direction $\theta$. 
From this perspective, a uniform (constant) likelihood function would result in a maximally uncertain (diffuse) event.
At the other extreme, an impulsive likelihood function would result in a minimally uncertain event.
Various measures can be used in this context.
This paper uses a modified version of the circular variance~\cite{sena2017localization,mardia2009directional}.
This measure is drawn from the field of directional statistics~\cite{mardia2009directional}, and has values between $0$, which is associated to an impulsive function, and $1$, which is associated to a constant function.
Calculating the modified circular variance involves normalising $f\left(\theta|\nitd,\nild\right)$ so that it sums to one (e.g. by adjusting the $K$ constant), and calculating the (modified) first cosine and sine moments:
\begin{align}
    \alpha&=\int_{0}^{2\pi}f\left(2\theta|\nitd,\nild\right)\cos(\theta)d\theta,\\
    \beta&=\int_{0}^{2\pi}f\left(2\theta|\nitd,\nild\right)\sin(\theta)d\theta~, 
\end{align}
where the modification consists of using $f\left(2\theta|\nitd,\nild\right)$ instead of $f\left(\theta|\nitd,\nild\right)$. 
This modification is motivated by the fact that the likelihood function only takes values in $\left[-\frac{\pi}{2},\frac{\pi}{2}\right]$, which would not lead to a circular variance of $1$ in case of a constant likelihood function. 
Then, the circular variance is given by~\cite{mardia2009directional}
\begin{equation}
   \text{H}\left(\nitd,\nild\right)=1-\sqrt{\alpha^2+\beta^2}~.
\end{equation}
 
 % One of them is Wiener entropy (also known as spectral flatness), 
% which is defined as the ratio of the geometric mean over the arithmetic mean ~\cite{johnston1988transform}:
% \begin{equation} 
%     \text{H}\left(\nitd,\nild\right)=\frac{\sqrt[\leftroot{-3}\uproot{3}K]{\prod_{k=1}^K f\left(\theta_k|\nitd,\nild\right)}}{\frac{1}{K}\sum_{k=1}^K f\left(\theta_k|\nitd,\nild\right)}~,
%     \label{eq:flatness}
% \end{equation}
% where $\theta_k$ are the angles of the considered free-field sources, and $K$ is their cardinality.

The final step is to normalize the value of $\text{H}$.
Indeed, since $f\left(\theta|\nitd,\nild\right)$ is never impulsive (even for free-field sources), the value of $\text{H}$ is always larger than zero.
For instance, for $p=0.7$, the minimum $\text{H}$ is $0.49$. 
The following normalization yields values close to zero for estimates associated to free-field sources:
\begin{equation}
    \overline{\text{H}}\left(\nitd,\nild\right)=\frac{\text{H}\left(\nitd,\nild\right)-\text{H}_{\text{min}}}{1-\text{H}_{\text{min}}}
    \label{eq:end}
\end{equation}
where $\overline{\text{H}}\left(\nitd,\nild\right)$ denotes the final localization uncertainty estimates of the proposed model and $\text{H}_{\text{min}}=\min_{\theta}\text{H}\left(\nfild(\theta),\nfitd(\theta)\right)$.

It should be noted that, in practice, all calculations above are made using a discretization of the angles $\theta$. 
The simulations in the remainder of this paper use a resolution of $5^\circ$.

% It is noted that alternative approaches have also been considered, including the information-theoretical definition of entropy, the Wiener entropy, or the $95\%$ confidence interval of the \gls{ml} estimator.
% These approaches have not been further pursued either because they do not yield a good performance or because they do not take into account the circular nature of the likelihood function. 

% but is considered here slightly less elegant as it requires the additional normalization step. 
% Another approach stems from the result that the \gls{ml} estimator of $\theta$ is asymptotically distributed as $\hat\theta\sim\mathcal{N}\left(\theta, I^{-1}(\theta)\right)$, where $I(\theta)$ is the Fisher information evaluated in $\theta$~\cite[p. 167]{kay1993fundamentals}. 
% The $95\%$ confidence interval of the \gls{ml} estimator is therefore $\hat\theta\pm\frac{1.96}{\sqrt{Q}I(\theta)}$, where $Q$ is the number of observations.
% This measure does not perform well as an estimate of localization uncertainty, 
% which is likely due to the fact that the model has access to a single observation (i.e. $Q=1$) and thus the asymptotic assumption is not valid. 

\subsection{Model validation}
\label{sec:experiment}
A formal listening test with 19 subjects was carried out in~\cite{desena2013} using a modified MUSHRA test~\cite{mushra}.
The subjects answered the question 
``How certain are you of the direction of the source?"
by giving a score on a continuous scale from 0 to 100.
%\begin{figure}[tb]
%\centering
%\includegraphics[width=5.8cm]{figures/experiment_mushra.pdf}
%\caption{The test setup for the locatedness experiment. The
%five uniformly spaced white loudspeakers form the reproduction system.
%The three dark loudspeakers in front of the listener are
%the acoustic pointers.
%Two listening position are considered. 
%One in the centre looking in
%the midline direction between $L_L$ and $L_R$.
%The second is $30$~cm off-centre, and more specifically $30/\sqrt{2}$~cm behind
%and to the left of the central position. 
%%The look direction for the off-centre position is parallel to the one in the centre.
%Adapted from~\cite{desena2013}.}
%\label{fig:setup}
%\end{figure}
The test was carried out in an audio booth using four synthesized $5$-channel surround sound methods: (a) pair-wise tangent panning law~\cite{Bernfeld:1973wd} (equivalent to horizontal \gls{vbap}); (b) near-field corrected second-order Ambisonics 
with mode-matching decoding at low frequency and
maximum-energy decoding at high frequency
~\cite{daniel2003spatial, Poletti:2000dk, daniel1998ambisonics};
(c) second-order Ambisonics with in-phase decoding~\cite{daniel2000reprsentation};
and (d) the quasi-coincident microphone array proposed in~\cite{desena2013}. 
The test was run for three sound source directions 
and two seating positions, and included both a reference (a loudspeaker in the intended direction) and an anchor (an approximately diffuse soundfield). 
The rendered virtual source directions were in front of the listener and $\pm{}18^{\circ}$ to the right and left of the front direction. 
Three excerpts (female speech, african bongo and cello) were used as representatives of common program material, and the resulting subjective scores were averaged across the three excerpts.
Details of the experiment are available in~\cite{desena2013}.

In order to validate the proposed model, the experiment was replicated here through simulations.
The stimuli used in the simulations were (a) long white noise burst ($500$~ms), (b) short white noise burst ($50$~ms), (c) short pink noise burst ($50$~ms), and (d) impulsive sound. 
All stimuli (except for the impulsive sound) were multiplied by a Tukey window with a $5\%$ taper parameter.

The loudspeaker signals for each of the $5$-channel surround sound methods mentioned above were then calculated. 
Notice that the experiment and the simulations employed multichannel sound reproduction where more than two loudspeakers can be substantially active in some of the methods used (i.e.~Ambisonics). 
While the model will be used in a stereophonic context in this paper, it is independent from the number of loudspeakers that the input ear signals originate from. 

The acoustic path from each loudspeaker to the head was simulated in free-field conditions using appropriate time delays and distance attenuation (inverse square law) that depend on the position of the head.
The ear pressure signals were obtained using the Kemar mannequin \glspl{hrtf}.
The so-obtained ear pressure signals associated with multiple loudspeakers were added together. 
Finally, the resulting ear pressure signals were fed to the proposed model, and localization uncertainty estimates were obtained using (\ref{eq:end}).

\figurename~\ref{fig:pearson} shows the absolute Pearson correlation between the subjective scores and 
the localization uncertainty estimates, as a function of the parameter $p$ for different stimuli. 
All the types of stimuli result in an absolute Pearson correlation stronger than $-0.98$ for $p\in[0.7,0.9]$, 
confirming that the model predictions are strongly correlated with the experimental data.
The reason why the correlation is negative is that subjects' answers to the question ``How certain are you of the direction of the source?" are merely inverted with respect to localization uncertainty.

All types of random noise perform particularly well. 
This indicates that the choice of stimulus does not appear to be critical when aiming to predict the localization uncertainty of an experiment involving a varied programme material like the one in~\cite{desena2013} (which included speech, bongos and cellos).
Furthermore, since the free-field pairs $\{\nfitd_i(\theta), \nfild_i(\theta)\}$ were obtained using short white noise bursts, the model appears to be robust to a mismatch between the stimulus used for the free-field pairs and the one used in the acoustic scene under analysis (in a machine learning context, these would be akin to training and testing, respectively).

% This also confirms that the 
% The Pearson correlation is approximately $-0.95$ for $p\in[0.25,0.35]$, 
% confirming that the model predictions are strongly correlated with the experimental data.

% Similar results are obtained (a) if the spherical head model is replaced with measured \glspl{hrtf}~\cite{gardner1994hrft}, (b) if instead of the exponential function in (\ref{eq:inversefunc}) one uses $\max_\theta (d_i(\theta))-d_i(\theta)$ as in~\cite{De-Sena:2013} and (c) if instead of the Wiener entropy in (\ref{eq:flatness}), one uses the information-theoretical entropy after normalization.

Fine tuning of the $p$-norm distance is also not critical as long as contradicting \gls{itd}-\gls{ild} cues lead to a an approximately bimodal likelihood function.
For the short white noise burst, values of $p$ between $0.5$ and $1.0$ all give correlation coefficients of approximately $-0.99$. 
% A correlation of $-0.83$ is obtained with the Manhattan distance ($p=1$).
%\footnote{It can be verified that for $p=2$, the distribution in~(\ref{eq:likestat}) becomes a mixture of zero-mean bivariate normal distributions~\cite{goodman1973multivariate}.}
The Euclidean distance ($p=2$), on the other hand, yields a much weaker correlation of $-0.84$.
At the other extreme, a small $p$ also leads to weak correlations, indicating that norms closer to the $l_0$-norm do not provide an effective measure in this context.

\begin{figure}
\centering 
    \newlength\figW
    \newlength\figH
    \setlength{\figW}{0.45\columnwidth}
    \setlength{\figH}{0.45\columnwidth}
\definecolor{mycolor1}{rgb}{0.00000,0.44700,0.74100}%
\definecolor{mycolor2}{rgb}{0.85000,0.32500,0.09800}%
\begin{tikzpicture}

\begin{axis}[%
width=0.951\figW,
height=\figH,
at={(0\figW,0\figH)},
scale only axis,
xmin=50,
xmax=90,
xlabel style={font=\color{white!15!black}},
xlabel={Subjective scores},
ymin=0.2,
ymax=1,
ylabel style={font=\color{white!15!black}},
ylabel={Model estimates},
axis background/.style={fill=white},
xmajorgrids,
ymajorgrids
]
\addplot [color=mycolor1, forget plot]
  table[row sep=crcr]{%
50	0.906578786673723\\
90	0.225140039061745\\
};
\addplot[only marks, mark=o, mark options={}, mark size=1.5000pt, draw=mycolor2, forget plot] table[row sep=crcr]{%
x	y\\
81.4444	0.351354\\
83.6667	0.372829\\
80.4074	0.373022\\
83.1481	0.321041\\
82.3333	0.383437\\
80.7778	0.358197\\
78.037	0.405843\\
83.9259	0.316886\\
73.7241	0.528557\\
65.8966	0.596204\\
73.6207	0.523467\\
73.2069	0.529843\\
64.8333	0.689003\\
51.6389	0.862396\\
62.1111	0.678035\\
58.5556	0.787002\\
};
\end{axis}
\end{tikzpicture}%
\caption{Scatter plot comparing the subjective scores of the experiment presented in~\cite{desena2013} and the estimates of the proposed model for $p=0.7$ and for short white noise burst. The blue line indicates the best-fitting line. The Pearson correlation is $-0.99$.}
\label{fig:corrmodel}
\end{figure}
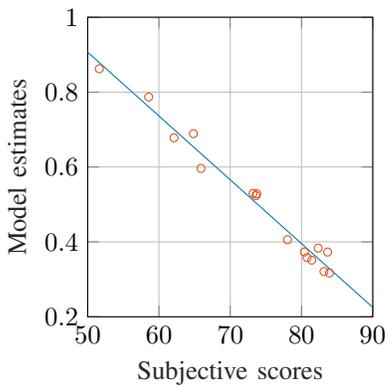

The simulations in the remainder of this paper use the short white noise burst, and a distance function with $p=0.7$ (an approximate midpoint of the $p\in[0.5,1.0]$ interval with~$-0.99$ correlation). 
\figurename~\ref{fig:corrmodel} shows the scatter plot comparing the model estimates with the subjective scores.

Note that the selection of an earlier set of results from an experiment using a five channel setup for the validation of the proposed model is deliberate. 
While the derivation of the model is based on two-channel stereophony as the simplest possible spatial audio reproduction system, its predictive power extends beyond that as is shown.

Note also that the experiment did not use any head or position tracking and the listeners were instructed to keep as still as possible. 
Such a listening scenario is not entirely ecologically valid. 
However, since the proposed model is neither dynamic nor does it incorporate any mechanism to account for listener motion, the results from the experiment are indicative of the predictive power of the model. 

\begin{figure}
\centering 
\begin{tikzpicture}
\begin{axis}[
    width=\columnwidth,
    height=4cm,
	grid=major,
	xmin=0,
	xmax=2,
	ymax=1,
	ymin=0.8,
	cycle list name=exotic,
    xlabel={$p$},
    ylabel={Absolute correlation}]
    \draw[dashed](0.7,0) -- (0.7,1);

\addplot coordinates { % white noise
(0.05,0.269255)
(0.075,0.701248)
(0.1,0.868561)
(0.125,0.927652)
(0.15,0.952472)
(0.175,0.965472)
(0.2,0.973072)
(0.25,0.979701)
(0.3,0.982028)
(0.4,0.984116)
(0.5,0.985789)
(0.6,0.987392)
(0.7,0.98861)
(0.8,0.989102)
(0.9,0.988583)
(1,0.986821)
(1.25,0.975872)
(1.5,0.954383)
(1.75,0.922287)
(2,0.881172)};
\label{white}

\addplot coordinates { % pink noise
(0.05,0.314888)
(0.075,0.716641)
(0.1,0.873759)
(0.125,0.933155)
(0.15,0.958751)
(0.175,0.970817)
(0.2,0.976714)
(0.25,0.981117)
(0.3,0.982165)
(0.4,0.982866)
(0.5,0.984258)
(0.6,0.98627)
(0.7,0.988352)
(0.8,0.990033)
(0.9,0.990955)
(1,0.99085)
(1.25,0.984978)
(1.5,0.970146)
(1.75,0.946618)
(2,0.915957)};
\label{pink}

\addplot coordinates { % 0.5 sec long white noise
(0.05,0.253757)
(0.075,0.701906)
(0.1,0.871371)
(0.125,0.930226)
(0.15,0.955124)
(0.175,0.967908)
(0.2,0.975109)
(0.25,0.98143)
(0.3,0.983489)
(0.4,0.985435)
(0.5,0.987155)
(0.6,0.988805)
(0.7,0.990021)
(0.8,0.990457)
(0.9,0.989829)
(1,0.987914)
(1.25,0.976462)
(1.5,0.954452)
(1.75,0.922002)
(2,0.880821)};
\label{longwhite}

\addplot coordinates { % impulsive sound
(0.05,0.045344)
(0.075,0.53543)
(0.1,0.793597)
(0.125,0.888738)
(0.15,0.9256)
(0.175,0.942518)
(0.2,0.951843)
(0.25,0.961767)
(0.3,0.966707)
(0.4,0.971257)
(0.5,0.973521)
(0.6,0.975035)
(0.7,0.976023)
(0.8,0.976386)
(0.9,0.975925)
(1,0.974408)
(1.25,0.9643)
(1.5,0.941856)
(1.75,0.90346)
(2,0.847472)};
\label{impulse}

\end{axis}
\end{tikzpicture}
\caption{Absolute Pearson correlation between the model predictions and the testing set as a function of the distance functional parameter $p$ for different stimuli: short white noise burst~(\ref{white}), long white noise burst~(\ref{longwhite}), impulse~(\ref{impulse}) and pink noise~(\ref{pink}). The short white noise burst curve~(\ref{white}) is not visible because it coincides almost exactly with the long white noise burst curve~(\ref{longwhite}). The dashed vertical line denotes the value chosen in the remainder of the simulations in this paper ($p=0.7$).}
\label{fig:pearson}
\end{figure}
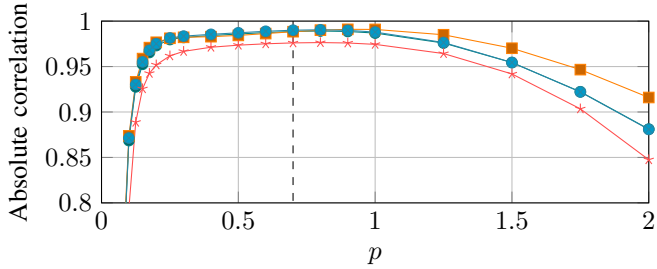

%p=0.1, pearson correlation: -0.90764
%p=0.2951, pearson correlation: -0.94852

% Note that the time of  arrival difference between loudspeakers $L_{R}$ and $L_{SL}$
% in the off-centre position is approximately $1.6$ ms, 
% which is above the $1$ ms threshold where 
% the {law of the first front} is in effect~\cite{BlauertBook,Litovsky:1999hc}. 
% The signal emitted by $L_{SL}$ for the HOA system is not negligible.
% The predicted uncertainty is accurate in this case as well, 
% although the proposed model (similarly to \cite{pulkki2005localization})
% does not incorporate inhibitory mechanisms~\cite{BlauertBook}, 
% which are thought to be at the base of the {law of the first front}~\cite{lindemann1986extension}.
% Notice how, in the general case, the predictions of the proposed model
% should be not considered reliable arrival delays beyond $1$ ms.

\section{Localization Uncertainty in Stereophonic Reproduction}
\label{sec:ti}

\subsection{Localization uncertainty in the center of the sweet-spot}
\label{sec:center}

\begin{figure}[t]
\centering
\includegraphics[width=0.95\columnwidth]{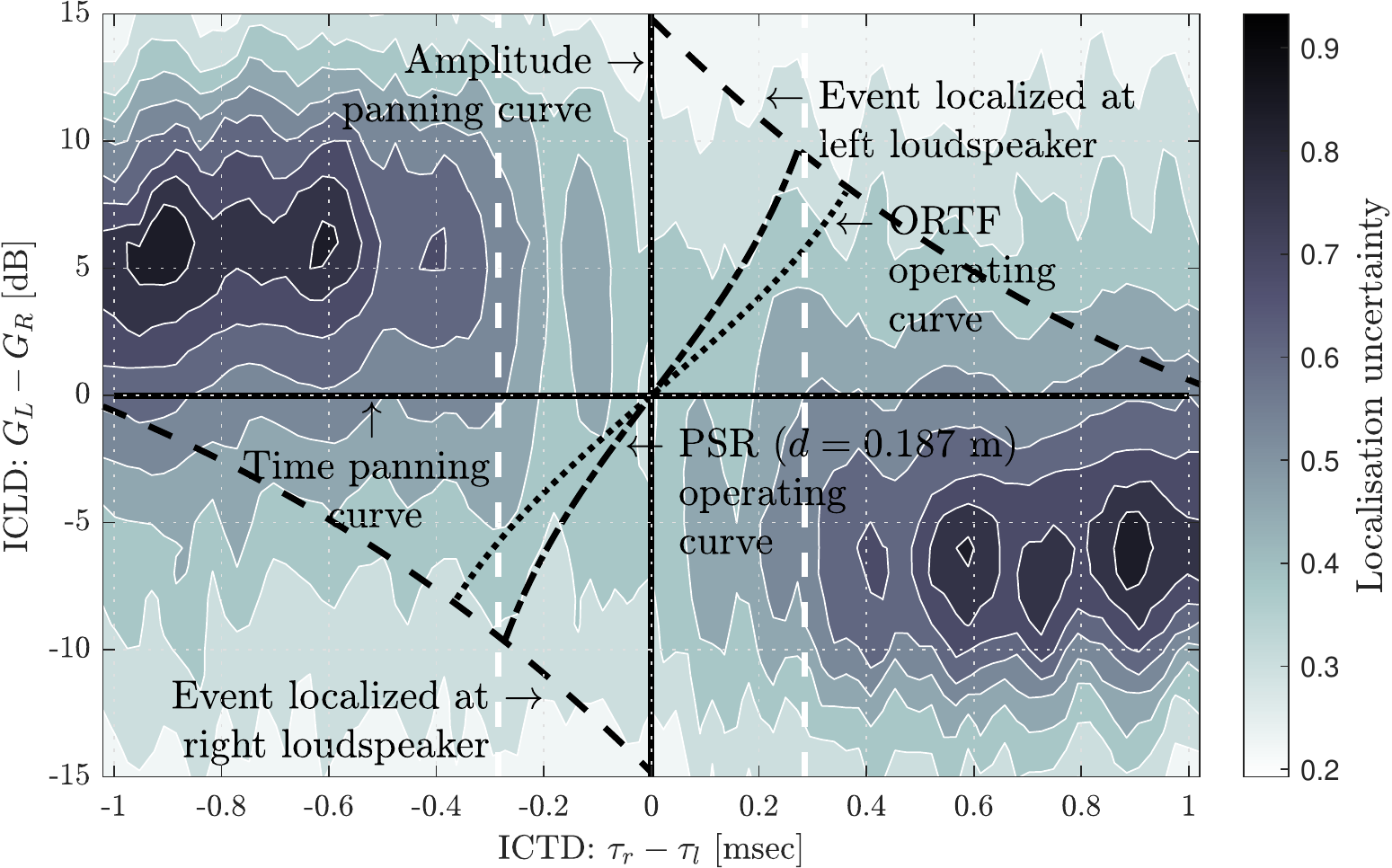}
\caption{Contour plot of localization uncertainty as a function of \gls{icld} and \gls{ictd}
in the center of the sweet-spot.
The dashed white lines denote delays that result in the loudspeaker signals arriving at the same time at one of the two ears.
Overlaid are the Williams' curves and the time-amplitude panning curve associated to the \gls{psr} method with $d=18.7$~cm.}
\label{fig:hictdicldcentre}
\end{figure}

\figurename~\ref{fig:hictdicldcentre} shows the localization uncertainty 
produced by the proposed model 
for a listener in the center of the sweet-spot and for a stereophonic reproduction system with base angle $\phi_0=60^\circ$ and $r_l=2$~m (see~\figurename~\ref{fig:refsystemrelative}).
Two areas in the second and fourth quadrants have a significant localization uncertainty.
These areas correspond to cases where \gls{icld} and \gls{ictd} provide inconsistent information: one 
loudspeaker is leading in terms of \gls{icld} (i.e. it is louder)
while the other is leading in terms of \gls{ictd} (i.e. it arrives earlier).
This indicates that inconsistent \gls{ictd}-\gls{icld} pairs somehow 
translate to unnatural \gls{itd}-\gls{ild} cues,
which is in agreement with the experimental findings of Leakey in~\cite{leakey1959some}.
It may also be observed in \figurename~\ref{fig:hictdicldcentre} 
that large \gls{icld} values (outside $\approx \pm 13$ dB) result in a
low localization uncertainty for all \glspl{ictd}.
Here, one loudspeaker signal is masking the other.

Finally, it can be  observed that for  \glspl{ictd} around $\pm 0.3$~ms
the localization uncertainty increases even in the first and third quadrants, where  \glspl{ictd}-\glspl{icld} pairs are consistent.
At  \glspl{ictd} of approximately $\pm 0.3$~ms, the two loudspeaker signals arrive at the same time at one of the two ears (more specifically, at the left ear for $\ictd\approx+0.3$~ms and at the right ear for  $\ictd\approx-0.3$~ms). 
Appendix~\ref{sec:appoverlap} proves that these \glspl{ictd} can be approximated as
\begin{equation}
    \tau_o \approx\pm \frac{r_h}{c}\left[\cos\left(\theta_e-\frac{\phi_0}{2}\right)+\frac{\phi_0}{2}+\theta_e-\frac{\pi}{2}\right]~,
    \label{eq:tauinter}
\end{equation}
where $r_h$ denotes the head radius and $\theta_e$ denotes the angle between the forward-looking direction and the ear.
For $r_h=9$~cm, $\theta_e=100^\circ$ and $\phi_0=60^\circ$, then $\tau_o=\pm0.27$~ms.
If signals of both loudspeakers arrive at the same time at one of the ears,  that ear effectively receives  one instance of the rendered acoustic event, whereas the other ear receives two instances. 
A number of psychoacoustic studies investigated effects of presenting three coherent stimuli, two to one ear and the third to the other ear. 
It was found that the perceived event had ``complex spatial structure" including cases where, depending on relative delays between the three stimuli, two distinct acoustic sources could be perceived~\cite{BlauertBook}. 
The results of the proposed model are in agreement with the findings of these experiments.
In fact, simulations not shown here for space reasons, confirm that as the head radius $r_h$ changes, the areas with higher \glspl{ictd} in the first and third quadrants move in accordance to~(\ref{eq:tauinter}).
If one wishes to avoid these areas, the \gls{ictd} values should be restricted to the open interval $\ictd\in]-\tau_o, \tau_o[$. 
% The proposed model is well aligned with this phenomenon too.
In \figurename~\ref{fig:hictdicldcentre}, notice how the \gls{psr} operating curve associated to $d=18.7$~cm avoids the areas with higher localization uncertainty around $\ictd \approx \pm 0.3$~ms. 
This is so by construction, as will be discussed later in Section~\ref{sec:timeinteffect}.

\subsection{Localization uncertainty in off-center positions}

\begin{figure}[tb]
\centering
     \begin{subfigure}{0.95\columnwidth}
         \centering
         \includegraphics[width=\textwidth]{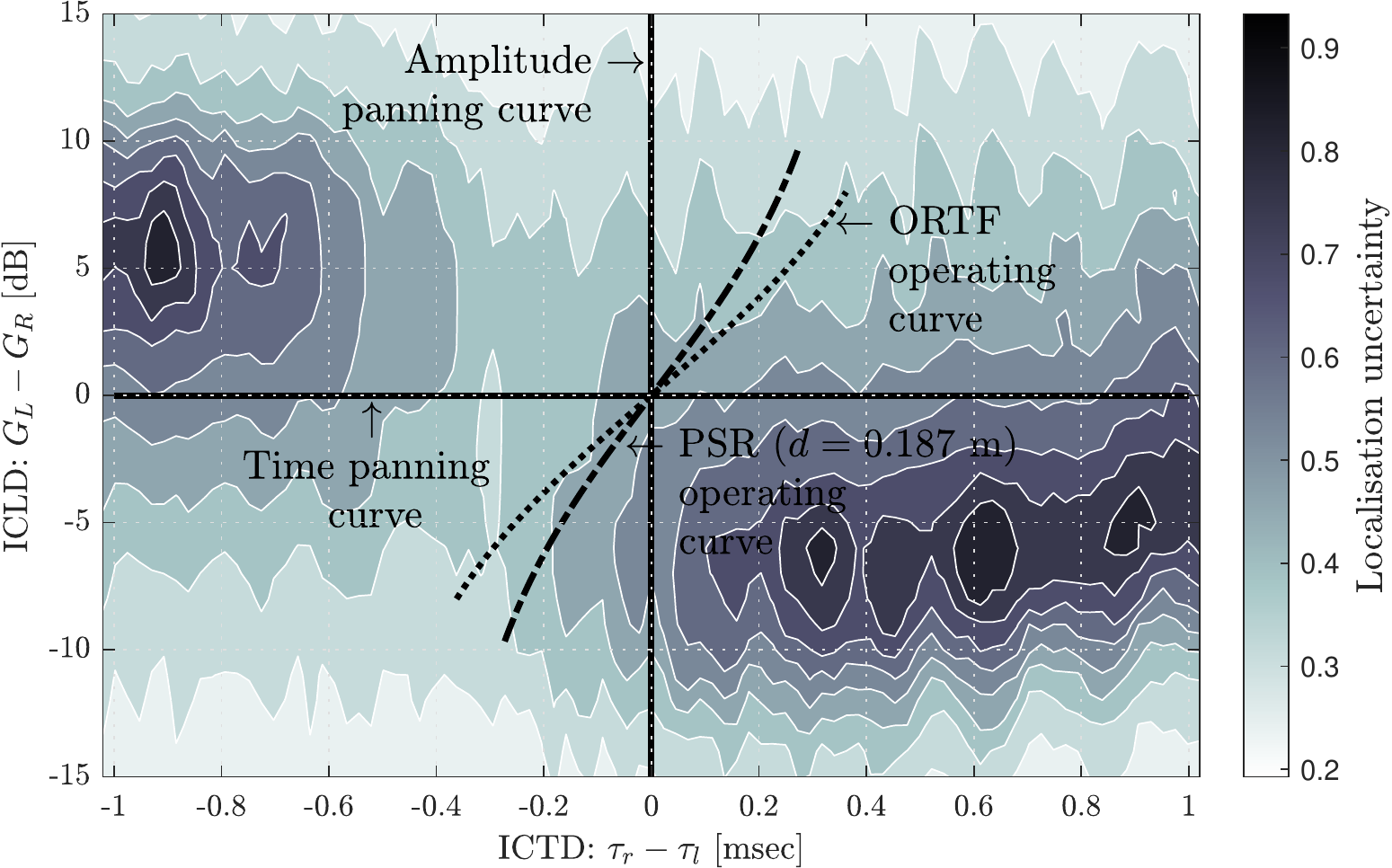}
         \caption{Left off-center position $\mathbf{x}=\{-10,0\}$~cm}
         \label{fig:hictdicldoffcentrel10}
     \end{subfigure}
     \vspace{0.5cm}
     
     \begin{subfigure}{0.95\columnwidth}
         \includegraphics[width=\textwidth]{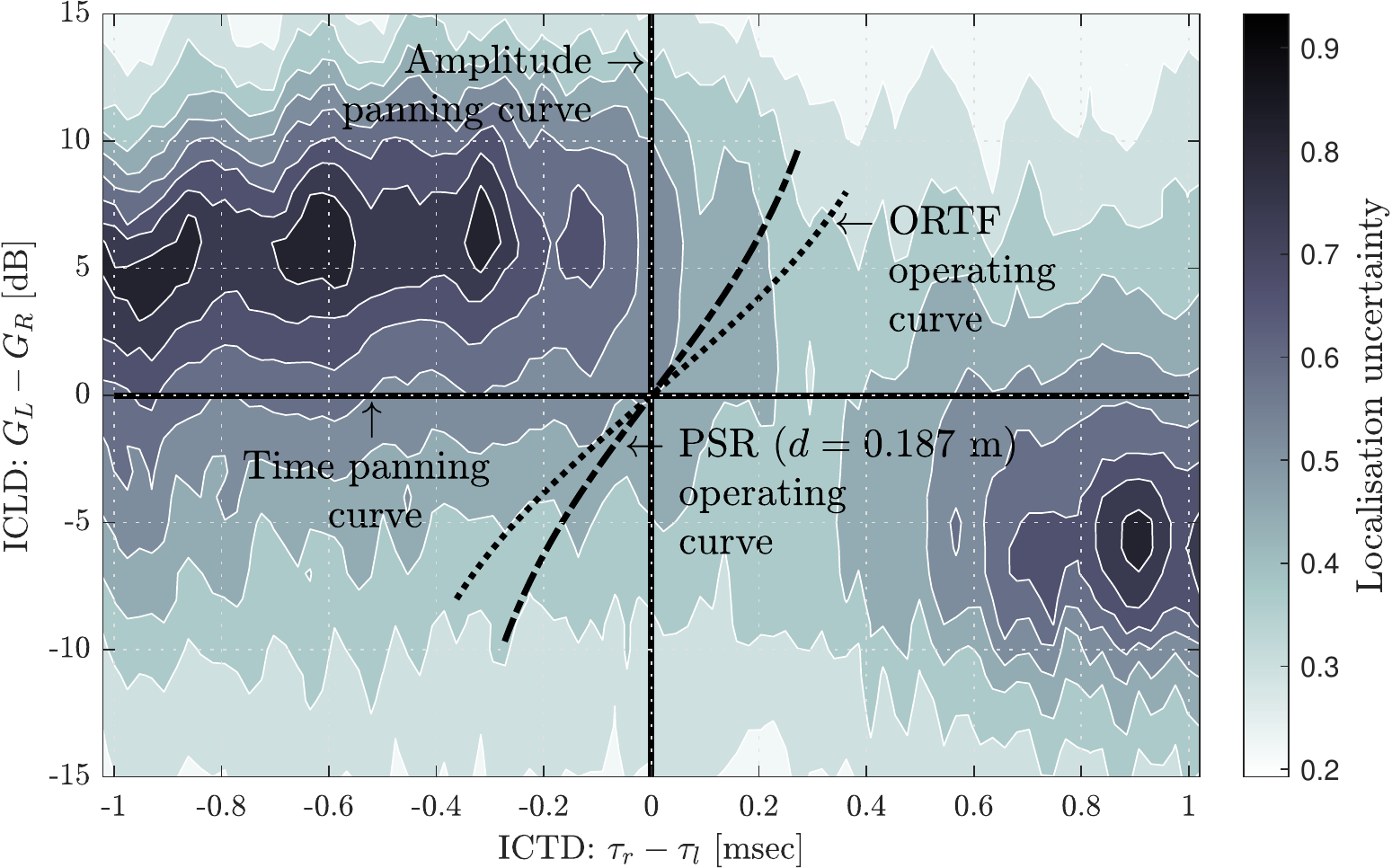}
         \caption{Right off-center position $\mathbf{x}=\{10,0\}$~cm}
         \label{fig:hictdicldoffcentre10}
     \end{subfigure}
     \vspace{0.5cm}
     
     \begin{subfigure}{0.95\columnwidth}
         \includegraphics[width=\columnwidth]{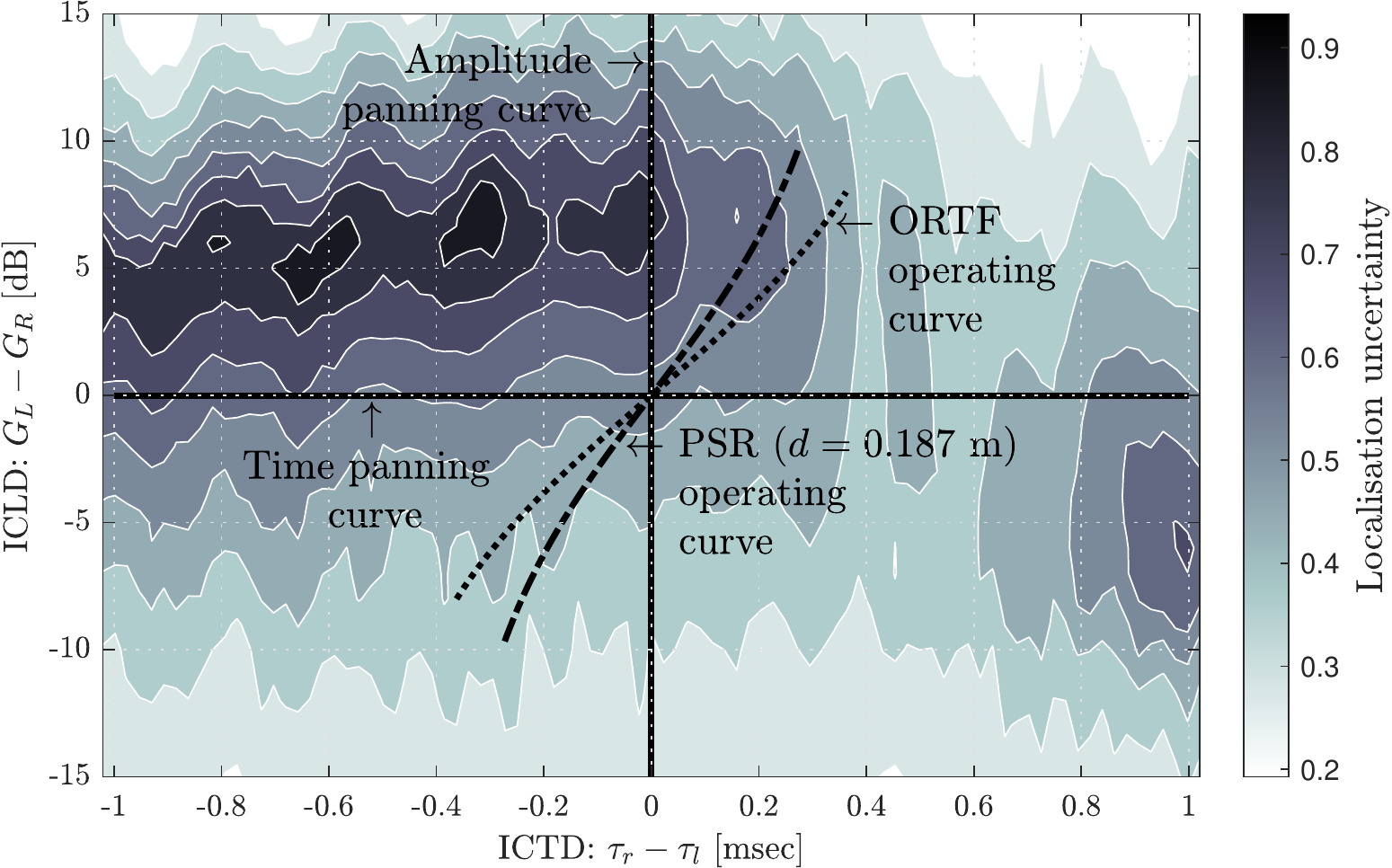}
         \caption{Right off-center position $\mathbf{x}=\{20,0\}$~cm}
         \label{fig:hictdicldoffcentre20}
     \end{subfigure}
     
\caption{Contour plot of localization uncertainty as a function of \gls{icld} and \gls{ictd} for a listener in the off-center positions $\mathbf{x}=\{-10,0\}$~cm,  $\mathbf{x}=\{10,0\}$~cm and $\mathbf{x}=\{10,0\}$~cm. %(see \figurename~\ref{fig:refsystemrelative}). 
Overlaid are the time-amplitude panning curve associated to the \gls{psr} method with $d=18.7$~cm and to the ORTF microphone pair.
}
%The shaded area indicates the region 
%where the predictions produced by the proposed model 
%are unreliable.
\label{fig:hictdicldoffcentre}
\end{figure}

% \begin{figure}[tb]
% \centering
% \includegraphics[width=0.95\columnwidth]{figures/plot_11_offcentre_20cm-eps-converted-to.pdf}
% \caption{Contour plot of localization uncertainty as a function of \gls{icld} and \gls{ictd}
% in the off-center position $\mathbf{x}=\{0.2,0\}$~m (see \figurename~\ref{fig:refsystemrelative}).
% Overlaid is the time-amplitude panning curve associated to the \gls{psr} method with $r_m=0.187$~m.}
% \label{fig:hictdicldoffcentre20}
% \end{figure}

% Shifting the listener to an off-center position affects 
% (a) the relative \gls{ictd}, (b) the \gls{ricld}, and (c) the direction of the loudspeakers relative to the listener. 
% For instance, a listener at a position $20$ cm to the left (i.e. $\{x,y\}=\{-0.2,0\}$ will result in $5$ degrees. 
% 24.79 degrees for the left loudspeaker and 34.71 degrees for the right loudspeaker

Figure~\ref{fig:hictdicldoffcentre} shows the localization uncertainty 
for a listener in a position $10$~cm to the left, and in a position $10$~cm and $20$~cm to the right of the sweet-spot.
The listener is still looking ahead, in a direction parallel to the y-axis.
It may be observed that these plots are almost identical to the on-center plot of~\figurename~\ref{fig:hictdicldcentre} but shifted horizontally.
This implies that the dominant effect is the change in the \glspl{ictd} observed by the listener as a consequence of having moved closer to the right loudspeaker.
The change in observed \glspl{icld}, on the other hand, appears to have a minor effect. 
Likewise, the relative change of direction of the loudspeakers also appears to have a minor effect (in the $0.2$ m position, the two loudspeakers appear at $25^\circ$ and $-35^\circ$ with respect to the listener, compared to $\pm 30^\circ$ for the central position). 

% This confirms that the the effect of \gls{rictd} is dominant with respect to \gls{ricld}. 
% It also indicates that the relative direction change of the loudspeakers has a minor effect (in the $20$ cm position, the two loudspeakers appear at $25^\circ$ and $35^\circ$ with respect to the listener, compared to $\pm 30^\circ$ for the central position). 

Notice how amplitude panning methods, which are associated to the line $\ictd=0$ ms (corresponding to the y-axis), lie in an area with increased localization uncertainty for these off-center positions. 
For the rightward positions in \figurename~\ref{fig:hictdicldoffcentre10} and~\ref{fig:hictdicldoffcentre20}, the localization uncertainty is particularly high for positive \gls{icld} values, which are meant to render phantom sources between the midline and the left loudspeaker. % (i.e. the far loudspeaker in this case).
Viceversa, for the leftward position \figurename~\ref{fig:hictdicldoffcentrel10}, the localization uncertainty is high for negative \gls{icld} values, 
which are meant to render phantom sources between the midline and the right loudspeaker. 
% In other words, a high localization uncertainty is observed whenever the listener moves to the side opposite to the rendered phantom source. 
% or the leftward position  position, 
Notice also how the \gls{psr} panning curve associated to $d=18.7$~cm largely avoids areas with higher localization uncertainty in both leftward and rightward positions $10$~cm off-center. 

% Finally, notice how the bottom half of the plane has a lower overall localization uncertainty. 
% This is the area of the plot where one renders phantom sources in between the midline and the right loudspeaker (the near loudspeaker).

% For this area, relative to the listener, the right loudspeaker leads both in amplitude and in time. 
%This is to a certain degree inevitable. 

\subsection{Definition of relative inter-channel time and level differences}

In order to aid the interpretation of Figures~\ref{fig:hictdicldoffcentre}, it is useful to define two new quantities: the \emph{relative} \gls{ictd} and \gls{icld}, i.e. equivalent inter-channel time and level difference as they are observed at a point away from the center of the loudspeaker array.
These quantities will be referred to as \gls{rictd} and \gls{ricld}, respectively.
% This section also shows that changes in \gls{rictd} are more significant than changes of \gls{ricld} for a given observation point displacement. 

Let the observation point be positioned at $(x,y)$ (see \figurename~\ref{fig:refsystemrelative}).
Appendix~\ref{sec:rictld} provides a closed-form approximation for the \gls{ricld} and \gls{rictd} under the assumption of small displacement compared to the loudspeaker distance: %(i.e. small $|x|/r_l$ and $|y|/r_l$):
\begin{align}
\begin{split}
\ricld & \approx \icld - \frac{x}{r_l} \frac{20\sin\left(\frac{\phi_0}{2}\right)}{\log_e(10)}~, \\
\rictd & \approx \ictd-x\frac{2}{c}\sin\left(\frac{\phi_0}{2}\right)~.
\end{split}
\label{eq:ricltd}
\end{align}

\subsection{Considerations on the effect of \protect\gls{rictd} and \protect\gls{ricld} in amplitude panning}
\label{sec:amplitude}

Consider now again the $\mathbf{x}=\{10,0\}$~cm off-center plot in \figurename~\ref{fig:hictdicldoffcentre10}.
The point $\ictd=0$~ms and $\icld=5$ dB (i.e. the left loudspeaker leads in amplitude) is associated to $\rictd=-0.29$~ms and $\ricld=4.7829$ dB, which have opposite sign and thus provide inconsistent information. 
The \gls{ricld} is now smaller than $\icld=5$~dB because the right loudspeaker is closer to the listener.
The change, however, is small ($0.22$~dB), and the left loudspeaker still leads in amplitude. 
The \gls{rictd}, on the other hand, has changed significantly with respect to \gls{ictd}, and the right loudspeaker is now leading in time. 
To summarise, the left loudspeaker is louder and thus leads in terms of amplitude, but the right loudspeaker is now closer and thus leads in terms of time.
Even though the \gls{ictd} and \gls{icld} were consistent, the horizontal shift of the listener caused the \gls{rictd} and \gls{ricld} to become inconsistent, which, in turn, led to a high localization uncertainty. 

Negative values of \gls{icld}, on the other hand, are less problematic in terms of localization uncertainty for a listener in this position. 
For instance, $\ictd=0$~ms and $\icld=-5$ dB is associated to relative values $\rictd=-0.29$~ms and $\ricld=-5.2171$ dB, which have the same sign and provide consistent information. 

Similar arguments can be made for a listener moving to the left (i.e. $x<0$), as shown in \figurename~\ref{fig:hictdicldoffcentrel10}. 
In this case, however, the plots shift horizontally to the \emph{left} and the critical area of the plot becomes the bottom one.

In summary, amplitude methods lead to higher localization uncertainty whenever the listener moves to one side but aims to render sound sources in directions closer to the opposite side.
Section~\ref{sec:ti} will show that time-amplitude methods provide a lower localization uncertainty in positions away from the center of the sweet-spot. 

\subsection{Further considerations on \protect\gls{rictd} and \protect\gls{ricld}}
Notice how the \gls{ricld} is a function of $\frac{x}{r_l}$, while the \gls{rictd} is independent from $r_l$. 
This implies that for large loudspeaker arrays, movement of the observation point away from the center of the sweet-spot causes a larger change of \gls{rictd} compared to \gls{ricld}.
This holds also even for living-room-sized arrays.
Consider for instance the \gls{rictd} that will result in full perceived shift of the phantom source to one loudspeaker, i.e.  $\rictd=1$ ms.
Assuming $\ictd=0$~ms, $\icld=0$~dB and $\phi_0=\pi/3$ ($60$ degrees), this is already achieved at $x=\frac{0.001 c}{2\sin\left(\phi_0/2\right)}=0.34$ m. 
% %x=0.34$ cm ($x\frac{2}{c} \sin\left(\frac{\phi_0}{2}\right)=1~\text{ms}\Rightarrow{}
% Assuming $\phi_0=\pi/3$ ($60$ deg) and $d=2$ m one 
In that position, the \gls{ricld}, on the other hand, is only $1.5$ dB.
In other words, a phantom source that appears halfway between loudspeakers for a listener in the center of the sweet-spot, collapses onto one of the loudspeakers when the listener moves to a position $34$~cm on either side. 
This compares to approximately $3$~cm on either side for binaural cross-talk cancellation using two  loudspeakers~\cite{rose2002sweet}. 
It should be noted that the degradation is gradual in both cases, which makes defining quantitatively the sweet-spot size difficult.
% A formal comparison of sweet-spot size is beyond the scope of this paper. 
% This value increases with decreasing base angle.
% For instance, for $\phi_0=\pi/9$ ($20$ degrees) $x\approx 1 $~m. 

\begin{figure}[tb]
     \centering
     \includegraphics[width=0.9\columnwidth]{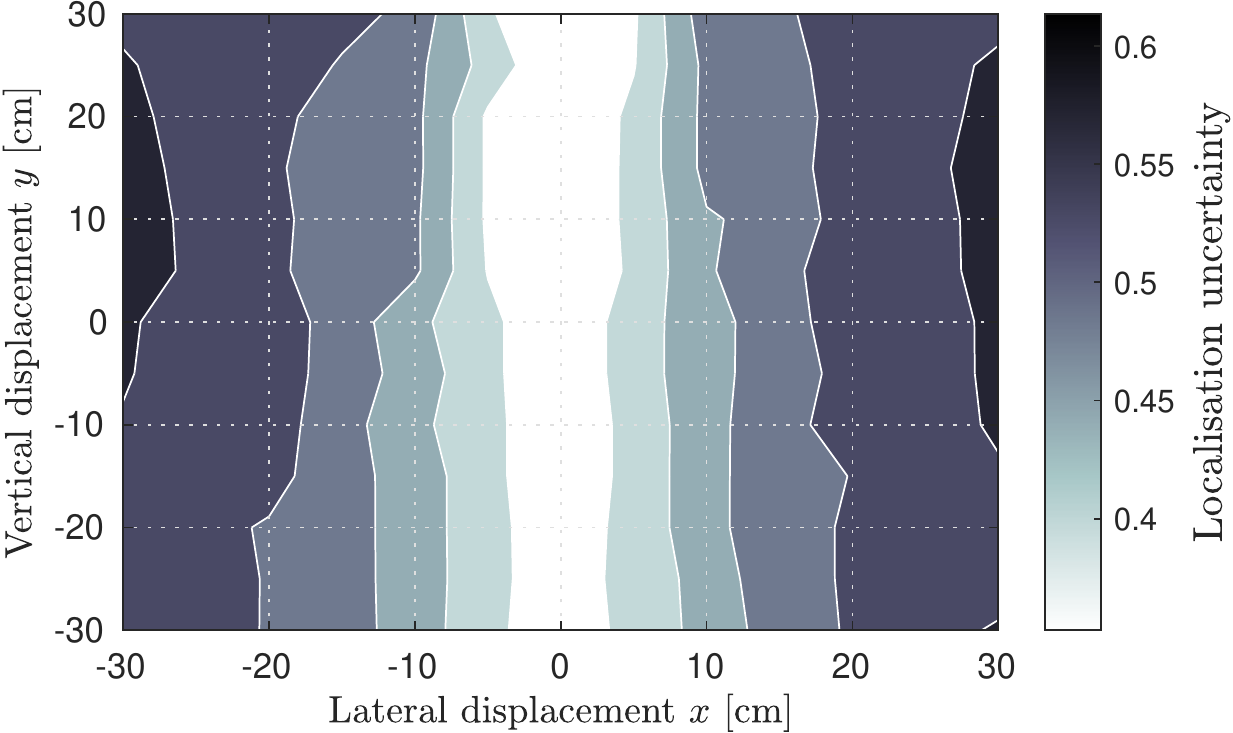}
     \caption{Localization accuracy of PSR for $d=0$~cm (i.e. coincident PSR) as a function of the listener position around the center of the sweet spot.}
     \label{fig:spatial0radius}
\end{figure}

Vertical displacement is less significant than horizontal displacement.
This can be inferred by the fact that the first-order approximation~(\ref{eq:ricltd}) does not  depend on $y$ (it can be shown that $y$ appears starting from the second-order term). 
This is confirmed by \figurename~\ref{fig:spatial0radius}, which shows the localization uncertainty as a function of listener position for PSR panning/recording associated to $d=0$~cm. 
The plot shows that the localization uncertainty varies almost exclusively as a function of the lateral displacement, $x$. 
% This indicates that vertical displacement has a weaker effect than horizontal displacement.
For this reason, the simulations in the remainder of the paper will focus on listener displacement along the $x$ axis ($y=0$).

\section{Time-Amplitude Panning/Recording Methods}
\label{sec:titrading}

% Before proceeding further, let the panning curve be defined by the two parametric functions, $\icld(\theta_s)$ and $\ictd(\theta_s)$.
% It is convenient to think of the parameter $\theta_s$ as the intended direction of the phantom source (although this is not necessary).
% It is reasonable to choose them as continuous and monotonic functions of $\theta_s$. 
% Other than these two conditions, one can choose them as desired. 

This section will show how small \glspl{ictd} allow to render consistent \gls{ricld}-\gls{rictd} values (and thus a lower localization uncertainty) in a slightly larger area around the center of the sweet-spot. 
The starting point of the analysis will be PSR panning/recording, which enables to vary the time-amplitude trade-off using the the inter-microphone distance, $d$, as a free parameter. 
Then, the performance of popular microphone arrangements will be assessed.
The section concludes with some comments on how the results extend to the multichannel case.

\subsection{Effect of time-amplitude trading on localization uncertainty using PSR panning/recording}
\label{sec:timeinteffect}

\begin{figure}[tb]
     \centering
     \begin{subfigure}{\columnwidth}
         \centering
         \includegraphics[width=\textwidth]{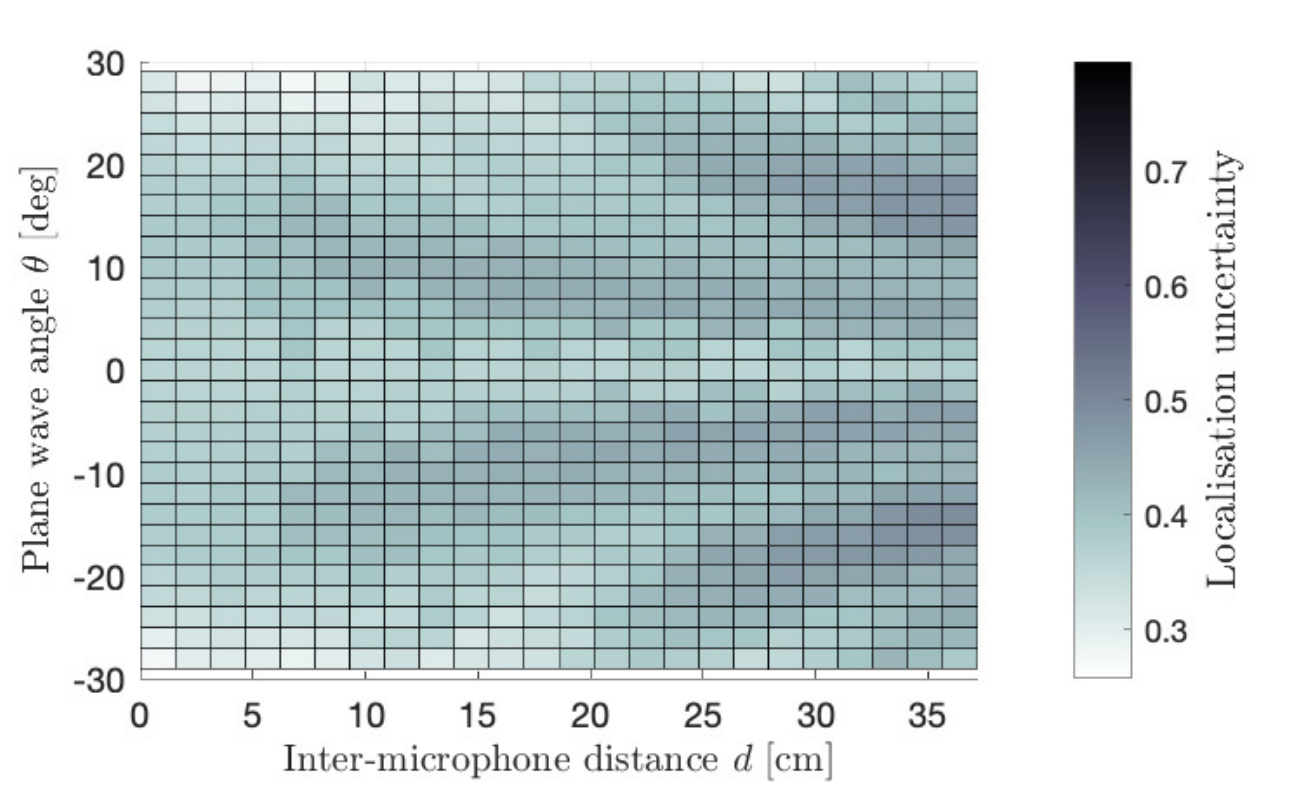}
         \caption{On-center}
         \label{fig:meshoncenter}
     \end{subfigure}
     
     \begin{subfigure}{\columnwidth}
         \centering
         \includegraphics[width=\textwidth]{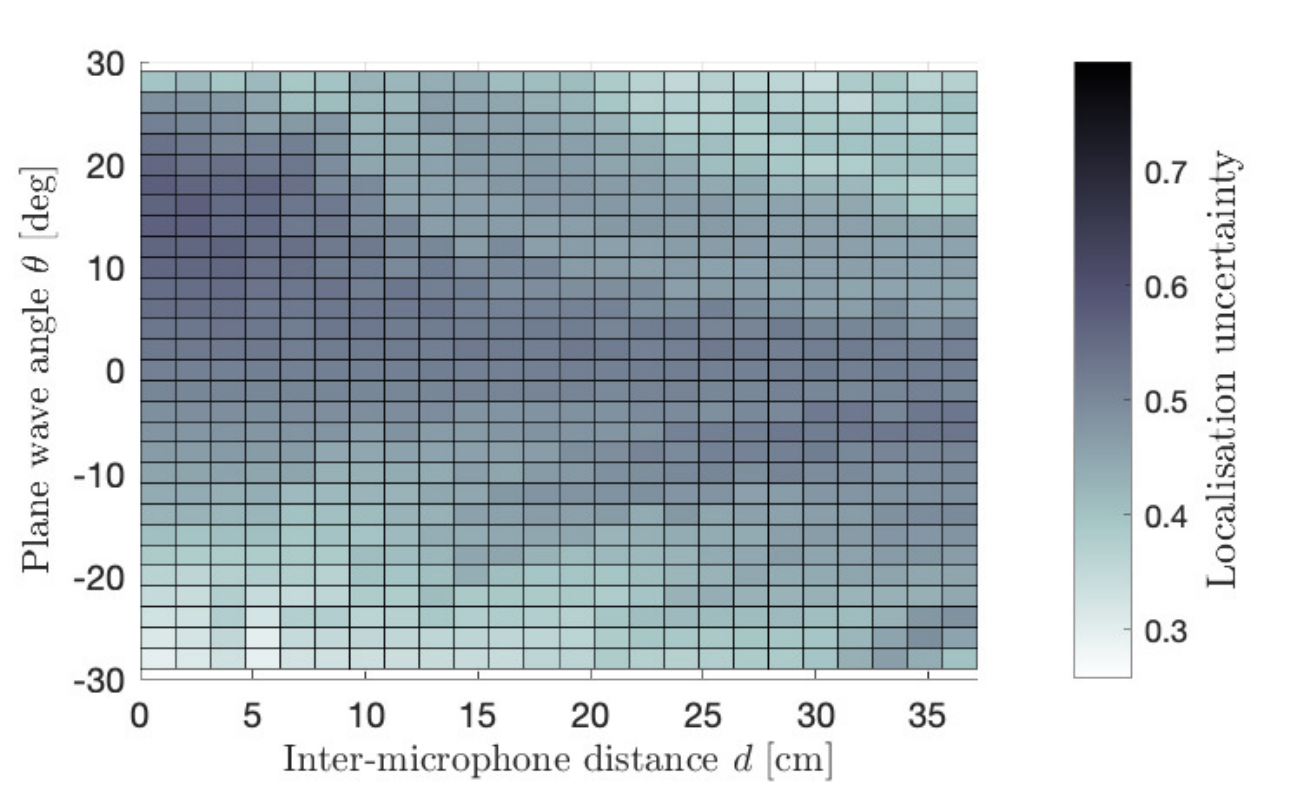}
         \caption{Off-center by $10$~cm}
         \label{fig:mesh10cm}
     \end{subfigure}
     
     \begin{subfigure}{\columnwidth}
         \centering
         \includegraphics[width=\textwidth]{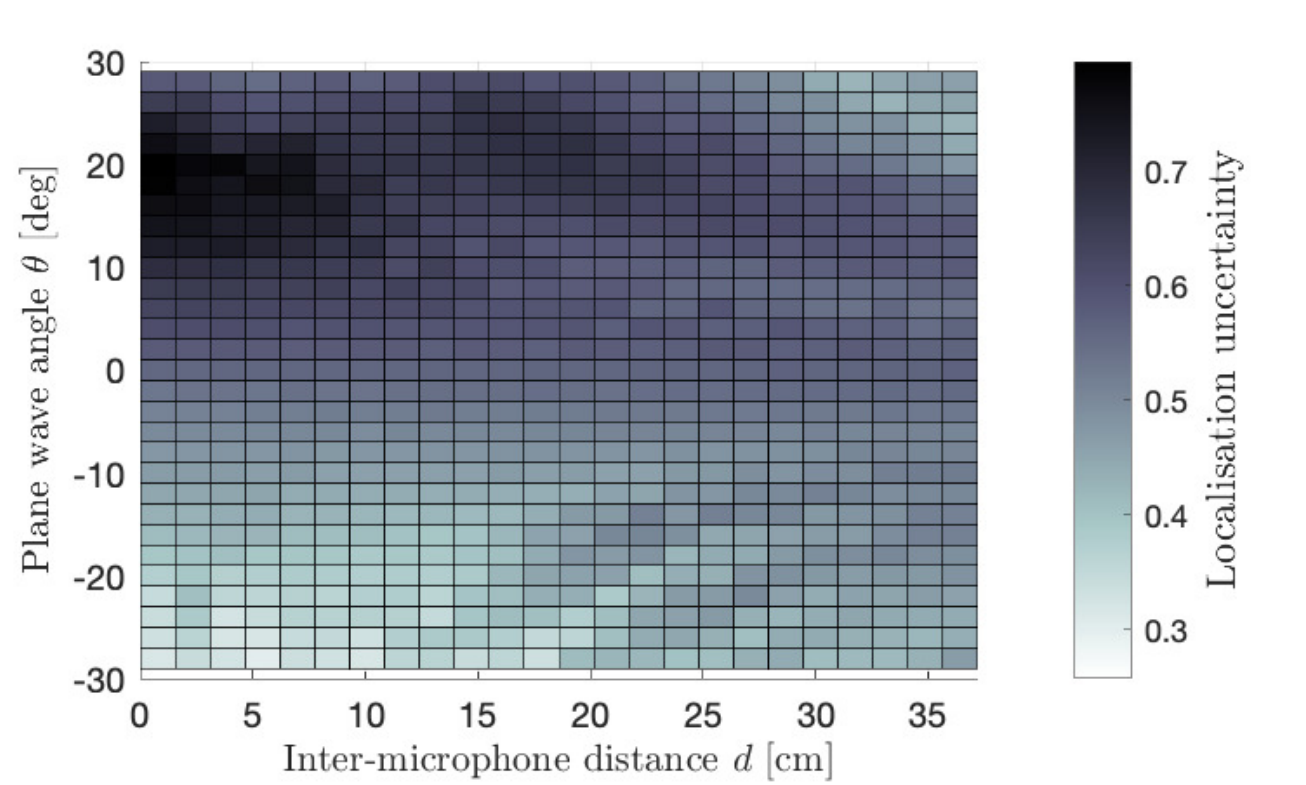}
         \caption{Off-center by $20$~cm}
         \label{fig:mesh20cm}
     \end{subfigure}
        \caption{Localization uncertainty as a function of inter-microphone distance, $d$, and plane wave angle, $\theta_s$, in (a) the center position, (b) in a position $10$~cm off-center to the right, i.e. $(x,y)=(10, 0)$~cm, and (c) in a position $20$~cm off-center to the right, i.e. $(x,y)=(20, 0)$~cm. The colorbar is the same for all three figures to facilitate a comparison between the three figures.}
        \label{fig:radiusangle}
\end{figure}

\figurename~\ref{fig:radiusangle} shows the localization uncertainty for PSR panning/recording as a function of inter-microphone distance $d$ and plane wave angle $\theta_s$ for three different observation positions. 
A number of observations can be made:
\begin{itemize}[leftmargin=*]
    \item As the distance from the center of the sweet spot increases, so does the maximum localization uncertainty. 
    \item In the center position (\figurename~\ref{fig:meshoncenter}), the localization uncertainty is lowest around $d=0$ (i.e. amplitude methods). Two ``sidebands'' appear in the plot beyond $d=20$~cm. These sidebands correspond to the vertical stripes observed in  \figurename~\ref{fig:hictdicldcentre} at $\pm\tau_o$.
    \item At $x=10$~cm (\figurename~\ref{fig:mesh10cm}), the localization uncertainty is higher for inter-microphone distance $d=0$~cm, i.e. amplitude methods.
    It remains high in plane wave directions around the midline, while it reduces for other directions, especially around the $\theta_s\in[10^\circ, 20^\circ]$ range.
    This can be observed also in \figurename~\ref{fig:hictdicldoffcentre10}, which shows the panning curve associated to $d=18.7$~cm. 
    Here, the panning curves avoids the areas with higher localization uncertainty coming in from the left.
    \item At $x=20$~cm (\figurename~\ref{fig:mesh20cm}), the localization uncertainty tends to reduce for higher $d$, where it also has a lower variability as a function of source angles $\theta_s$. 
\end{itemize}

\begin{figure}[tb]
\small
    \centering
    % \newlength\figH
    % \newlength\figW
    \setlength{\figH}{5cm}
    \setlength{\figW}{0.8\columnwidth}
\definecolor{mycolor1}{rgb}{0.00000,0.44700,0.74100}%
\definecolor{mycolor2}{rgb}{0.85000,0.32500,0.09800}%
\definecolor{mycolor3}{rgb}{0.92900,0.69400,0.12500}%
\begin{tikzpicture}

\begin{axis}[%
width=0.951\figW,
height=\figH,
at={(0\figW,0\figH)},
scale only axis,
xmin=0,
xmax=37.2,
xlabel style={font=\color{white!15!black}},
xlabel={Inter-microphone distance $d$ [cm]},
ymin=0.35,
ymax=0.5,
ylabel style={font=\color{white!15!black}},
ylabel={Localization uncertainty},
axis background/.style={fill=white},
xmajorgrids,
ymajorgrids,
legend style={legend cell align=left, align=left, draw=white!15!black}
]
\addplot [color=mycolor1, mark=x, mark options={solid, mycolor1}]
  table[row sep=crcr]{%
0	0.358660539215686\\
1.55	0.351188725490196\\
3.1	0.354917892156863\\
4.65	0.36347181372549\\
6.2	0.368109068627451\\
7.75	0.365964460784314\\
9.3	0.369942401960784\\
10.85	0.368801470588235\\
12.4	0.375819852941176\\
13.95	0.375492647058824\\
15.5	0.376355392156863\\
17.05	0.382660539215686\\
18.6	0.384345588235294\\
20.15	0.392546568627451\\
21.7	0.397316176470588\\
23.25	0.400998774509804\\
24.8	0.400566176470588\\
26.35	0.403080882352941\\
27.9	0.405324754901961\\
29.45	0.416041666666667\\
31	0.416982843137255\\
32.55	0.421536764705882\\
34.1	0.422264705882353\\
35.65	0.424251225490196\\
37.2	0.428109068627451\\
};
\addlegendentry{On-center (x=0 cm)}

\addplot [color=mycolor2, mark=o, mark options={solid, mycolor2}]
  table[row sep=crcr]{%
0	0.437096145276292\\
1.55	0.432830548128342\\
3.1	0.430067067736185\\
4.65	0.42553453654189\\
6.2	0.42401559714795\\
7.75	0.421992758467023\\
9.3	0.420120209447415\\
10.85	0.417607620320856\\
12.4	0.416632575757576\\
13.95	0.413791555258467\\
15.5	0.412330102495544\\
17.05	0.410160427807487\\
18.6	0.409578319964349\\
20.15	0.408244875222816\\
21.7	0.409336007130125\\
23.25	0.410055369875223\\
24.8	0.411062277183601\\
26.35	0.411401626559715\\
27.9	0.413787321746881\\
29.45	0.415368426916221\\
31	0.417672682709447\\
32.55	0.421278966131907\\
34.1	0.423117758467023\\
35.65	0.425423908199643\\
37.2	0.429678364527629\\
};
\addlegendentry{$\text{Averaged x}\in\text{[0,5] cm}$}

\addplot [color=mycolor3, mark=diamond, mark options={solid, mycolor3}]
  table[row sep=crcr]{%
0	0.478316253063726\\
1.55	0.474140548406863\\
3.1	0.470401118259804\\
4.65	0.466030790441177\\
6.2	0.463879901960784\\
7.75	0.460489276960784\\
9.3	0.455851179534314\\
10.85	0.451495404411765\\
12.4	0.448521905637255\\
13.95	0.443562806372549\\
15.5	0.441304993872549\\
17.05	0.43702650122549\\
18.6	0.435516773897059\\
20.15	0.432397977941177\\
21.7	0.431607230392157\\
23.25	0.430365042892157\\
24.8	0.429470741421569\\
26.35	0.42783509497549\\
27.9	0.427975566789216\\
29.45	0.427351026348039\\
31	0.428095741421569\\
32.55	0.43041659007353\\
34.1	0.430014859068627\\
35.65	0.431385186887255\\
37.2	0.433890318627451\\
};
\addlegendentry{$\text{Averaged x}\in\text{[0,15] cm}$}

\end{axis}
\end{tikzpicture}%
    \caption{Localization uncertainty as a function of inter-microphone distance~$d$. Results are averaged across angles between the midline and the left loudspeaker.}
    \label{fig:cumulative}
\end{figure}
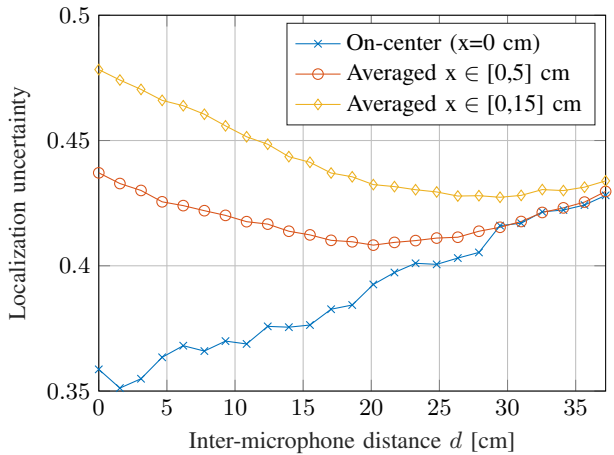

\figurename~\ref{fig:cumulative} shows the localization uncertainty as a function of inter-microphone distance $d$. % averaged across directions $\theta_s$. 
% The top curve considers only the central position.
Here the results are averaged across angles between the midline and the left loudspeaker, which are more difficult to render when the listener moves to the right of the center.
The curve associated to the center position shows that, in that position, the inter-microphone distance with the lowest uncertainty is close to $d=0$~cm, as expected.
% Around $d=0.2$~m the uncertainty has a dip before increasing further, which is due to the sidebands of \figurename~\ref{fig:hictdicldcentre}.
In typical use cases, one would like to minimize the uncertainty as it is observed in a whole region around the center of the sweet spot, instead of specific positions. 
\figurename~\ref{fig:cumulative} shows the localization uncertainty averaged for listener displacements between $x=0$~cm and $x=5$~cm (with a $1$~cm resolution). 
Here, the absolute minimum is achieved at $d=20$~cm. 
\figurename~\ref{fig:cumulative} also shows the localization uncertainty averaged for displacements between $x=0$~cm and $x=15$~cm. 
In this case, the average uncertainty increases significantly at small $d$, and the absolute minimum moved to $d\approx 30$~cm.

In summary, the results suggest that smaller inter-microphone distances (amplitude methods) are preferable for positions close to the center, while larger inter-microphone distances result in lower localization uncertainty at larger distances. 
% In other words, the more the listener moves away from the center, the more advantageous it is to rely on \glspl{ictd}.
There is thus a trade off between lower localization uncertainty in the center of the sweet spot and away from the center.

% Notice that the upper limit for the array radius cannot be probed using the proposed model as it does not account for precedence effect, which is critical for larger \gls{rictd}.

Notice that if one can assume that only one listener is present, and that some form of user tracking is available, then the results here suggest that it would be beneficial to correct the $\ictd$s such that the $\rictd=0$. 
In other words, if the location of the (single) listener is known, one can correct the time delays such that the listener experiences amplitude panning. 

If, on the other hand, the listener's position is not being tracked, it is typically difficult to predict how far the listener/s will move away from the center of the sweet spot. 
A possible compromise is to use the inter-microphone distance that results in the largest \gls{ictd} to be no larger than $\tau_o$, and thus avoid the areas with slightly higher localization uncertainty in the first and third quadrant shown in \figurename~\ref{fig:hictdicldcentre}.
One can find this inter-microphone distance by equating $\ictd_\text{max}$ with $\tau_o$, and isolating $d$:
% \begin{equation}
%     d= r_h \frac{\cos\left(\theta_e-\frac{\phi_0}{2}\right)+\frac{\phi_0}{2}+\theta_e-\frac{\pi}{2}}{2\sin^2\left(\frac{\phi_0}{2}\right)}~.
%     \label{eq:optrm}
% \end{equation}
\begin{equation}
    d= r_h \frac{\cos\left(\theta_e-\frac{\phi_0}{2}\right)+\frac{\phi_0}{2}+\theta_e-\frac{\pi}{2}}{\sin\left(\frac{\phi_0}{2}\right)}~.
    \label{eq:optrm}
\end{equation}
For instance, the inter-microphone distance for $\theta_e=100^\circ$ and $r_h=9$~cm, is $d=18.7$~cm, which is the one used for the panning curve shown in \figurename~\ref{fig:hictdicldcentre} and \ref{fig:hictdicldoffcentre}.

\subsection{Comparison with popular stereophonic microphone arrangements}

This subsection presents the results of a comparison between three versions of PSR for $d=0, 18.7, 37.2$~cm, and the popular stereophonic microphone arrangements summarized in Table~\ref{table:mics}. 
The comparison involves choosing which source angles $\theta_s$ to simulate. 
% The considered source angles $\theta_s$ are uniformly spread across the coverage angle. 
While it is clear that for PSR the source angles should be within the microphone base-angle ($60^\circ$), the choice for other microphone arrangements is ambiguous. 
Indeed, sound engineers typically make a choice of the relative position of the sources of interest (e.g. an orchestra) often based on artistic considerations. %sound sources and reflections will impinge on the microphone pair from all directions.
In order to make a more unambiguous choice here, the range of angles is chosen according to which source angles results in $\ictd$-$\icld$ pairs within the Williams curves.
The resulting ranges are termed here \emph{coverage angles}, and are reported in Table~\ref{table:mics} for the popular microphone arrangements.
In the simulations that follow, a set of thirty $\theta_s$ are uniformly taken within the coverage angle for each microphone arrangement.

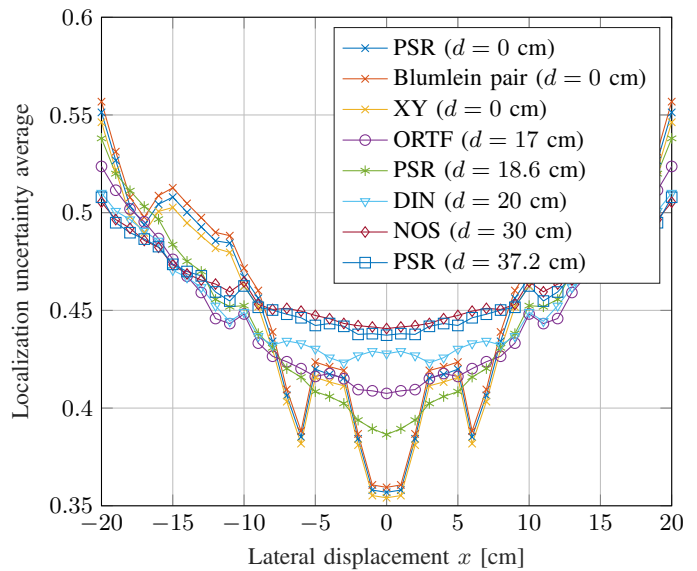
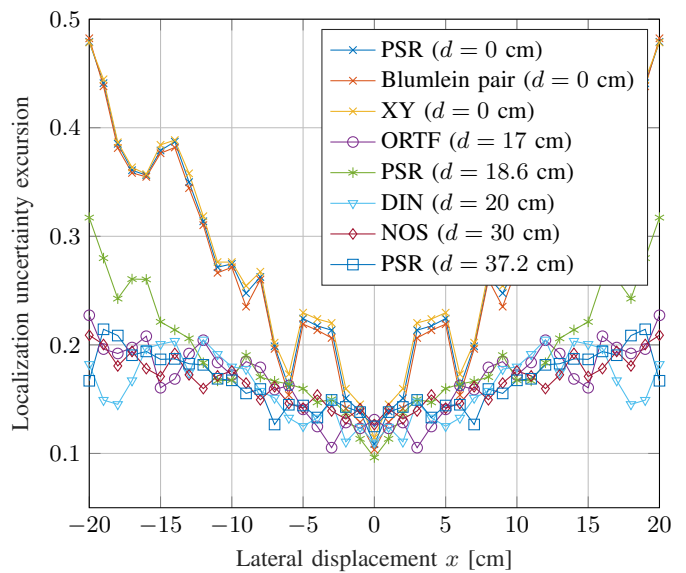
\begin{figure}
\small
     \centering
     \begin{subfigure}{\columnwidth}
         \centering
    \setlength{\figH}{6.5cm}
    \setlength{\figW}{0.9\columnwidth}
\definecolor{mycolor1}{rgb}{0.00000,0.44700,0.74100}%
\definecolor{mycolor2}{rgb}{0.85000,0.32500,0.09800}%
\definecolor{mycolor3}{rgb}{0.92900,0.69400,0.12500}%
\definecolor{mycolor4}{rgb}{0.49400,0.18400,0.55600}%
\definecolor{mycolor5}{rgb}{0.46600,0.67400,0.18800}%
\definecolor{mycolor6}{rgb}{0.30100,0.74500,0.93300}%
\definecolor{mycolor7}{rgb}{0.63500,0.07800,0.18400}%
\begin{tikzpicture}

\begin{axis}[%
width=0.951\figW,
height=\figH,
at={(0\figW,0\figH)},
scale only axis,
xmin=-20,
xmax=20,
xlabel style={font=\color{white!15!black}},
xlabel={Lateral displacement $x$ [cm]},
ymin=0.35,
ymax=0.6,
ylabel style={font=\color{white!15!black}},
ylabel={Localization uncertainty average},
axis background/.style={fill=white},
xmajorgrids,
ymajorgrids,
legend style={legend cell align=left, align=left, draw=white!15!black}
]
\addplot [color=mycolor1, mark=x, mark options={solid, mycolor1}]
  table[row sep=crcr]{%
-20	0.551311764705882\\
-19	0.526745098039216\\
-18	0.503332026143791\\
-17	0.49378954248366\\
-16	0.504324183006536\\
-15	0.507949019607843\\
-14	0.500176470588235\\
-13	0.492696732026144\\
-12	0.485566013071895\\
-11	0.484445098039216\\
-10	0.466967973856209\\
-9	0.456211764705882\\
-8	0.434133986928105\\
-7	0.406494117647059\\
-6	0.385235294117647\\
-5	0.419883006535948\\
-4	0.417230718954248\\
-3	0.415461437908497\\
-2	0.384271895424837\\
-1	0.35803137254902\\
0	0.356908496732026\\
1	0.35803137254902\\
2	0.384271895424837\\
3	0.415461437908497\\
4	0.417230718954248\\
5	0.419883006535948\\
6	0.385235294117647\\
7	0.406494117647059\\
8	0.434133986928105\\
9	0.456211764705882\\
10	0.466967973856209\\
11	0.484445098039216\\
12	0.485566013071895\\
13	0.492696732026144\\
14	0.500176470588235\\
15	0.507949019607843\\
16	0.504324183006536\\
17	0.49378954248366\\
18	0.503332026143791\\
19	0.526745098039216\\
20	0.551311764705882\\
};
\addlegendentry{PSR ($d=0$ cm)}

\addplot [color=mycolor2, mark=x, mark options={solid, mycolor2}]
  table[row sep=crcr]{%
-20	0.556802614379085\\
-19	0.531205882352941\\
-18	0.507819607843137\\
-17	0.497661437908497\\
-16	0.508703921568627\\
-15	0.512713725490196\\
-14	0.50472614379085\\
-13	0.497441176470588\\
-12	0.489903921568627\\
-11	0.488224836601307\\
-10	0.471603921568627\\
-9	0.460157516339869\\
-8	0.438952941176471\\
-7	0.409530718954248\\
-6	0.388269281045752\\
-5	0.423535947712418\\
-4	0.421169281045752\\
-3	0.419258823529412\\
-2	0.386782352941176\\
-1	0.360681699346405\\
0	0.359432026143791\\
1	0.360681699346405\\
2	0.386782352941176\\
3	0.419258823529412\\
4	0.421169281045752\\
5	0.423535947712418\\
6	0.388269281045752\\
7	0.409530718954248\\
8	0.438952941176471\\
9	0.460157516339869\\
10	0.471603921568627\\
11	0.488224836601307\\
12	0.489903921568627\\
13	0.497441176470588\\
14	0.50472614379085\\
15	0.512713725490196\\
16	0.508703921568627\\
17	0.497661437908497\\
18	0.507819607843137\\
19	0.531205882352941\\
20	0.556802614379085\\
};
\addlegendentry{Blumlein pair ($d=0$ cm)}

\addplot [color=mycolor3, mark=x, mark options={solid, mycolor3}]
  table[row sep=crcr]{%
-20	0.546278431372549\\
-19	0.521824836601307\\
-18	0.498067320261438\\
-17	0.489619607843137\\
-16	0.500628104575163\\
-15	0.502660784313726\\
-14	0.494708496732026\\
-13	0.488086274509804\\
-12	0.48179477124183\\
-11	0.479648366013072\\
-10	0.462387581699346\\
-9	0.451841830065359\\
-8	0.430075163398693\\
-7	0.403301307189542\\
-6	0.381815032679739\\
-5	0.415698039215686\\
-4	0.413338562091503\\
-3	0.411396078431373\\
-2	0.381060784313725\\
-1	0.355226797385621\\
0	0.354107189542484\\
1	0.355226797385621\\
2	0.381060784313725\\
3	0.411396078431373\\
4	0.413338562091503\\
5	0.415698039215686\\
6	0.381815032679739\\
7	0.403301307189542\\
8	0.430075163398693\\
9	0.451841830065359\\
10	0.462387581699346\\
11	0.479648366013072\\
12	0.48179477124183\\
13	0.488086274509804\\
14	0.494708496732026\\
15	0.502660784313726\\
16	0.500628104575163\\
17	0.489619607843137\\
18	0.498067320261438\\
19	0.521824836601307\\
20	0.546278431372549\\
};
\addlegendentry{XY ($d=0$ cm)}

\addplot [color=mycolor4, mark=o, mark options={solid, mycolor4}]
  table[row sep=crcr]{%
-20	0.523656209150327\\
-19	0.511476470588235\\
-18	0.501914379084967\\
-17	0.495333333333333\\
-16	0.48677385620915\\
-15	0.476107843137255\\
-14	0.467441830065359\\
-13	0.459152941176471\\
-12	0.445816339869281\\
-11	0.443232026143791\\
-10	0.448143137254902\\
-9	0.433233986928104\\
-8	0.426513071895425\\
-7	0.423724183006536\\
-6	0.420190196078431\\
-5	0.416186274509804\\
-4	0.417684967320261\\
-3	0.415014379084967\\
-2	0.409452287581699\\
-1	0.408835294117647\\
0	0.407503921568627\\
1	0.408835294117647\\
2	0.409452287581699\\
3	0.415014379084967\\
4	0.417684967320261\\
5	0.416186274509804\\
6	0.420190196078431\\
7	0.423724183006536\\
8	0.426513071895425\\
9	0.433233986928104\\
10	0.448143137254902\\
11	0.443232026143791\\
12	0.445816339869281\\
13	0.459152941176471\\
14	0.467441830065359\\
15	0.476107843137255\\
16	0.48677385620915\\
17	0.495333333333333\\
18	0.501914379084967\\
19	0.511476470588235\\
20	0.523656209150327\\
};
\addlegendentry{ORTF ($d=17$ cm)}

\addplot [color=mycolor5, mark=asterisk, mark options={solid, mycolor5}]
  table[row sep=crcr]{%
-20	0.538022222222222\\
-19	0.519953594771242\\
-18	0.511241830065359\\
-17	0.503178431372549\\
-16	0.496661437908497\\
-15	0.48361045751634\\
-14	0.475103267973856\\
-13	0.469952287581699\\
-12	0.455601960784314\\
-11	0.451862091503268\\
-10	0.452424183006536\\
-9	0.438098039215686\\
-8	0.430866013071895\\
-7	0.420201960784314\\
-6	0.415857516339869\\
-5	0.408412418300654\\
-4	0.405932679738562\\
-3	0.402401307189543\\
-2	0.393828758169935\\
-1	0.389433333333333\\
0	0.386458169934641\\
1	0.389433333333333\\
2	0.393828758169935\\
3	0.402401307189543\\
4	0.405932679738562\\
5	0.408412418300654\\
6	0.415857516339869\\
7	0.420201960784314\\
8	0.430866013071895\\
9	0.438098039215686\\
10	0.452424183006536\\
11	0.451862091503268\\
12	0.455601960784314\\
13	0.469952287581699\\
14	0.475103267973856\\
15	0.48361045751634\\
16	0.496661437908497\\
17	0.503178431372549\\
18	0.511241830065359\\
19	0.519953594771242\\
20	0.538022222222222\\
};
\addlegendentry{PSR ($d=18.6$ cm)}

\addplot [color=mycolor6, mark=triangle, mark options={solid, rotate=180, mycolor6}]
  table[row sep=crcr]{%
-20	0.510045751633987\\
-19	0.50080522875817\\
-18	0.496467973856209\\
-17	0.490545751633987\\
-16	0.485562091503268\\
-15	0.47057908496732\\
-14	0.466828104575163\\
-13	0.461775163398693\\
-12	0.452313725490196\\
-11	0.444384967320261\\
-10	0.449069281045752\\
-9	0.43674183006536\\
-8	0.432273202614379\\
-7	0.434333333333333\\
-6	0.433050980392157\\
-5	0.430269281045752\\
-4	0.425769934640523\\
-3	0.423203921568627\\
-2	0.426884967320261\\
-1	0.429050980392157\\
0	0.427834640522876\\
1	0.429050980392157\\
2	0.426884967320261\\
3	0.423203921568627\\
4	0.425769934640523\\
5	0.430269281045752\\
6	0.433050980392157\\
7	0.434333333333333\\
8	0.432273202614379\\
9	0.43674183006536\\
10	0.449069281045752\\
11	0.444384967320261\\
12	0.452313725490196\\
13	0.461775163398693\\
14	0.466828104575163\\
15	0.47057908496732\\
16	0.485562091503268\\
17	0.490545751633987\\
18	0.496467973856209\\
19	0.50080522875817\\
20	0.510045751633987\\
};
\addlegendentry{DIN ($d=20$ cm)}

\addplot [color=mycolor7, mark=diamond, mark options={solid, mycolor7}]
  table[row sep=crcr]{%
-20	0.505378431372549\\
-19	0.496477777777778\\
-18	0.492035947712418\\
-17	0.485783006535948\\
-16	0.48229477124183\\
-15	0.47330522875817\\
-14	0.468645098039216\\
-13	0.465529411764706\\
-12	0.463467973856209\\
-11	0.459495424836601\\
-10	0.464654248366013\\
-9	0.453074509803922\\
-8	0.450477124183007\\
-7	0.451028758169935\\
-6	0.449607189542484\\
-5	0.447397385620915\\
-4	0.445749673202614\\
-3	0.443312418300654\\
-2	0.442170588235294\\
-1	0.441500653594771\\
0	0.440694117647059\\
1	0.441500653594771\\
2	0.442170588235294\\
3	0.443312418300654\\
4	0.445749673202614\\
5	0.447397385620915\\
6	0.449607189542484\\
7	0.451028758169935\\
8	0.450477124183007\\
9	0.453074509803922\\
10	0.464654248366013\\
11	0.459495424836601\\
12	0.463467973856209\\
13	0.465529411764706\\
14	0.468645098039216\\
15	0.47330522875817\\
16	0.48229477124183\\
17	0.485783006535948\\
18	0.492035947712418\\
19	0.496477777777778\\
20	0.505378431372549\\
};
\addlegendentry{NOS ($d=30$ cm)}

\addplot [color=mycolor1, mark=square, mark options={solid, mycolor1}]
  table[row sep=crcr]{%
-20	0.507885620915033\\
-19	0.494853594771242\\
-18	0.489749019607843\\
-17	0.486483660130719\\
-16	0.482635294117647\\
-15	0.473556209150327\\
-14	0.46988954248366\\
-13	0.467648366013072\\
-12	0.45941045751634\\
-11	0.454771895424837\\
-10	0.462490849673203\\
-9	0.451579738562091\\
-8	0.450169934640523\\
-7	0.448030718954248\\
-6	0.44622091503268\\
-5	0.442278431372549\\
-4	0.44326862745098\\
-3	0.44183660130719\\
-2	0.437713725490196\\
-1	0.438073202614379\\
0	0.437309803921569\\
1	0.438073202614379\\
2	0.437713725490196\\
3	0.44183660130719\\
4	0.44326862745098\\
5	0.442278431372549\\
6	0.44622091503268\\
7	0.448030718954248\\
8	0.450169934640523\\
9	0.451579738562091\\
10	0.462490849673203\\
11	0.454771895424837\\
12	0.45941045751634\\
13	0.467648366013072\\
14	0.46988954248366\\
15	0.473556209150327\\
16	0.482635294117647\\
17	0.486483660130719\\
18	0.489749019607843\\
19	0.494853594771242\\
20	0.507885620915033\\
};
\addlegendentry{PSR ($d=37.2$ cm)}

\end{axis}
\end{tikzpicture}%
    \caption{Localization uncertainty averaged across $\theta_s$ angles.}
    \label{fig:latdispl}
    \end{subfigure}
    \vspace{0.3cm}
    
     \begin{subfigure}{\columnwidth}
         \centering
    \setlength{\figH}{6.5cm}
    \setlength{\figW}{0.9\columnwidth}
\definecolor{mycolor1}{rgb}{0.00000,0.44700,0.74100}%
\definecolor{mycolor2}{rgb}{0.85000,0.32500,0.09800}%
\definecolor{mycolor3}{rgb}{0.92900,0.69400,0.12500}%
\definecolor{mycolor4}{rgb}{0.49400,0.18400,0.55600}%
\definecolor{mycolor5}{rgb}{0.46600,0.67400,0.18800}%
\definecolor{mycolor6}{rgb}{0.30100,0.74500,0.93300}%
\definecolor{mycolor7}{rgb}{0.63500,0.07800,0.18400}%
\begin{tikzpicture}

\begin{axis}[%
width=0.951\figW,
height=\figH,
at={(0\figW,0\figH)},
scale only axis,
xmin=-20,
xmax=20,
xlabel style={font=\color{white!15!black}},
xlabel={Lateral displacement $x$ [cm]},
ymin=0.05,
ymax=0.5,
ylabel style={font=\color{white!15!black}},
ylabel={Localization uncertainty excursion},
axis background/.style={fill=white},
xmajorgrids,
ymajorgrids,
legend style={legend cell align=left, align=left, draw=white!15!black}
]
\addplot [color=mycolor1, mark=x, mark options={solid, mycolor1}]
  table[row sep=crcr]{%
-20	0.479137254901961\\
-19	0.441039215686274\\
-18	0.385235294117647\\
-17	0.360843137254902\\
-16	0.356313725490196\\
-15	0.379058823529412\\
-14	0.387196078431372\\
-13	0.349901960784314\\
-12	0.313843137254902\\
-11	0.271509803921569\\
-10	0.274588235294118\\
-9	0.247588235294118\\
-8	0.262549019607843\\
-7	0.198705882352941\\
-6	0.161039215686275\\
-5	0.22421568627451\\
-4	0.21743137254902\\
-3	0.21356862745098\\
-2	0.150764705882353\\
-1	0.138745098039216\\
0	0.108588235294118\\
1	0.138745098039216\\
2	0.150764705882353\\
3	0.21356862745098\\
4	0.21743137254902\\
5	0.22421568627451\\
6	0.161039215686275\\
7	0.198705882352941\\
8	0.262549019607843\\
9	0.247588235294118\\
10	0.274588235294118\\
11	0.271509803921569\\
12	0.313843137254902\\
13	0.349901960784314\\
14	0.387196078431372\\
15	0.379058823529412\\
16	0.356313725490196\\
17	0.360843137254902\\
18	0.385235294117647\\
19	0.441039215686274\\
20	0.479137254901961\\
};
\addlegendentry{PSR ($d=0$ cm)}

\addplot [color=mycolor2, mark=x, mark options={solid, mycolor2}]
  table[row sep=crcr]{%
-20	0.482392156862745\\
-19	0.438196078431373\\
-18	0.381313725490196\\
-17	0.358411764705882\\
-16	0.354764705882353\\
-15	0.376509803921569\\
-14	0.381823529411765\\
-13	0.344372549019608\\
-12	0.310274509803922\\
-11	0.266392156862745\\
-10	0.271450980392157\\
-9	0.23521568627451\\
-8	0.260078431372549\\
-7	0.196333333333333\\
-6	0.153764705882353\\
-5	0.219058823529412\\
-4	0.213490196078431\\
-3	0.206392156862745\\
-2	0.141254901960784\\
-1	0.128823529411765\\
0	0.103470588235294\\
1	0.128823529411765\\
2	0.141254901960784\\
3	0.206392156862745\\
4	0.213490196078431\\
5	0.219058823529412\\
6	0.153764705882353\\
7	0.196333333333333\\
8	0.260078431372549\\
9	0.23521568627451\\
10	0.271450980392157\\
11	0.266392156862745\\
12	0.310274509803922\\
13	0.344372549019608\\
14	0.381823529411765\\
15	0.376509803921569\\
16	0.354764705882353\\
17	0.358411764705882\\
18	0.381313725490196\\
19	0.438196078431373\\
20	0.482392156862745\\
};
\addlegendentry{Blumlein pair ($d=0$ cm)}

\addplot [color=mycolor3, mark=x, mark options={solid, mycolor3}]
  table[row sep=crcr]{%
-20	0.478549019607843\\
-19	0.444607843137255\\
-18	0.386588235294118\\
-17	0.363549019607843\\
-16	0.357294117647059\\
-15	0.38421568627451\\
-14	0.388607843137255\\
-13	0.357960784313726\\
-12	0.318490196078432\\
-11	0.276274509803922\\
-10	0.276235294117647\\
-9	0.254352941176471\\
-8	0.267372549019608\\
-7	0.202039215686275\\
-6	0.172921568627451\\
-5	0.229647058823529\\
-4	0.22378431372549\\
-3	0.22021568627451\\
-2	0.159588235294118\\
-1	0.145176470588235\\
0	0.114509803921569\\
1	0.145176470588235\\
2	0.159588235294118\\
3	0.22021568627451\\
4	0.22378431372549\\
5	0.229647058823529\\
6	0.172921568627451\\
7	0.202039215686275\\
8	0.267372549019608\\
9	0.254352941176471\\
10	0.276235294117647\\
11	0.276274509803922\\
12	0.318490196078432\\
13	0.357960784313726\\
14	0.388607843137255\\
15	0.38421568627451\\
16	0.357294117647059\\
17	0.363549019607843\\
18	0.386588235294118\\
19	0.444607843137255\\
20	0.478549019607843\\
};
\addlegendentry{XY ($d=0$ cm)}

\addplot [color=mycolor4, mark=o, mark options={solid, mycolor4}]
  table[row sep=crcr]{%
-20	0.227235294117647\\
-19	0.196117647058824\\
-18	0.192078431372549\\
-17	0.197549019607843\\
-16	0.207980392156863\\
-15	0.160509803921569\\
-14	0.168627450980392\\
-13	0.192019607843137\\
-12	0.204294117647059\\
-11	0.18378431372549\\
-10	0.172098039215686\\
-9	0.184588235294118\\
-8	0.179294117647059\\
-7	0.158843137254902\\
-6	0.162588235294118\\
-5	0.140607843137255\\
-4	0.124803921568627\\
-3	0.105313725490196\\
-2	0.128470588235294\\
-1	0.123117647058824\\
0	0.131098039215686\\
1	0.123117647058824\\
2	0.128470588235294\\
3	0.105313725490196\\
4	0.124803921568627\\
5	0.140607843137255\\
6	0.162588235294118\\
7	0.158843137254902\\
8	0.179294117647059\\
9	0.184588235294118\\
10	0.172098039215686\\
11	0.18378431372549\\
12	0.204294117647059\\
13	0.192019607843137\\
14	0.168627450980392\\
15	0.160509803921569\\
16	0.207980392156863\\
17	0.197549019607843\\
18	0.192078431372549\\
19	0.196117647058824\\
20	0.227235294117647\\
};
\addlegendentry{ORTF ($d=17$ cm)}

\addplot [color=mycolor5, mark=asterisk, mark options={solid, mycolor5}]
  table[row sep=crcr]{%
-20	0.31721568627451\\
-19	0.279921568627451\\
-18	0.242607843137255\\
-17	0.260607843137255\\
-16	0.260313725490196\\
-15	0.221196078431372\\
-14	0.213803921568627\\
-13	0.206019607843137\\
-12	0.183549019607843\\
-11	0.166862745098039\\
-10	0.167509803921569\\
-9	0.190529411764706\\
-8	0.170372549019608\\
-7	0.166490196078431\\
-6	0.163921568627451\\
-5	0.159745098039216\\
-4	0.147039215686275\\
-3	0.149803921568627\\
-2	0.137196078431372\\
-1	0.11343137254902\\
0	0.0962549019607845\\
1	0.11343137254902\\
2	0.137196078431372\\
3	0.149803921568627\\
4	0.147039215686275\\
5	0.159745098039216\\
6	0.163921568627451\\
7	0.166490196078431\\
8	0.170372549019608\\
9	0.190529411764706\\
10	0.167509803921569\\
11	0.166862745098039\\
12	0.183549019607843\\
13	0.206019607843137\\
14	0.213803921568627\\
15	0.221196078431372\\
16	0.260313725490196\\
17	0.260607843137255\\
18	0.242607843137255\\
19	0.279921568627451\\
20	0.31721568627451\\
};
\addlegendentry{PSR ($d=18.6$ cm)}

\addplot [color=mycolor6, mark=triangle, mark options={solid, rotate=180, mycolor6}]
  table[row sep=crcr]{%
-20	0.182137254901961\\
-19	0.149117647058823\\
-18	0.14521568627451\\
-17	0.167137254901961\\
-16	0.196882352941177\\
-15	0.200588235294118\\
-14	0.203470588235294\\
-13	0.174176470588235\\
-12	0.204666666666667\\
-11	0.191352941176471\\
-10	0.179803921568628\\
-9	0.177686274509804\\
-8	0.15743137254902\\
-7	0.151117647058824\\
-6	0.132686274509804\\
-5	0.125039215686274\\
-4	0.13278431372549\\
-3	0.147666666666667\\
-2	0.110725490196078\\
-1	0.124352941176471\\
0	0.109843137254902\\
1	0.124352941176471\\
2	0.110725490196078\\
3	0.147666666666667\\
4	0.13278431372549\\
5	0.125039215686274\\
6	0.132686274509804\\
7	0.151117647058824\\
8	0.15743137254902\\
9	0.177686274509804\\
10	0.179803921568628\\
11	0.191352941176471\\
12	0.204666666666667\\
13	0.174176470588235\\
14	0.203470588235294\\
15	0.200588235294118\\
16	0.196882352941177\\
17	0.167137254901961\\
18	0.14521568627451\\
19	0.149117647058823\\
20	0.182137254901961\\
};
\addlegendentry{DIN ($d=20$ cm)}

\addplot [color=mycolor7, mark=diamond, mark options={solid, mycolor7}]
  table[row sep=crcr]{%
-20	0.208901960784314\\
-19	0.200333333333333\\
-18	0.180607843137255\\
-17	0.194\\
-16	0.178156862745098\\
-15	0.170901960784313\\
-14	0.192137254901961\\
-13	0.172411764705882\\
-12	0.159745098039216\\
-11	0.171450980392157\\
-10	0.176\\
-9	0.164803921568627\\
-8	0.14943137254902\\
-7	0.16143137254902\\
-6	0.146098039215686\\
-5	0.141686274509804\\
-4	0.153882352941176\\
-3	0.139137254901961\\
-2	0.131745098039216\\
-1	0.14121568627451\\
0	0.126509803921568\\
1	0.14121568627451\\
2	0.131745098039216\\
3	0.139137254901961\\
4	0.153882352941176\\
5	0.141686274509804\\
6	0.146098039215686\\
7	0.16143137254902\\
8	0.14943137254902\\
9	0.164803921568627\\
10	0.176\\
11	0.171450980392157\\
12	0.159745098039216\\
13	0.172411764705882\\
14	0.192137254901961\\
15	0.170901960784313\\
16	0.178156862745098\\
17	0.194\\
18	0.180607843137255\\
19	0.200333333333333\\
20	0.208901960784314\\
};
\addlegendentry{NOS ($d=30$ cm)}

\addplot [color=mycolor1, mark=square, mark options={solid, mycolor1}]
  table[row sep=crcr]{%
-20	0.166941176470588\\
-19	0.214352941176471\\
-18	0.208588235294117\\
-17	0.190470588235294\\
-16	0.193980392156863\\
-15	0.186647058823529\\
-14	0.187098039215686\\
-13	0.182627450980392\\
-12	0.181607843137255\\
-11	0.16878431372549\\
-10	0.167352941176471\\
-9	0.155294117647059\\
-8	0.159333333333333\\
-7	0.126666666666667\\
-6	0.144862745098039\\
-5	0.144137254901961\\
-4	0.133078431372549\\
-3	0.149274509803922\\
-2	0.142803921568627\\
-1	0.138078431372549\\
0	0.124882352941177\\
1	0.138078431372549\\
2	0.142803921568627\\
3	0.149274509803922\\
4	0.133078431372549\\
5	0.144137254901961\\
6	0.144862745098039\\
7	0.126666666666667\\
8	0.159333333333333\\
9	0.155294117647059\\
10	0.167352941176471\\
11	0.16878431372549\\
12	0.181607843137255\\
13	0.182627450980392\\
14	0.187098039215686\\
15	0.186647058823529\\
16	0.193980392156863\\
17	0.190470588235294\\
18	0.208588235294117\\
19	0.214352941176471\\
20	0.166941176470588\\
};
\addlegendentry{PSR ($d=37.2$ cm)}

\end{axis}
\end{tikzpicture}%
    \caption{Localization uncertainty excursion across $\theta_s$ angles calculated as the Difference between maximum and minimum localization uncertainty.}
    \label{fig:latdisplvar}
    \end{subfigure}
\caption{Comparison of different PSR versions with popular stereophonic microphone arrangements.}
\label{fig:comparison}
\end{figure}

\figurename~\ref{fig:comparison} shows the localization uncertainty as a function of lateral displacement $x$. 
\figurename~\ref{fig:latdispl} shows the average across source angles, while \figurename~\ref{fig:latdisplvar} shows the excursion across source angles, defined as the difference between the largest and the smallest localization uncertainty across source angles.
% In general, a small excursion is 
% Depending on the/
A number of observations can be made:
\begin{itemize}[leftmargin=*]
\item The amplitude methods, i.e. Blumlein pair, PSR with $d=0$~cm, and cardioid XY pair (all of which are indicated by a $\times$ marker) have very similar performance; the difference between them stems from the slightly different distribution of $\icld$s across source angles $\theta_s$.
\item In comparison to time-amplitude methods, the average uncertainty of the amplitude methods is lower for $|x|<2$~cm but higher beyond $|x|>10$~cm.
\item The uncertainty excursion of the amplitude methods is significantly higher than  time-amplitude methods beyond $|x|>10$~cm, reaching almost $0.5$ at $|x|=20$~cm.
This is associated to the difficulty in rendering sound sources associated to direction opposite to the lateral movement of the listener, as discussed in Section~\ref{sec:amplitude}, and it means that that sound sources at different angles will be perceived differently.
\item In the center ($x=0$~cm), the average uncertainty is lowest for the amplitude methods, and then, in order, for PSR ($d=18.6$~cm), ORTF ($d=17$~cm), DIN ($d=20$~cm), PSR ($d=37.2$~cm) and NOS ($d=30$~cm). PSR ($d=18.6$~cm) is lowest among time-amplitude methods because it is the only one that does not cross the $\pm\tau_o$ threshold (see \figurename~ \ref{fig:hictdicldcentre}).
\item The PSR with $d=18.6$~cm has a reasonably low average uncertainty across the lateral displacement, a low excursion within $|x|<15$~cm, but a higher excursion compared to other time-amplitude methods beyond $|x|>15$~cm.
\item The NOS and PSR version with $d=37.2$~cm have a higher average uncertainty in the center, but a smaller variation of the average uncertainty across $x$, and a low overall excursion. 
\item The PSR ($d=18.6$~cm), ORTF ($d=17$~cm), DIN ($d=20$~cm), and NOS ($d=30$~cm) all have low excursion, with little discernible difference between them across $x$. 
\end{itemize}

Overall, these results further confirm that amplitude methods have a better performance at the center of the sweet-spot, while time-amplitude methods have a lower overall uncertainty and a lower excursion away from the sweet-spot. 
A lower excursion means that whenever multiple sound sources are present at different angles, they are rendered with a similar localization uncertainty, which is an appealing property. 

A further advantage of \gls{psr} over the other time-amplitude methods, is that it has been shown to have a high localization \emph{accuracy}, meaning that subjects would perceive sound sources in directions close to the intended $\theta_s$ ~\cite{desena2013}.
This is also related to the fact that in PSR the coverage angle is identical to the loudspeaker base-angle, while the other methods will have (at least) a compression/decompression of angles from the coverage angle to the loudspeaker base-angle. 
% The authors are not aware of listening experiments formally assessing the localization 
An additional advantage of having the coverage angle identical to the loudspeaker base-angle (and microphone base-angle) is that the extension to the multichannel case is straightforward. 
% The other 

\subsection{Comments on the relation to multichannel PSR}
The PSR stereophonic setup considered thus far can be seen as a subset of a multichannel circular array. 
In~\cite{desena2013}, it was shown that having only the two loudspeakers closest to $\theta_s$ as active yields a lower spatial fluctuation of the active intensity vector field and thus a larger sweet spot. 
This turns the overall multichannel design problem into a number of stereophonic problems. 
% Since the coverage angle is identical to the microphone base-angle, adjacent stereophonic pairs can easily cover adjacent and non-overlapping coverage angle. 

Consider the case of $M$ microphones distributed uniformly around the circle with $\phi_m=\phi_0=2\pi/M$.
Each of these microphones is connected to a loudspeaker in the same angular configuration, without mixing.
Johnston and Lam proposed in \cite{Johnston:2000oe} a microphone array with $M=5$, resulting in a base angle of $\phi_0=2\pi/5=72^\circ$, and radius $r_m=15.5$~cm. 
This array radius is very close to the value needed to avoid \gls{ictd} larger than $\tau_o=\pm0.3$~ms. 
Indeed, using (\ref{eq:drm}) and (\ref{eq:optrm}) for $\phi_0=2\pi/5=72^\circ$ yields $r_m=16.2$~cm. % (notice that the difference with $d=0.187$~m used elsewhere in this paper is due to the different base angles). 
%\footnote{If one assumes $r_h=0.0865$~m instead of $r_h=0.09$~m, the  value of $r_m$ is exactly $0.155$~m.}
Johnston and Lam stated that $r_m=15.5$~cm would conserve \gls{itd} cues that the listener would have experienced in the recording space, but without providing a rigorous explanation (it could indeed be argued, as Bernfeld did in~\cite{Bernfeld:1973wd}, that correct \gls{itd} cues can also be achieved by using \glspl{icld} alone).
This paper shows that Johnston and Lam's choice of array radius is the one that leads to a good compromise between low localization uncertainty in central and non-central listener positions.
The same choice was made in~\cite{desena2013}, where the directional accuracy was also improved significantly. 

% This paper provides a rigorous explanation as to why this choice is indeed optimal.
% This paper justifies a posteriori the choice as the one that leads to a compromise of localization uncertainty between central and non-central listener positions.

% Consider now the stereophonic setup as a subset of the uniform circular array with $N$ microphones and inter-microphone separation of $2\pi/N$.
% As $N$ increases, $\phi_0$ decreases and the optimal array radius increase.
% Although in the limit $N\rightarrow \infty$ one should converge to WFS, and thus $r_l=r_m$, this is not what happens according to \ref{eq:optrm}.

\section{Summary and Conclusions}
\label{sec:conclusions}
% A computational model for the prediction 
% of subjective locatedness judgements was proposed. 
% A novel way of jointly using the \gls{ild} and \gls{itd} cues was suggested. 
% The deviation between the natural \gls{ild}-\gls{itd} pairs 
% and the measured ones was calculated as the distance associated to the $0.5$-norm, 
% and this choice was motivated on psychoacoustic grounds. 
% On the basis of this distance, a score function was calculated, 
% which was interpreted as the likelihood that an 
% auditory event is perceived in a given direction. 
% The locatedness prediction was then calculated 
% as the entropy of the normalised score function.
% The model was validated by comparison with 
% the results of the locatedness test 
% presented in Section~\ref{sec:locatedtest}.
% It was found that the correlation between the model 
% predictions and the experimental data 
% was very strong (0.94).

Two-channel stereophony remains to this day the most commonly used audio reproduction system.
% The paper analysed the localization uncertainty associated to  two-channel stereophony, which remains the most common reproduction system to this day.
% the most common reproduction system in use today remains the two-channel stereophonic system.
% More specifically, it 
This paper focused on the effect of inter-channel time and level differences on perceived localization uncertainty. 
A computational model was proposed, based on calculating a distance functional between the observed \gls{ild}-\gls{itd} values and the ones associated to free-field sound sources.
The distance functional was chosen as the $0.7$-norm which is capable of modelling the splitting of auditory events observed experimentally in case of contradicting \gls{ild}-\gls{itd} cues.
The model predictions had a high correlation with results of formal listening experiments.
The model can also be used to predict the perceived localization angle, but this is left to future research. %goes beyond the scope of this paper and

The model was then used to predict the localization uncertainty under stereophonic reproduction.
It was found that contradicting \gls{ictd}-\gls{icld} pairs were associated to a high localization uncertainty. % and (b) \glspl{ictd} resulting in the loudspeaker signals overlapping in time at either ear. 
Closed form approximations of the \glspl{ictd} and \glspl{icld} relative to the listener (denoted as \gls{rictd} and \gls{ricld}) were presented.
% It was observed that \gls{rictd} changes are perceptually dominant with respect to changes of \gls{ricld}, especially for larger loudspeaker arrays.
% This meant that, when a listener moves to the side, the localization uncertainty profiles would shift 
It was observed that when a listener moves away from the center of the sweet-spot, the \gls{ricld} remains largely unchanged, while the \gls{rictd} changes even for small displacements.
In off-center positions, then, one can obtain contradicting \gls{ricld}-\gls{rictd} pairs even if the original \gls{icld}-\gls{ictd} pairs were not.
It was then explained that non-zero \glspl{ictd} (i.e. time-amplitude methods) will delay the onset of contradicting \gls{ricld} and \gls{rictd}. 
More significant use of \glspl{ictd} results in a higher uncertainty in the center of the sweet-spot, but a relatively lower one in off-center positions. 
A comparison of amplitude methods (Blumlein, XY) and time-amplitude methods (\gls{psr}, ORTF, DIN and NOS) confirms that time-amplitude methods have a lower localization uncertainty in off-center positions, and that the localization uncertainty has a lower variability as a function of source angle. 

These results reveal that near-coincident microphone pairs, which are often used for their sense of spaciousness deriving from lower inter-channel decorrelation~\cite{rumsey2001spatial}, also have an advantage in terms of localization uncertainty.
% This result may help reconcile the long-standing differences between the audio researchers, who generally tend to adopt amplitude methods (e.g. \gls{vbap}, Ambisonics, SDM) based largely on physically-motivated arguments, and the audio recording practitioners, who in many cases are likely to prefer time-amplitude methods (e.g. ORTF~\cite{eargle2004microphone,rumsey2001spatial}) over amplitude methods, based on evaluative listening. 

In this paper it was not possible to analyze widely spaced microphone arrays such as the ones criticized by  Lipshitz~\cite{lipshitz1986stereo} or time panning as criticized by Lee and Rumsey~\cite{lee2013level}, since they exceed the $1$~ms \gls{ictd} limit beyond which one has to also model the law of the first front. 
This paper is therefore not in contradiction with the findings in~\cite{lipshitz1986stereo} and \cite{lee2013level}, as the conclusions made here are limited to coincident and near-coincident microphone arrays. 
The computational model could be modified to account for the law of the first front, e.g. by means of inhibitions mechanisms~\cite{lindemann1986extension}, thus allowing analysis of widely spaced microphone arrays. 
% This is a matter for future work.

Throughout this paper, it was implicitly assumed that a low localization uncertainty is desirable. 
This is motivated by the fact that an actual plane wave is well localized spatially, and that 
to render it accurately it needs to have a low localization uncertainty. 
% It can be posited that 
% If a soundfield can be expressed as a summation of plane waves (e.g. room reflections in the far field) and each is rendered correctly, then, the sense of envelopment in the target soundfield can also be recreated.
This may not be necessarily the case e.g. in an artistic context, where one may prefer to render sound sources as difficult to localize.
The results in this paper are useful in this context too as they can also be read with an implicit preference for high uncertainty. 
% This paper is also useful within this context. %as results can also be read with a preference for high uncertainty. 
% The results in this paper are also useful with this objective in mind.

% Future work will involve finding optimal \gls{icld}-\gls{ictd} pairs as a function of frequency. 
% This will bring it closer to 

\begin{appendices}

\section{}
\label{sec:appoverlap}
This appendix provides a closed-form approximation of the \gls{ictd} that results in the two loudspeaker signals to arrive at the same time at the left ear (the corresponding case for the right ear will follow).
Let the head radius be denoted by $r_h$ and the angle between the forward-looking direction and the ear  be denoted by $\theta_e$.
Let $\theta_{0}=\cos^{-1}\left(\frac{r_h}{r_l}\right)$ be the angle of tangential incidence on the spherical head. 
Assuming that $\left|\frac{r_h}{r_l}\right|<<1$, i.e. the head radius is much smaller than the loudspeaker distance, then $\theta_0\approx \frac{\pi}{2}$.

The angle between the left loudspeaker and the left ear is $\theta_{LL}=\theta_e-\frac{\phi_0}{2}$ 
and under most conditions, $\theta_{LL}<\theta_{0}$. 
The distance between the left loudspeaker and the left ear is then:
\begin{align}
    d_{LL}&=\sqrt{r_l^2+r_h^2-2rr_h\cos(\theta_{LL})}~\\
    &=r_l\sqrt{1+\frac{r_h^2}{r_l^2}-2\frac{r_h}{r_l}\cos\left(\theta_e-\frac{\phi_0}{2}\right)}\\
    &\approx r_l-r_h\cos\left(\theta_e-\frac{\phi_0}{2}\right)~,
    \label{eq:approxdll}
\end{align}
where in the last step, the quadratic term $\frac{r^2_h}{r^2_l}$ is ignored and the square root is approximated using the first-order Taylor series approximation.

The angle between the right loudspeaker and the left ear is $\theta_{RL}=\frac{\phi_0}{2}+\theta_e$ 
and under most conditions, $\theta_{RL}>\theta_{0}$. 
The distance between the right loudspeaker and the left ear, then, is given by the summation of the distance between the loudspeaker and the point of tangential incidence, and the remaining angular sector:
\begin{equation}
    d_{RL}=\sqrt{r_l^2+r_h^2}+r_h(\theta_{RL}-\theta_{0})~.
\end{equation}
For small $\frac{r_h}{r_l}$, the quadratic term $(\frac{r_h}{r_l})^2$ can be ignored:
\begin{equation}
    d_{RL}\approx r_l+r_h\left(\frac{\phi_0}{2}+\theta_e-\frac{\pi}{2}\right)~.
    \label{eq:approxdrl}
\end{equation}

The delay applied to the left loudspeaker, $\tau_o$, that makes the two loudspeaker signals arrive at the same time at the left ear satisfies the equation
$
    \frac{d_{LL}}{c} + \tau_o=\frac{d_{RL}}{c}~.
$
Replacing equations (\ref{eq:approxdll}) and (\ref{eq:approxdrl}), and solving for $\tau_o$ yields
\begin{equation}
    \tau_o\approx \frac{r_h}{c}\left[\cos\left(\theta_e-\frac{\phi_0}{2}\right)+\frac{\phi_0}{2}+\theta_e-\frac{\pi}{2}\right]~.
    \label{eq:tauo}
\end{equation}

Due to the symmetry of the problem, the $\ictd$ that results in the loudspeaker signals arriving at the same time at the \emph{right} ear is $-\tau_o$.
% Due to the symmetry of the problem, if one applies the same delay $\tau_o$ multiplied by $-1$, then the loudspeaker signals will arrive at the same time at the right ear. 
% Since the delays are applied to two different loudspeakers, their corresponding \gls{ictd} values have opposite sign.

\section{}
\label{sec:rictld}
This appendix provides a closed-form approximation of the \gls{rictd} and \gls{ricld}. 
Towards this end, approximations of the relative distance between an observation point $(x,y)$ and the two loudspeakers at a distance $r_l$ from the center of the sweet-spot $(0,0)$ are sought first. 
It will be assumed  that $\left|x/r_l\right|<<1$ and $\left|y/r_l\right|<<1$.
The distance between the observation point $(x,y)$ and the right loudspeaker can be approximated as 
\begin{align}
    d_R^2&=\left[r_l\sin\left(\frac{\phi_0}{2}\right)-x\right]^2+\left[r_l\cos\left(\frac{\phi_0}{2}\right)-y\right]^2\\
%         &=d^2\left[\sin\left(\frac{\phi_0}{2}\right)-\left(\frac{x}{r_l}\right)^2+\cos\left(\frac{\phi_0}{2}\right)-\left(\frac{y}{r_l}\right)^2\right]\\
         &\approx r_l^2\left[1-2\frac{x}{r_l}\sin\left(\frac{\phi_0}{2}\right)-2\frac{y}{r_l}\cos\left(\frac{\phi_0}{2}\right)\right]~,
\end{align}
where in the second step it is assumed that $\left(x/r_l\right)^2\approx 0$, and $\left(y/r_l\right)^2\approx 0$.
It is useful to define $a=2\frac{x}{r_l}\sin\left(\phi_0/2\right)$ and $b=2\frac{y}{r_l}\cos\left(\phi_0/2\right)$, so that $d_R$ can be rewritten as
$d_R\approx r_l\sqrt{1-(a+b)}$.
The first-order Taylor series approximation of the square root yields:
\begin{equation}
    d_R\approx r_l\left(1-\frac{a+b}{2}\right)
    \label{eq:dr}
\end{equation}

Using similar steps, $d_L$ can be approximated by
$
    d_L\approx  r_l\left[1+(a-b)/2\right]~.
$
The time delay between the two loudspeakers observed in $(x,y)$ can now be approximated as
\begin{equation}
    \frac{d_R-d_L}{c}\approx -\frac{r_l a}{c}=-x\frac{2}{c} \sin\left(\frac{\phi_0}{2}\right)~.\label{eq:deltaT}
\end{equation}
% Notice how the dependence on the $y$-displacement is not present in the first-order approximation (it does appear in the second-order term).

The relative level observed in $(x,y)$ for the two point-like loudspeakers can now be approximated as 
\begin{equation}
    20\log_{10}\left(\frac{\frac{1}{d_L}}{\frac{1}{d_R}}\right)
    \approx 20\log_{10}\frac{1-\frac{x}{2r_l}\sin\left(\frac{\phi_0}{2}\right)-\frac{y}{2r_l}\cos\left(\frac{\phi_0}{2}\right)}{1+\frac{x}{2r_l}\sin\left(\frac{\phi_0}{2}\right)-\frac{y}{2r_l}\cos\left(\frac{\phi_0}{2}\right)}~,
\end{equation}
and, using a first-order Taylor series approximation:
\begin{equation}
    20\log_{10}\left(\frac{\frac{1}{d_L}}{\frac{1}{d_R}}\right)\approx -\frac{x}{r_l} \frac{20}{\log_e(10)} \sin\left(\frac{\phi_0}{2}\right)~,\label{eq:deltaL}
\end{equation}
% Having assumed that $\frac{y}{r_l}<<1$, and observing that $\frac{y}{r_l}\cos\left(\frac{\phi_0}{2}\right)$ has the same sign at numerator and denominator, its contribution is ignored in the following:
% \begin{align}
%     \text{\gls{ricld}}&\approx 20\log_{10}\left(\frac{1+\frac{x}{r_l}\sin\left(\frac{\phi_0}{2}\right)}{1-\frac{x}{r_l}\sin\left(\frac{\phi_0}{2}\right)}\right)\\
%     &=\frac{20}{\log_e(10)}\log_{e}\left(\frac{1+\frac{x}{r_l}\sin\left(\frac{\phi_0}{2}\right)}{1-\frac{x}{r_l}\sin\left(\frac{\phi_0}{2}\right)}\right)\\
%     &\approx \frac{x}{r_l} \frac{20}{\log_e(10)} \sin\left(\frac{\phi_0}{2}\right)~.
% \end{align}
% Notice how the dependency on the $y$-displacement is not present in the  first-order approximation (it does appear in the second-order term).
Using (\ref{eq:deltaT}) and (\ref{eq:deltaL}), one obtains $\ricld \approx \icld - \frac{x}{r_l}$ and $\rictd \approx \ictd-x\frac{2}{c}\sin\left(\phi_0/2\right)$.
% \begin{align}
% \begin{split}
% \ricld & \approx \icld - \frac{x}{r_l} \frac{20\sin\left(\frac{\phi_0}{2}\right)}{\log_e(10)}~, \\
% \rictd & \approx \ictd-x\frac{2}{c}\sin\left(\frac{\phi_0}{2}\right)~.
% \end{split}
% \end{align}

\end{appendices}

\section*{Acknowledgments}
The authors would like to thank Niccol\`{o} Antonello for pointing out that the proposed computational model has a statistical interpretation. 

\bibliographystyle{IEEEtran}
\bibliography{references}

\begin{IEEEbiography}[{\includegraphics[width=1in,height=1in,clip,keepaspectratio]{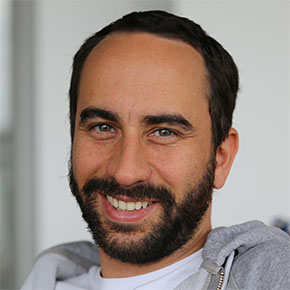}}]{Enzo De Sena}
(S'11-M'14-SM'19) is a Lecturer in Audio at the Institute of Sound Recording at the University of Surrey. He received the M.Sc. degree (cum laude) in 2009 from the Università degli Studi di Napoli ``Federico II" in Telecommunication Engineering and the Ph.D. degree in Electronic Engineering in 2013 from King’s College London. Between 2012 and 2013 he was a Teaching Fellow at King's College London. Between 2013 and 2016 he was a Postdoctoral Research Fellow at the Department of Electrical Engineering at KU Leuven. He held visiting  positions at Stanford University (2013), Aalborg University (2014-2015), Imperial College London (2016), and King's College London (2018-current). He is a former Marie Curie Fellow. His current research interests include room acoustics modelling, multichannel audio, microphone beam forming, and binaural modelling. For more information see desena.org.
\end{IEEEbiography}
\vskip -2\baselineskip plus -1fil
\begin{IEEEbiography}[{\includegraphics[width=1in,height=1.25in,clip,keepaspectratio]{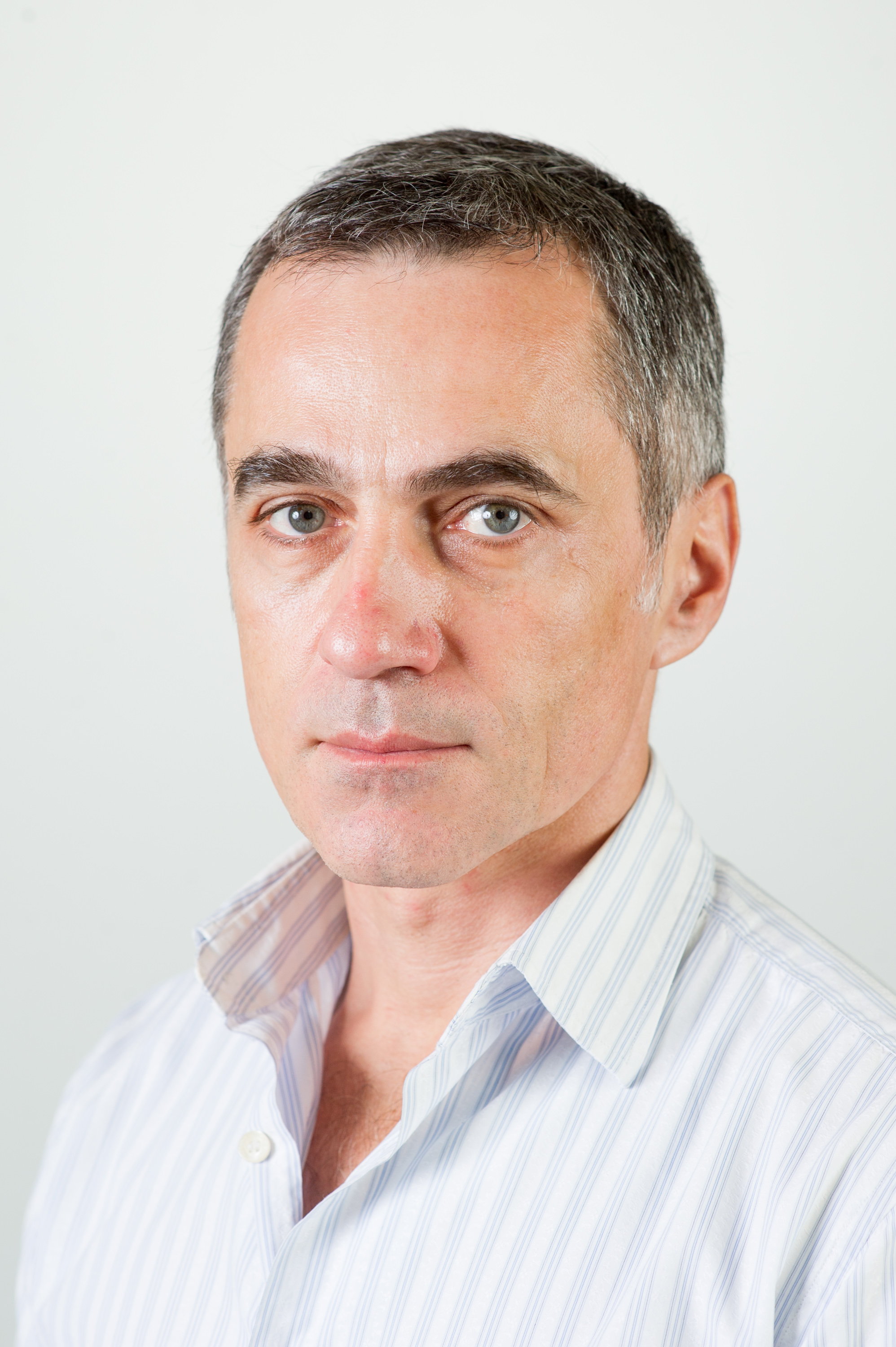}}]{Zoran Cvetkovi\'c} (SM'04) is Professor of Signal Processing at King’s College London. He received his Dipl. Ing. and Mag. degrees from the University of Belgrade, Yugoslavia, the M.Phil. from Columbia University, and the Ph.D. in electrical engineering from the University of California, Berkeley. He held research positions at EPFL, Lausanne, Switzerland (1996), and at Harvard University (2002-04). Between 1997 and 2002, he was a member of the technical staff of AT\&T Shannon Laboratory. His research interests are in the broad area of signal processing, ranging from theoretical aspects of signal analysis to applications in audio and speech technology, and neuroscience. He served as an Associate Editor of IEEE Transactions on Signal Processing. 
\end{IEEEbiography}
\vskip -2\baselineskip plus -1fil
\begin{IEEEbiography}[{\includegraphics[width=1in,height=1.25in,clip,keepaspectratio]{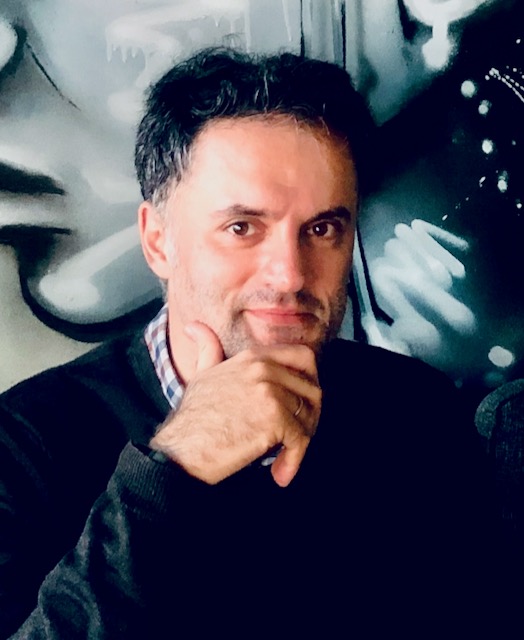}}]{H\"{u}seyin Hac\i{}habibo\u{g}lu} (S'96-M'00-SM'12) is an Associate Professor of Signal Processing at Graduate School of Informatics, Middle East Technical University, Ankara, Turkey. He received the B.Sc. (honors) degree from the Middle East Technical University (METU), Ankara, Turkey, in 2000, the M.Sc. degree from the University of Bristol, Bristol, U.K., in 2001, both in electrical and electronic engineering, and the Ph.D. degree in computer science from Queen's University Belfast, Belfast, U.K., in 2004. He held research positions at University of Surrey, Guildford, U.K.\ (2004-2008) and King's College London, London, U.K.\ (2008-2011). His research interests include audio signal processing, room acoustics, multichannel audio systems, psychoacoustics of spatial hearing, microphone arrays, and game audio. Dr.\ Hac\i{}habibo\u{g}lu is a Senior Member of the IEEE, Member of the IEEE Signal Processing Society, Member of UKRI International Development Peer Review College, Audio Engineering Society (AES), Turkish Acoustics Society (TAD), and the European Acoustics Association (EAA) and an associate editor of \textit{IEEE/ACM Transactions on Audio, Speech, and Language Processing}. 
\end{IEEEbiography}
\vskip -2\baselineskip plus -1fil
\begin{IEEEbiography}[{\includegraphics[width=1in,height=1.25in,clip,keepaspectratio]{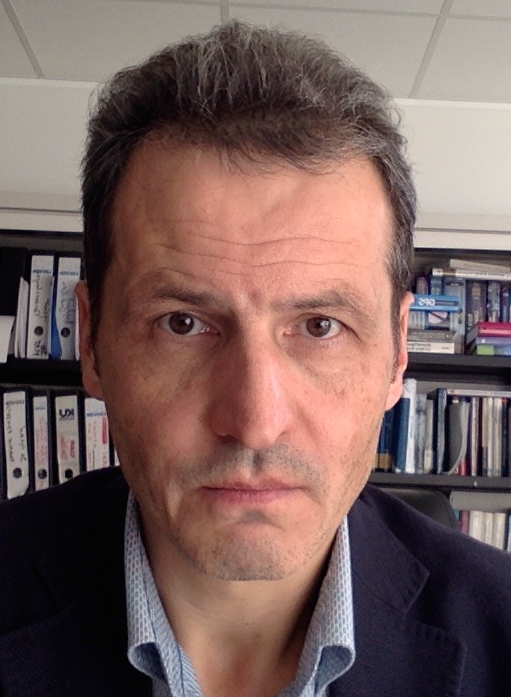}}]{Marc Moonen}
 (M'94, SM'06, F'07) is a Full Professor at the Electrical Engineering
Department of KU Leuven, where he is heading a research team working in the area of numerical algorithms and
signal processing for digital communications, wireless communications,
DSL and audio signal processing.

He is a Fellow of the IEEE (2007) and a Fellow of EURASIP (2018).
He received the 1994 KU Leuven Research Council Award, the 1997 Alcatel
Bell (Belgium) Award (with Piet Vandaele), the 2004 Alcatel Bell
(Belgium) Award (with Raphael Cendrillon), and was a 1997 Laureate of
the Belgium Royal Academy of Science. He received journal best paper
awards from the IEEE Transactions on Signal Processing (with Geert Leus and with Daniele Giacobello)
and from Elsevier Signal Processing (with Simon Doclo).

He was chairman of the IEEE Benelux Signal Processing Chapter
(1998-2002), a member of the IEEE Signal Processing Society Technical Committee on Signal Processing for Communications, and President of EURASIP (European Association for Signal Processing, 2007-2008 and 2011-2012).
He has served as Editor-in-Chief for the EURASIP Journal on Applied
Signal Processing (2003-2005), Area Editor for Feature Articles in IEEE Signal Processing Magazine (2012-2014), and has been a member of the editorial board of Signal Processing, IEEE Transactions on Circuits and Systems II, IEEE Signal Processing Magazine, Integration-the VLSI Journal, EURASIP Journal on Wireless Communications and Networking and EURASIP Journal on Advances in Signal Processing.
\end{IEEEbiography}
\vskip -2\baselineskip plus -1fil
\begin{IEEEbiography}[{\includegraphics[width=1in,height=1.25in,clip,keepaspectratio]{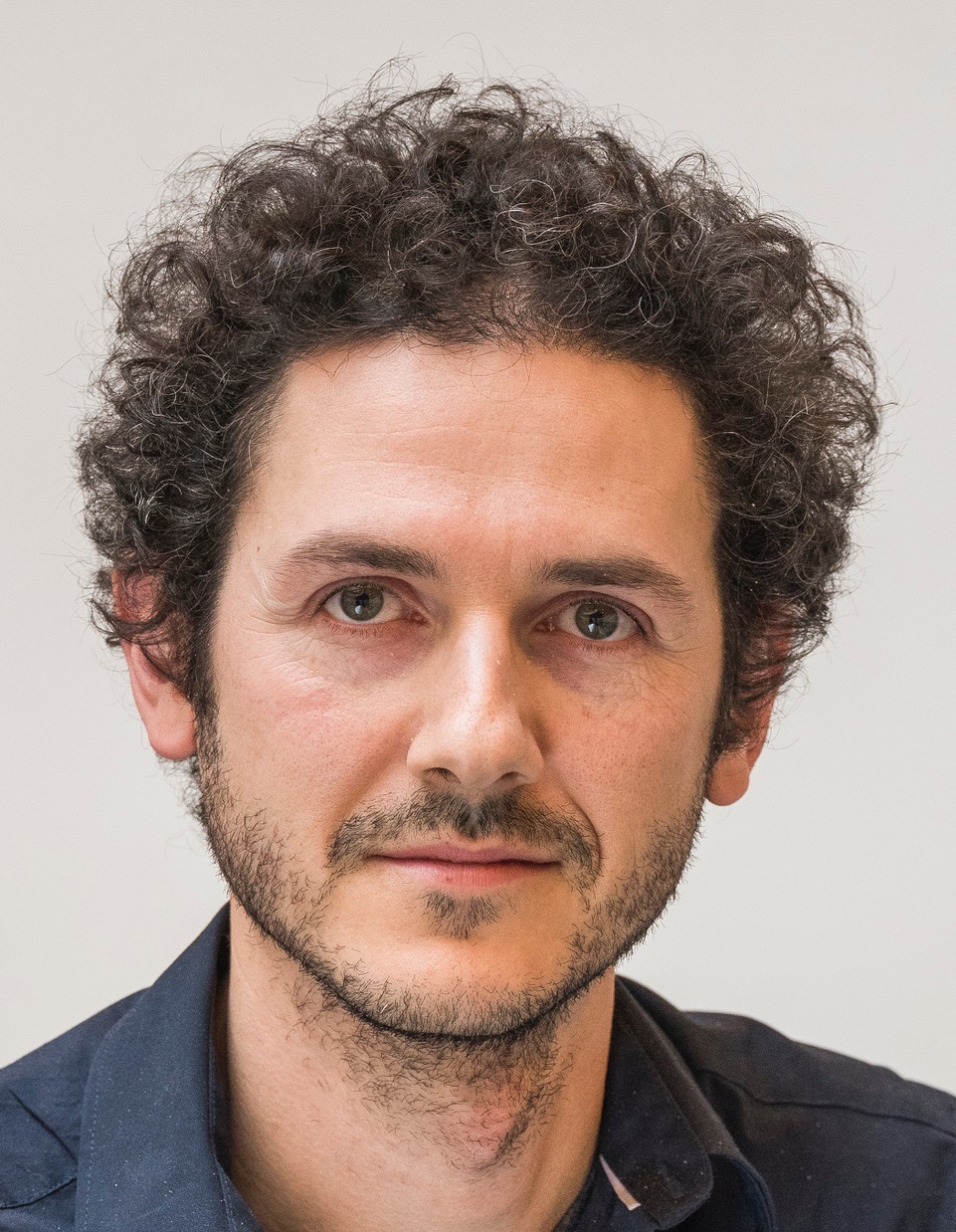}}]{Toon van Waterschoot}(S'04, M'12) received MSc (2001) and PhD (2009) degrees in Electrical Engineering, both from KU Leuven, Belgium, where he is currently an Associate Professor and Consolidator Grantee of the European Research Council (ERC). He has previously also held teaching and research positions at Delft University of Technology in The Netherlands and the University of Lugano in Switzerland. His research interests are in signal processing, machine learning, and numerical optimization, applied to acoustic signal enhancement, acoustic modeling, audio analysis, and audio reproduction. 

He has been serving as an Associate Editor for the Journal of the Audio Engineering Society and for the EURASIP Journal on Audio, Music, and Speech Processing, and as a Guest Editor for Elsevier Signal Processing. He is a Director of the European Association for Signal Processing (EURASIP), a Member of the IEEE Audio and Acoustic Signal Processing Technical Committee, a Member of the EURASIP Special Area Team on Acoustic, Speech and Music Signal Processing, and a Founding Member of the EAA Technical Committee in Audio Signal Processing. He was the General Chair of the 60th AES International Conference in Leuven, Belgium (2016), and has been serving on the Organizing Committee of the European Conference on Computational Optimization (EUCCO 2016), the IEEE Workshop on Applications of Signal Processing to Audio and Acoustics (WASPAA 2017), and the 28th European Signal Processing Conference (EUSIPCO 2020). He is a member of EURASIP, IEEE, ASA, and AES.
\end{IEEEbiography}

\end{document}